\documentclass[11pt]{article}
\usepackage{amssymb,amsmath}
\usepackage{bm} 
\usepackage{booktabs} 
\usepackage{array}
\usepackage{latexsym}
\usepackage{graphicx}
\usepackage{color}
\usepackage{datetime}
\usepackage[nosort]{cite}
\usepackage{verbatim}
\usepackage{enumerate}
\usepackage{chngpage} 
\usepackage{mathrsfs}
\usepackage{euscript}
\usepackage{psfrag}
\usepackage{subfig}
\usepackage{anyfontsize}
\usepackage{braket}

\usepackage{blindtext}%
\usepackage[theorems,breakable]{tcolorbox}%

\usepackage{multirow,graphicx}
\usepackage{makecell}
\usepackage{float}
\usepackage{tikz}
\usetikzlibrary{decorations.text}
\usepackage{enumitem}
\setlist{itemsep=0pt}
\usetikzlibrary{math}
\usepackage{setspace}

\usepackage{mciteplus}

\usepackage[colorlinks=true,      linkcolor=blue,      urlcolor=blue,      
            filecolor=blue,      citecolor=blue,       pdfstartview=FitH,     
						pdfpagemode=UseNone,      bookmarksopen=true]{hyperref}  
\usepackage[all]{hypcap}     

\usepackage{tikz}
\usepackage{pgfplots}
\usetikzlibrary{decorations.text,intersections, pgfplots.fillbetween}
\usetikzlibrary{math}
 \usetikzlibrary{3d,shapes.geometric,shadows.blur}
\usepackage{tikz-3dplot}

\usepackage{blindtext}%
\usepackage[theorems,breakable]{tcolorbox}%

\newtcbtheorem{mytheo}{}%
{colback=gray!5,colframe=gray!35!black,fonttitle=\bfseries}{th}

\newtcbtheorem{lem}{}{%
        colback=green!5,%
        colframe=green!35!black,%
        fonttitle=\bfseries,title %
    }{lem}

\definecolor{amaranthred}{rgb}{0.83,0.13,0.18}
\definecolor{amazon}{rgb}{0.23,0.48,0.34}
\definecolor{bdazzledblue}{rgb}{0.18,0.35,0.58}
\definecolor{absolutezero}{rgb}{0.0,0.28,0.73}
\definecolor{bitterlemon}{rgb}{0.79,0.88,0.05}
\definecolor{byzantine}{rgb}{0.74,0.2,0.64}
\definecolor{turquoise}{rgb}{0.19, 0.84, 0.78}
\definecolor{burgundy}{rgb}{0.5, 0.0, 0.13}
\definecolor{airforceblue}{rgb}{0.36, 0.54, 0.66}
\definecolor{arsenic}{rgb}{0.23, 0.27, 0.29}

\newcommand{\comm}[1]{} 

\def\({\left(}
\def\){\right)}
\def\[{\left[}
\def\]{\right]}

\def\One{{\hbox{ 1\kern-.8mm l}}}

\def\barray{\begin{array}}
\def\earray{\end{array}}
\def\be{\begin{equation}}
\def\ee{\end{equation}}
\def\bea{\begin{eqnarray}}
\def\eea{\end{eqnarray}}
\def\bal{\begin{align}}
\def\eal{\end{align}}
\def\nn{\nonumber}

\def\-{\,-\,}
\def\={\,=\,}
\def\+{\,+\,}
\def\equi{\,\equiv\,}

\numberwithin{equation}{section} 

\definecolor{cardinal}{rgb}{0.6,0,0}
\definecolor{darkgreen}{rgb}{0,0.4,0}
\definecolor{golden}{rgb}{0.92, 0.7, 0}
\definecolor{midnight}{rgb}{0, 0, 0.5}
\definecolor{darkblue}{rgb}{0, 0, 0.7}
\definecolor{purple}{rgb}{0.5, 0, 0.5}

\definecolor{amaranthred}{rgb}{0.83,0.13,0.18}
\definecolor{amazon}{rgb}{0.23,0.48,0.34}
\definecolor{bdazzledblue}{rgb}{0.18,0.35,0.58}
\definecolor{absolutezero}{rgb}{0.0,0.28,0.73}
\definecolor{bitterlemon}{rgb}{0.79,0.88,0.05}
\definecolor{byzantine}{rgb}{0.74,0.2,0.64}
\definecolor{turquoise}{rgb}{0.19, 0.84, 0.78}
\definecolor{burgundy}{rgb}{0.5, 0.0, 0.13}

\def\IR{\mathbb{R}}

\def\cA{{\cal A}}

\def\cF{{\cal F}}

\def\cJ{{\cal J}}

\def\cO{{\cal O}}

\def\cT{{\cal T}}

\usetikzlibrary{positioning}

\begin{document}

\phantom{AAA}
\vspace{-10mm}

\begin{flushright}
%
%
\end{flushright}

\vspace{2cm}

\begin{center}

{\fontsize{19}{23}\selectfont{\bf Building the Blocks of Schwarzschild }
}

\vspace{1.5cm}

{\large{\bf {Rapha\"el Dulac$^{1}$ and Pierre Heidmann$^{2,3}$}}}

\vspace{7mm}

\centerline{$^1$ Ecole Normale Superieure,}
\centerline{45 Rue d'Ulm, 75005 Paris, France.}
\vspace{0.2cm}
\centerline{$^2$ Department of Physics,}
\centerline{$^3$ Center for Cosmology and AstroParticle Physics (CCAPP),}
\centerline{The Ohio State University,}
\centerline{191 W Woodruff Ave, Columbus, OH 43210, USA.}

\vspace{7mm} 

{\footnotesize\upshape\ttfamily heidmann.5@osu.edu,~ raphael.dulac@ens.fr } 

\vspace{15mm}
 
\textsc{Abstract}

\end{center}

\begin{adjustwidth}{5mm}{5mm} 
 
\noindent

\noindent

We demonstrate that the Schwarzschild black hole can be ``resolved'' into bound states of Reissner-Nordström black holes in four dimensions. These bound states closely resemble the Schwarzschild geometry from the asymptotic region up to an infinitesimal distance away from the Schwarzschild horizon. Below this scale, the horizon is replaced by novel spacetime structures supported by intense and entrapped electromagnetic flux.  The flux originates from collinear black holes that can be brought arbitrarily close to extremality. We find that the charge distribution follows a universal pattern,  with magnitudes scaling proportionally to the total mass and alternating in sign. Moreover, the bound states always have an entropy that constitutes a fraction of the Schwarzschild entropy.  Constructed in four dimensions, the black holes are kept apart by struts, for which we analyze tensions and energies.  These solutions pave the way for analogous constructions in supergravity and for a brane/anti-brane description of the Schwarzschild black hole in string theory.

\bigskip

\end{adjustwidth}

\vspace{8mm}
 

\thispagestyle{empty}

\newpage



\tableofcontents

\newpage

\section{Introduction}

Black holes inevitably break down matter into its most fundamental components, whether viewed in general relativity or as a strongly-coupled quantum entity in quantum gravity.  In string theory,  this is often described in terms of strings and branes. In the 1990s, pivotal studies provided a microscopic formulation of extremal black holes \cite{Strominger:1996sh,Sen:1995in,Maldacena:1997de,Dijkgraaf:1996cv}.  This involved counting brane configurations with identical charges to these black holes at weak coupling,  where gravity is turned off,  and relying on supersymmetry to extrapolate to the semiclassical regime where the brane configurations backreact and are described by the black hole geometry.

Despite these advancements, two significant questions persist. First, the fate of individual microstates upon transitioning to the black hole regime remains unclear,  necessitating an exact description of brane configurations at strong coupling.  While not all states are expected to admit a supergravity description,  the microstate geometry program \cite{Mathur:2005zp,Bena:2022ldq,Bena:2022rna} have made significant progress in interpreting the relevant degrees of freedom in supergravity to provide description of some atypical states.  These are brane bound states forming smooth horizonless geometries. In this frame, branes correspond to gravitational sources carrying electromagnetic charges,  and the smooth geometries resolve the horizon into novel topological structures supported by electromagnetic flux,  thereby introducing new physics at the horizon scale.

The second question concerns extending the microscopic description to non-extremal black holes.  Without supersymmetry, the number of states is not protected, and strong coupling effects ruin the extrapolation from the weakly-coupled stringy regime to the semiclassical regime, where the black hole picture prevails. Nonetheless,  studies near extremality or under the assumption of non-interacting brane systems suggest a brane/anti-brane formulation of black hole entropy away from extremality \cite{Callan:1996dv,Horowitz:1996fn,Breckenridge:1996sn,Horowitz:1996ay}.  Despite these advancements, a thorough description of brane/anti-brane configurations that can fully account for the entropy of non-extremal black holes, with the Schwarzschild black hole as the quintessential example, remains elusive.

In \cite{Heidmann:2023kry,Heidmann:2023thn}, a new approach has been proposed that combines insights from both aforementioned questions. On one hand,  it uses tools from the microstate geometry program to build neutral configurations of branes and anti-branes in supergravity. These configurations form regular clusters of extremal (BPS and anti-BPS) black holes,  with zero net charge, and their entropy accounts for a fraction of the Schwarzschild entropy \cite{Heidmann:2023kry}. On the other hand, the gravitational constraints on these geometries establish a link between their entropy and the microscopic counting of states in the weakly-coupled brane and anti-brane systems.  In essence,  they provide a microscopic description of a subset of (atypical) Schwarzschild states in terms of branes and anti-branes,  while also providing an illustration of the geometries in supergravity.

Remarkably, the solutions in \cite{Heidmann:2023kry,Heidmann:2023thn}, despite corresponding to wide distributions of charged black holes, yield a spacetime that is indistinguishable from a vacuum solution. Termed ``electromagnetic entrapment'' in \cite{Heidmann:2023thn}, this phenomenon arises from the property of ultra-compact configurations of self-gravitating charges to ``entrap'' their own electromagnetic field such that they are indistinguishable from a single monopole source. This property is crucial for describing non-extremal black holes in supergravity by inducing novel brane and anti-brane structures at their horizons without affecting the spacetime outside.

However, a major drawback of the solutions presented in \cite{Heidmann:2023kry,Heidmann:2023thn} is that they are not indistinguishable from a Schwarzschild black hole but rather a distorted version,  known as the $\delta=2$ Zipoy-Vorhees spacetime \cite{zipoy1966topology,voorhees1970static,stephani2009exact,Kodama:2003ch}. To validate the approach initiated in \cite{Heidmann:2023kry,Heidmann:2023thn}, it is necessary to demonstrate that electromagnetic degrees of freedom can be activated in the near-horizon region of Schwarzschild and that the latter can be resolved into bound states of charged black holes. By ``resolve, '' we mean that the bound states are neutral and cannot be distinguished from the Schwarzschild black hole from the asymptotic region up to an infinitesimal scale away from its horizon. Then, by embedding the bound states in supergravity and smoothing out the region between the black holes, these solutions would provide a microscopic origin for the Schwarzschild black hole in terms of branes and anti-branes that localize at its  horizon, replacing it with a novel spacetime structure supported by intense electromagnetic flux.\\

In this paper, we take the first essential step towards this objective.  We demonstrate that the Schwarzschild black hole can be resolved by activating relevant electromagnetic degrees of freedom in its vicinity.  We achieve this without relying on the full supergravity framework and field content, instead using a simpler four-dimensional theory of gravity coupled with an electromagnetic U(1) gauge field.

In this framework, we construct bound states of $N$ collinear Reissner-Nordström black holes that cannot be distinguished from a Schwarzschild black hole,  except in the near-horizon region.  Our solutions represent a specific subset of the broader class of solutions derived from the Ernst formalism and the Sibgatullin method in \cite{NoraBreton1998,Manko_1993,Ruiz:1995uh}.  To ensure indistinguishability,  a significant number of black holes is required, such that deviations manifest only at a scale below $2M(1+\epsilon)$, where $\epsilon=\cO(N^{-1})$. This approach aligns with the supergravity perspective, where the charged black holes correspond to configurations of branes/anti-branes: breaking down the Schwarzschild horizon into multiple sources is necessary for a comprehensive description of its microstructure.

Crucially, the extremality of the individual black holes does not affect our construction and can be arbitrarily adjusted. We demonstrate that the key criterion lies in the distribution of charges among the black holes. By alternating the signs,  we establish a universal distribution, $|Q_i| \sim 2 \sqrt{x_i(2M-x_i)}$, where $0<x_i<2M$ denotes the position of the $i^\text{th}$ black hole within the bound state of size $2M$. This distribution results in a spherical shape for the magnitude of charges, independent of the extremality properties of the black holes, with small values at the edges and a maximum of order $2M$ in the middle.

We conduct a detailed analysis of the thermodynamic properties of the bound states. First, we demonstrate that their entropy constitutes a fraction of the Schwarzschild entropy, reaching a minimum value of approximately $0.35 S_\text{Schw}$ when the black holes are extremal. Second, despite differing temperatures compared to the Schwarzschild black hole, the bound states exhibit an effective temperature derived from the mass/entropy formula that is comparable in magnitude.

Furthermore, due to the four-dimensional construction, the microscopic black holes are held apart by struts, which are strings with negative tension and energy \cite{Costa:2000kf}. Despite their singularity, these struts carry important physical characteristics of the bound state. We demonstrate that their properties, in terms of tensions and energy, remain within reasonable bounds despite the ultra-compact nature of the solutions. This suggests that they could be resolved into smooth topological bubbles in supergravity, as illustrated in previous studies \cite{Elvang:2002br,Bah:2021owp,Bah:2021rki,Heidmann:2021cms,Heidmann:2023kry,Heidmann:2023thn}.

 \begin{figure}[t]
\hspace{-0.5cm} \includegraphics[width=1.05 \textwidth]{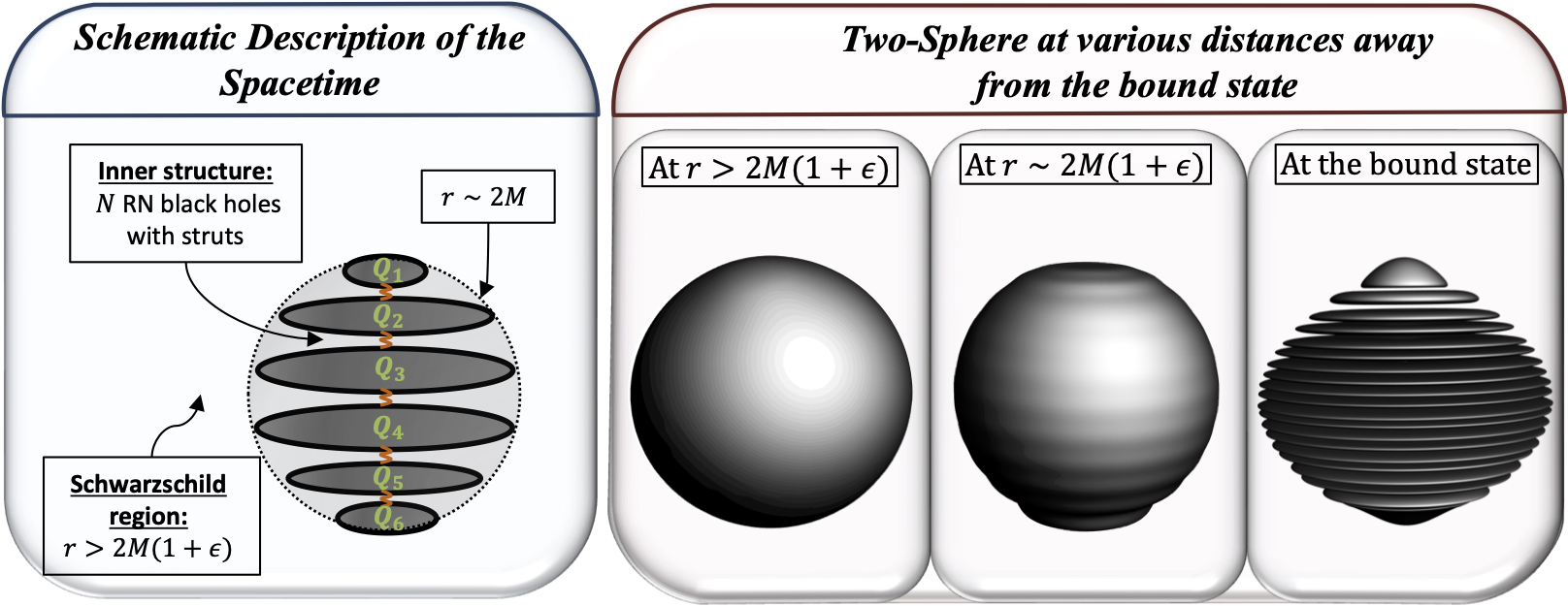}
\caption{The bound states of $N$ Reissner-Nordström black holes that are indistinguishable from the Schwarzschild black hole.  The left figure corresponds to a schematic description of the spacetime and how the geometries differ below $r<2M(1+\epsilon)$ where $\epsilon=\cO(N^{-1})$ is the resolutions scale.  On the right,  the figure corresponds to the two-sphere at three different radii for $N=20$ black holes.  The first is in a region where the geometries are indistinguishable and the sphere is round. The middle is in the intermediate region when deviations appear and the third is at the bound state locus corresponding $N$ microscopic horizons held apart by struts.}
\label{fig:Intro}
\end{figure}

In Fig.\ref{fig:Intro}, we summarize the main properties of the solutions. The left figure provides a schematic description of the resolution process, illustrating how the Schwarzschild horizon has been replaced by a cluster of $N \gg 1$ collinear charged black holes. The right figure highlights the deviation from Schwarzschild by plotting the two-sphere as we approach a bound state composed of $N=20$ black holes.\footnote{These plots represent the parametric two-dimensional surface $R(r,\theta)$, derived from the two-sphere line element $ds_2^2 = R(r,\theta)^2 \left( d\theta^2 + \sin^2 \theta \,d\phi^2 \right)$. } They clearly illustrate the transition from a Schwarzschild geometry at $r\gtrsim 2M(1+\cO(N^{-1}))$ (a round sphere) to a nontrivial electromagnetic spacetime structure. This structure consists of $N$ microscopic horizons squeezed tightly by gravitational contraction, and separated by ``empty'' regions where the struts are located. Remarkably, the entire structure maintains a spherical shape, crucial for preserving a total area that scales with $M^2$. \\

The results presented in this paper can be viewed as an effective four-dimensional description of  Schwarzschild microstructure. It paves the way for the next step, which involves embedding the solutions in supergravity, resolving the struts into topological bubbles to form regular bound states of branes and anti-branes. This approach promises a microscopic formulation of (a fraction of) the Schwarzschild entropy and insight into the manifestation of its microstructure at the horizon scale. Furthermore, the construction raises intriguing questions for future exploration. Firstly, the bound states exhibit an effective temperature not associated with Hawking radiation from the microscopic black holes but possibly arising from similar quantum pair creations in between the black holes. Secondly, the emerging structure in the near horizon region of the Schwarzschild black hole could provide new observables beyond the conventional picture of black holes in general relativity. In this regard, investigating the response under perturbation, gravitational waves, imaging, and particle scattering could reveal potential deviations from Schwarzschild. We aim to address these questions in future research.

In Section \ref{sec:ErnstGen}, we review some established results concerning the Ernst formalism in four dimensions and introduce a new concept of indistinguishability between Ernst solutions. Section \ref{sec:BSofNRN} provides a brief overview of the general solutions corresponding to $N$ collinear Reissner-Nordström black holes derived in \cite{NoraBreton1998}, along with the construction of a subclass of solutions that closely resemble a Schwarzschild black hole at large $N$. We analyze the spacetime structure of these bound states in Section \ref{sec:AnalysisLargeN}. Finally, in Section \ref{sec:AnalysisSmallN}, we refine the class of solutions to derive bound states that do not necessitate excessively large $N$ to resolve the Schwarzschild black hole.  Section \ref{sec:Conclusion} concludes with reflections and future outlook.

\section{Electrostatic solutions in four dimensions}
\label{sec:ErnstGen}

In this section, we lay the groundwork for constructing four-dimensional spacetimes composed of multiple  charged sources. First, we detail the four-dimensional classical theory of gravity,  alongside the static Ernst formalism. This formalism allows for analytic derivations of static and axially-symmetric solutions from an integrable system of equations. Subsequently, we provide an overview of Ernst solutions in four dimensions, and we develop key concepts such as ``axis data'' and the indistinguishability between solutions.

\subsection{Ernst formalism}

The Ernst formalism has been developed for solutions possessing two commuting Killing vectors in four-dimensional Einstein-Maxwell theory given by the action \cite{Ernst:1967wx,Ernst:1967by}:
\begin{equation}
S_4 \= \frac{1}{16\pi G_4} \,\int \,d^4 x \sqrt{-g}\,\left( R -\frac{1}{4} F_{\mu\nu} F^{\mu\nu}\right)\,,
\label{eq:4daction}
\end{equation}
where $G_4$ is the four-dimensional Newton constant, $R$ is the Ricci scalar, and $F$ is a two-form field strength. Its static formulation consists in restricting to static axially-symmetric solutions characterized by $\partial_t$ and $\partial_\phi$ as the timelike and spacelike isometries, and the fields depend on a two-dimensional plane in Weyl-Papapetrou coordinates, denoted as $(\rho,z)$:
\begin{equation}
\begin{split}
ds_4^2 &\= - \frac{dt^2}{Z^2}+ Z^2\left[e^{4\nu}\left( d\rho^2+dz^2\right) +\rho^2 d\phi^2 \right],\quad F \=  -2\cos \eta \, dA\wedge dt+ 2\sin \eta\, dH \wedge d\phi\,.
\end{split}
\label{eq:StaticErnstMetric4d}
\end{equation}
The functions $A$ and $H$ are the electric and magnetic potentials respectively,  $Z$ is the gravitational redshift function, and $\nu$ determines the three-dimensional base. The magnetic and electric potentials must be related by electromagnetic duality:
\begin{equation}
    \star_2 dH = \rho Z^2 dA\,,
    \label{eq:MagDualErnst}
\end{equation}
where $\star_2$ is the Hodge operator in the flat $(\rho,z)$ space. As such,  $F$ corresponds to a dyonic field with magnetic and electric sources, and $\eta$ serves as a dyonic parameter controlling the electric charges relative to the magnetic charges. Having only electric (resp. magnetic) charges yields $\eta=0$ (resp. $\eta=\pi/2$).\\

The Einstein-Maxwell equations lead to 
\begin{align}
& \Delta \log Z + Z^2 \,  \nabla A . \nabla A \,=\, 0\,, \qquad  \nabla .  \left( \rho Z^2 \nabla A\right) \,=\, 0 \,,\label{eq:EOMErnst}\\
&\frac{\partial_z \nu}{\rho} = \partial_\rho \log Z \,\partial_z \log Z - Z^2 \partial_\rho  A\partial_z  A, \quad \frac{2\partial_\rho \nu}{\rho} = \left( \partial_\rho \log Z\right)^2 - \left(\partial_z \log Z\right)^2 - Z^2  \left((\partial_\rho A)^2-(\partial_z A)^2 \right), \nonumber
\end{align}
where we have defined the gradient and Laplacian,  $\Delta \equi \frac{1}{\rho} \,\partial_\rho \left( \rho \,\partial_\rho\right) + \partial_z^2$ and $\nabla \equi (\partial_\rho,\partial_z)$.
Those equations correspond to the electrostatic limit of the Ernst equations with the following Ernst potentials: $\Psi= A$ and $\mathcal{E}= Z^{-2} - A^2$.\footnote{More precisely,  the equations \eqref{eq:EOMErnst} can be written in the generic Ernst form \begin{align}
&(\text{Re}(\mathcal{E}) + \Psi^* \Psi ) \,\Delta \mathcal{E} = (\nabla \mathcal{E}  +2 \Psi^* \nabla \Psi) \nabla \mathcal{E}\,,\quad (\text{Re}(\mathcal{E})+ \Psi^* \Psi ) \,\Delta \Psi = (\nabla \mathcal{E} +2 \Psi^* \nabla \Psi) \nabla \Psi\,, \nn\\
&(\text{Re}(\mathcal{E}) + \Psi^* \Psi ) \,R^{(3)}_{ij} = \frac{1}{2} \partial_{(i}\mathcal{E} \, \partial_{j)}\mathcal{E}^* + \Psi \,\partial_{(i} \mathcal{E} \,\partial_{j)} \Psi^* + \Psi^* \,\partial_{(i} \mathcal{E}^* \,\partial_{j)} \Psi - (\mathcal{E}+\mathcal{E}^*)\, \partial_{(i} \Psi \,\partial_{j)} \Psi^*\,,
\end{align}
where $R^{(3)}$ is the Ricci tensor of the three-dimensional base $ds_3^2= e^{4\nu}\left( d\rho^2+dz^2\right) +\rho^2 d\phi^2$.}

\subsection{Solutions and rod structure}

\begin{figure}
\centering
    \begin{tikzpicture}[dot/.style = {circle, fill, minimum size=#1,
              inner sep=0pt, outer sep=0pt},
dot/.default = 6pt  
                    ] 
\def\deb{-10} 
\def\inter{0.7} 
\def\ha{2.8} 
\def\zaxisline{5} 
\def\rodsize{1.7} 
\def\interrod{1} 
\def\fin{0.5} 
\def\posx{\deb+0.5+1.5*\rodsize}
\def\posy{\ha-1.1*\inter}




\draw (\fin+0.2,\ha-\zaxisline*\inter-0.3) node{$z$};

\draw[black,line width=0.3mm] (\deb-0.7,\ha-\zaxisline*\inter) -- (\deb+1.7*\rodsize+\interrod+0.5,\ha-\zaxisline*\inter);\draw[black,line width=0.3mm,dotted] (\deb+1.7*\rodsize+\interrod+0.5,\ha-\zaxisline*\inter) -- (\deb+1.7*\rodsize+\interrod+1,\ha-\zaxisline*\inter);

\draw[black,line width=0.3mm] (\deb+1.7*\rodsize+\interrod+1,\ha-\zaxisline*\inter) -- (\deb+1.7*\rodsize+\interrod+3,\ha-\zaxisline*\inter);\draw[black,line width=0.3mm,dotted] (\deb+1.7*\rodsize+\interrod+3,\ha-\zaxisline*\inter) -- (\deb+1.7*\rodsize+\interrod+3.5,\ha-\zaxisline*\inter);
\draw[black,->, line width=0.3mm] (\deb+1.7*\rodsize+\interrod+3.5,\ha-\zaxisline*\inter) -- (\fin+0.2,\ha-\zaxisline*\inter);


\draw[black,line width=1mm] (\deb,\ha-\zaxisline*\inter) -- (\deb+\rodsize,\ha-\zaxisline*\inter);
\draw[line width=0.3mm] (\deb+\rodsize,\ha-\zaxisline*\inter+0.1) -- (\deb+\rodsize,\ha-\zaxisline*\inter-0.1);
\draw[line width=0.3mm] (\deb,\ha-\zaxisline*\inter+0.1) -- (\deb,\ha-\zaxisline*\inter-0.1);

\draw[black,line width=1mm] (\deb+\rodsize+\interrod,\ha-\zaxisline*\inter) -- (\deb+1.7*\rodsize+\interrod,\ha-\zaxisline*\inter);
\draw[line width=0.3mm] (\deb+\rodsize+\interrod,\ha-\zaxisline*\inter+0.1) -- (\deb+\rodsize+\interrod,\ha-\zaxisline*\inter-0.1);
\draw[line width=0.3mm] (\deb+1.7*\rodsize+\interrod,\ha-\zaxisline*\inter+0.1) -- (\deb+1.7*\rodsize+\interrod,\ha-\zaxisline*\inter-0.1);

\draw[black,line width=1mm] (\deb+1.7*\rodsize+\interrod+1.5,\ha-\zaxisline*\inter) -- (\deb+2.3*\rodsize+\interrod+1.5,\ha-\zaxisline*\inter);
\draw[line width=0.3mm] (\deb+1.7*\rodsize+\interrod+1.5,\ha-\zaxisline*\inter+0.1) -- (\deb+1.7*\rodsize+\interrod+1.5,\ha-\zaxisline*\inter-0.1);
\draw[line width=0.3mm] (\deb+2.3*\rodsize+\interrod+1.5,\ha-\zaxisline*\inter+0.1) -- (\deb+2.3*\rodsize+\interrod+1.5,\ha-\zaxisline*\inter-0.1);

\draw[black,line width=1mm] (\fin-0.7-\rodsize,\ha-\zaxisline*\inter) -- (\fin-0.7,\ha-\zaxisline*\inter);
\draw[line width=0.3mm] (\fin-0.7-\rodsize,\ha-\zaxisline*\inter+0.1) -- (\fin-0.7-\rodsize,\ha-\zaxisline*\inter-0.1);
\draw[line width=0.3mm] (\fin-0.7,\ha-\zaxisline*\inter+0.1) -- (\fin-0.7,\ha-\zaxisline*\inter-0.1);

\draw[gray,<->] (\deb,\ha-\zaxisline*\inter-1) -- (\fin-0.7,\ha-\zaxisline*\inter-1) node[midway,below,gray] { $\ell$};


\node[dot=5pt,gray, label=above:{\small $(\rho,z)$}] at (\posx,\posy) {};
\draw[gray] (\deb+1.7*\rodsize+\interrod+1.5,\ha-\zaxisline*\inter) -- (\posx,\posy) node[midway,right,black] {{\small $r_{2i-1}$}};

\draw[gray,<->] (\posx,\ha-\zaxisline*\inter+0.08) -- (\posx,\posy-0.14) node[midway,left,gray] { $\rho$};


\draw (\deb,\ha-\zaxisline*\inter-0.5) node{{\small $\alpha_1$}};
\draw (\deb+\rodsize,\ha-\zaxisline*\inter-0.5) node{{\small $\alpha_2$}};

\draw (\deb+\rodsize+\interrod,\ha-\zaxisline*\inter-0.5) node{{\small $\alpha_3$}};
\draw (\deb+1.7*\rodsize+\interrod,\ha-\zaxisline*\inter-0.5) node{{\small $\alpha_4$}};

\draw (\deb+1.7*\rodsize+\interrod+1.5,\ha-\zaxisline*\inter-0.5) node{{\small $\alpha_{2i-1}$}};
\draw (\deb+2.3*\rodsize+\interrod+1.5,\ha-\zaxisline*\inter-0.5) node{{\small $\alpha_{2i}$}};

\draw (\fin-0.7-\rodsize,\ha-\zaxisline*\inter-0.5) node{{\small $\alpha_{2N-1}$}};
\draw (\fin-0.7,\ha-\zaxisline*\inter-0.5) node{{\small $\alpha_{2N}$}};

\end{tikzpicture}
\caption{Schematic description of $N$ rod sources placed on the $z$-axis in the $(\rho,z)$ plane.}
\label{fig:rodDiaGen}
\end{figure}
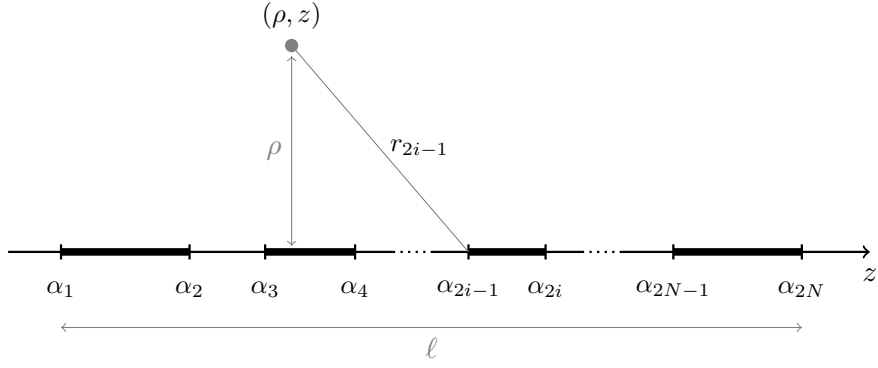

In Weyl-Papapetrou coordinates, solutions of static Ernst equations are sourced by segments on the $z$-axis ($\rho=0$), denoted as \emph{rods}. Each source induces a mass term in $Z$, and potentially an electric (resp. magnetic) charge in $A$ (resp. $H$). A generic rod diagram has been depicted in Fig.\ref{fig:rodDiaGen} with some conventions. 
The $i^\text{th}$ rod is delimited by the rod endpoints, $z=\alpha_{2i-1}$ and $z=\alpha_{2i}$. We have introduced $\ell$ as the length of the configuration,
\begin{equation}
    \ell \= \alpha_{2N}-\alpha_1\,,
\end{equation}
where $N$ corresponds to the total number of rods.
The distances to the rod endpoints, $r_i$ are given by
\begin{equation}
    r_i \equi \sqrt{\rho^2+\left(z-\alpha_{i}\right)^2}.
\end{equation}

Occasionally, we prefer to express the solutions in spherical coordinates centered around the configuration:\footnote{The transformation from Weyl-Papapetrou coordinates to spherical coordinates gives
 \begin{equation}
 r \equi \frac{r_1+r_{2N}+\ell}{2},\quad \cos \theta \equi \frac{r_1 - r_{2N}}{\ell},\quad d\rho^2 + dz^2 = \left( 1- \frac{\ell\cos^2 \frac{\theta}{2}}{r}\right)\left( 1- \frac{\ell\sin^2 \frac{\theta}{2}}{r}\right) \, \left[\frac{dr^2}{\left( 1- \frac{\ell}{r}\right)} +r^2 \,d\theta^2\right]\,. \nn
 \end{equation}}
\begin{equation}
\rho= \sqrt{r(r-\ell)}\,\sin \theta \,,\qquad z = \left(r-\frac{\ell}{2} \right) \cos \theta +\frac{\alpha_1+\alpha_{2N}}{2} \,.
\label{eq:SpherCoord}
\end{equation}

\begin{figure}[t]
\begin{center}
\includegraphics[width= 0.5 \textwidth]{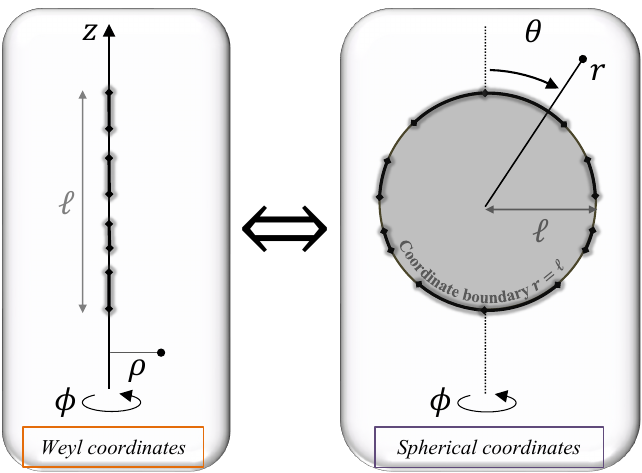}
\caption{Description of an Ernst solution of length $\ell$ in Weyl coordinates \eqref{eq:StaticErnstMetric4d} and in spherical coordinates \eqref{eq:GenericMetricSpher}. The  coordinates are related by \eqref{eq:SpherCoord}.}
\label{fig:WeylSpherCoor}
\end{center}
\end{figure}

In this coordinate system, we have $r\geq \ell$ and $0\leq \theta\leq \pi$, and the rods are located at $r=\ell$ for various ranges of $\theta$, $\tau_{2i-1}\leq \theta\leq \tau_{2i}$, where $\cos \tau_k = 1-\frac{2(\alpha_{2N}-\alpha_k)}{\ell}$. Moreover, the metric \eqref{eq:StaticErnstMetric4d} reads:
\begin{equation}
\begin{split}
ds_4^2 = - \frac{dt^2}{Z^2}+ Z^2\left(1-\frac{\ell}{r}\right) \left[G \left( \frac{dr^2}{1-\frac{\ell}{r}}+r^2 d\theta^2\right) +r^2 \sin^2\theta \, d\phi^2 \right],
\end{split}
\label{eq:GenericMetricSpher}
\end{equation}
where $G\equiv e^{4\nu}\,\left( 1- \frac{\ell\cos^2 \frac{\theta}{2}}{r}\right)\left( 1- \frac{\ell\sin^2 \frac{\theta}{2}}{r}\right)\left(1-\frac{\ell}{r}\right)^{-1}$.  We have depicted a typical profile of an Ernst solution in Weyl  or spherical coordinates in Fig.\ref{fig:WeylSpherCoor}.

\subsection{Axis data and indistinguishability}
\label{sec:AxisData}

One of the remarkable aspects of Ernst solutions is that their entire structure can be deduced from the Ernst data along the symmetry axis. This means that by fixing the values of the metric tensor and electromagnetic field tensor on the axis of symmetry, one can construct the solutions without needing further information. The Sibgatullin method \cite{Manko_1993,Ruiz:1995uh} is a powerful solution generating technique of Ernst's equations, allowing for the extraction of analytic solutions from what is called the \emph{axis data} of the Ernst potentials.

A particularly significant solution, corresponding to an axisymmetric configuration of $N$ Reissner-Nordström black holes, was derived using this method in \cite{NoraBreton1998}. This solution, which will be discussed later in this paper, has a complex form that makes its analysis quite intricate. However, it is uniquely defined by its axis data, which consist of two functions on the $z$-axis:
\begin{equation}
e(z) = 1+ \sum_{i=1}^N \frac{e_i}{z-\beta_i},\quad f(z) = \sum_{i=1}^N \frac{f_i}{z-\beta_i},
\label{eq:AxisData}
\end{equation}
where $(\beta_i,e_i,f_i)$ represent $3N$ parameters that determine indirectly the positions, charges and masses of the $N$ black holes. For example, the rod positions, $\alpha_k$, are the $2N$ zeroes of $e(z)+f(z)^2$.

\begin{figure}[t]
\begin{center}
\includegraphics[width= 0.9 \textwidth]{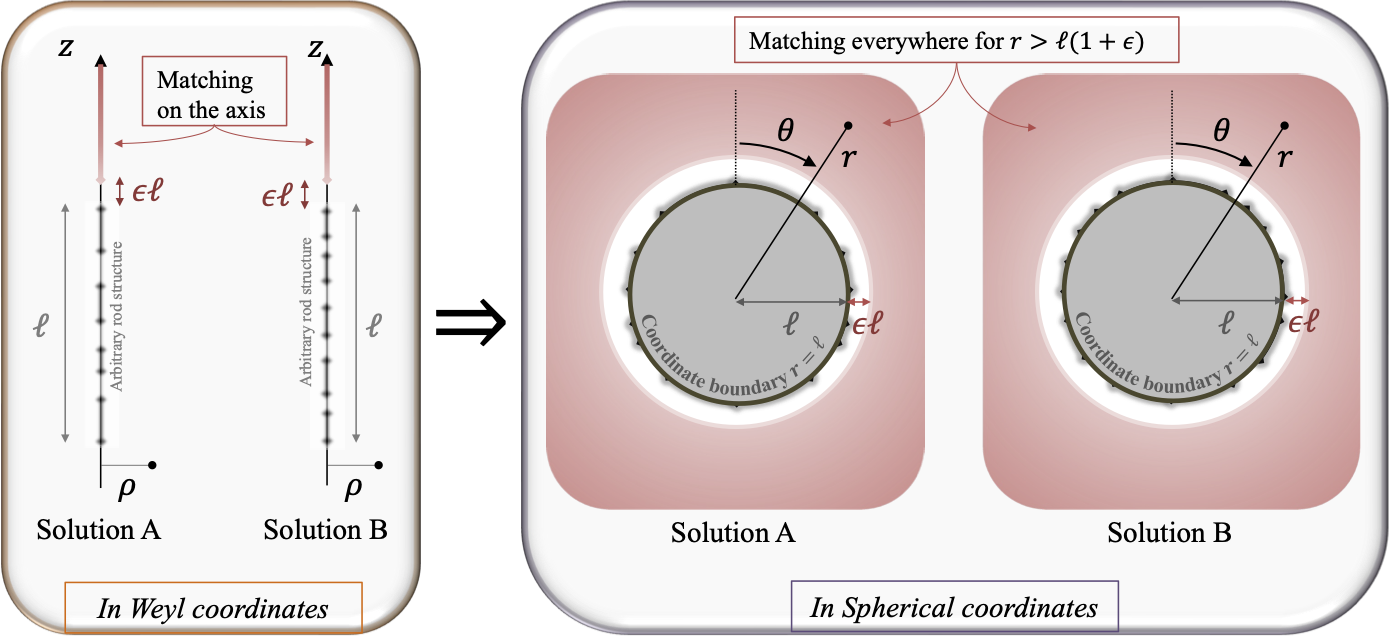}
\caption{Description of the property that two Ernst solutions, A and B, with the same configuration size, $\ell$, match everywhere for $r>\ell(1+\epsilon)$ if and only if they match on their symmetry axis for $z>\ell(1+\epsilon)$.}
\label{fig:MatchingAxis}
\end{center}
\end{figure}

The key properties of these axis data are:
\begin{itemize}
\item They correspond to the values of the gravitational and electromagnetic fields on the $z$-axis above the rod configuration:
\begin{equation}
Z(\rho=0,z\geq \alpha_{2N})^{-2} = e(z)+f(z)^2,\quad A(\rho=0,z\geq \alpha_{2N}) = f(z),\quad H(\rho=0,z\geq \alpha_{2N})=0.
\label{eq:AxisDatavsFields}
\end{equation}
\item There is a one-to-one correspondence between an Ernst solution, $(Z,A,H,\nu)$, and its axis data, $(e,f)$. Therefore, two solutions with identical axis data are necessarily identical throughout the whole space.
\item All gravitational and electromagnetic multipole moments of an axisymmetric solution can be obtained by expanding the axis data at large $z$ \cite{Simon:1983kz,Sotiriou:2004ud,Fodor:2020fnq}.
\end{itemize}
Thus, the axis data of an axisymmetric solution provide significant insights into the entire solution. This leads to the ``indistinguishability" property between Ernst solutions illustrated in Fig.\ref{fig:MatchingAxis}: \\

\underline{\emph{Indistinguishability property:}} if two Ernst solutions have the same configuration size, $\ell$, and their axis data match on the symmetry axis from the asymptotics up to a small scale above their last rod, then the solutions will match similarly everywhere. By ``similarly,'' we mean that there exists a small scale, $\epsilon \ll 1$, for which the two solutions match when $r\gtrsim \ell (1+\epsilon)$, where $r$ is the radial coordinate centered around the rod configuration \eqref{eq:SpherCoord}. \\

The property serves as a powerful tool, enabling us to establish a connection between the spacetimes of two Ernst solutions, A and B, through a simple analysis along their symmetry axis. Throughout the remainder of this paper, we will predominantly rely on this property, particularly when comparing a single-rod vacuum solution for solution A, the Schwarzschild black hole,  with a more complex solution B, featuring multiple charged sources.

\subsection{Schwarzschild black hole}

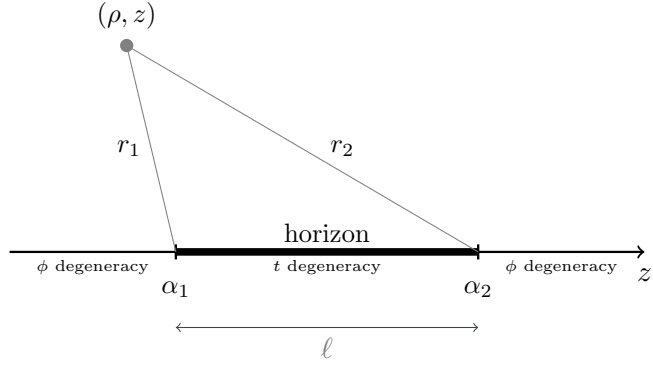
\begin{figure}
\centering
    \begin{tikzpicture}[dot/.style = {circle, fill, minimum size=#1,
              inner sep=0pt, outer sep=0pt},
dot/.default = 6pt  
                    ] 
\def\deb{-10} 
\def\inter{0.7} 
\def\ha{2.8} 
\def\zaxisline{5} 
\def\rodsize{1.7} 
\def\interrod{1} 
\def\fin{-2} 
\def\posx{\deb+0.5+0.5*\rodsize}
\def\posy{\ha-1.1*\inter}
\def\sizerod{0.75}
\def\sizerodi{0.25}




\draw (\fin+0.2,\ha-\zaxisline*\inter-0.3) node{$z$};

\draw[black,->, line width=0.3mm] (\deb-0.2,\ha-\zaxisline*\inter) -- (\fin+0.2,\ha-\zaxisline*\inter);


\draw[black,line width=1mm] (\sizerod*\deb+\sizerodi*\fin,\ha-\zaxisline*\inter) -- (\sizerodi*\deb+\sizerod*\fin,\ha-\zaxisline*\inter);
\draw[line width=0.3mm] (\sizerod*\deb+\sizerodi*\fin,\ha-\zaxisline*\inter+0.1) -- (\sizerod*\deb+\sizerodi*\fin,\ha-\zaxisline*\inter-0.1);
\draw[line width=0.3mm] (\sizerodi*\deb+\sizerod*\fin,\ha-\zaxisline*\inter+0.1) -- (\sizerodi*\deb+\sizerod*\fin,\ha-\zaxisline*\inter-0.1);

\draw[arsenic,<->] (\sizerodi*\deb+\sizerod*\fin,\ha-\zaxisline*\inter-1) -- (\sizerod*\deb+\sizerodi*\fin,\ha-\zaxisline*\inter-1) node[midway,below,gray] { $\ell$};


\node[dot=5pt,gray, label=above:{\small $(\rho,z)$}] at (\posx,\posy) {};
\draw[gray] (\sizerodi*\deb+\sizerod*\fin,\ha-\zaxisline*\inter) -- (\posx,\posy) node[midway,right,black] {{\small $\,\,\,\,r_{2}$}};
\draw[gray] (\sizerod*\deb+\sizerodi*\fin,\ha-\zaxisline*\inter) -- (\posx,\posy) node[midway,left,black] {{\small $r_{1}$}};


\draw (\sizerod*\deb+\sizerodi*\fin,\ha-\zaxisline*\inter-0.5) node{{\small $\alpha_{1}$}};
\draw (\sizerodi*\deb+\sizerod*\fin,\ha-\zaxisline*\inter-0.5) node{{\small $\alpha_{2}$}};

\draw (0.5*\deb+0.5*\sizerod*\deb+0.5*\sizerodi*\fin-0.1,\ha-\zaxisline*\inter-0.2) node{{\tiny $\phi$ degeneracy}};
\draw (0.5*\fin+0.5*\sizerodi*\deb+0.5*\sizerod*\fin+0.1,\ha-\zaxisline*\inter-0.2) node{{\tiny $\phi$ degeneracy}};
\draw (0.5*\deb+0.5*\fin,\ha-\zaxisline*\inter-0.2) node{{\tiny $t$ degeneracy}};
\draw (0.5*\deb+0.5*\fin,\ha-\zaxisline*\inter+0.25) node{{\small horizon}};

\end{tikzpicture}
\caption{Rod diagram of the Schwarzschild black hole in the $(\rho,z)$ plane.}
\label{fig:rodDiaSchw}
\end{figure}

The Schwarzschild metric can be derived as a single-rod solution of the static Ernst formalism with trivial electromagnetic gauge field:
\begin{equation}
\begin{split}
    Z^{-2} &\= \frac{r_2+r_1-\ell}{r_2+r_1+\ell} \= 1-\frac{\ell}{r} ,\qquad A=H=0\,,\\ e^{4\nu} &\= \frac{\left(r_2+r_1 \right)^2-\ell^2}{4 r_2 r_1}\= \frac{1-\frac{\ell}{r}}{\left( 1- \frac{\ell\cos^2 \frac{\theta}{2}}{r}\right)\left( 1- \frac{\ell\sin^2 \frac{\theta}{2}}{r}\right)},
\end{split}
\end{equation}
where $\ell$ is the length of the rod, $\ell=\alpha_2-\alpha_1$. The rod diagram of the Schwarzschild black hole has been depicted in Fig.\ref{fig:rodDiaSchw} where we have indicated which dimension is shrinking on the $z$-axis.

In Weyl-Papapetrou coordinates, the rod, at $\rho=0$ and $\alpha_1 \leq z\leq \alpha_2$, corresponds to the horizon of the Schwarzschild black hole, $r=\ell$. Note that in this coordinate system, the spherical symmetry of the solution is not manifest, and it only emerges once the coordinates are changed to the spherical coordinates \eqref{eq:GenericMetricSpher} with $G=1$.

The ADM mass, temperature and entropy of the black hole are (in units $G_4=1$):
\begin{equation}
    M \=\frac{\ell}{2}\,,\qquad S_\text{Schw} \= 4\pi M^2 \,,\qquad T_\text{Schw} \= \frac{1}{8\pi M}\,.
    \label{eq:SchwMassEntropyTemp}
\end{equation}

In terms of axis data \eqref{eq:AxisData}, the Schwarzschild solution is uniquely given by the functions:
\begin{equation}
    e(z)= \frac{z-\alpha_2}{z-\alpha_1}\,,\qquad f(z)=0\,,
    \label{eq:SchwAxisData}
\end{equation}
which corresponds to $\beta_1=\alpha_1$, $e_1=-\ell$ and $f_1=0$ in \eqref{eq:AxisData}.

\section{Bound states of Reissner-Nordström black holes}
\label{sec:BSofNRN}

In this section, we review the Ernst solutions corresponding to $N$ collinear Reissner-Nordström black holes \cite{NoraBreton1998}, and construct a subclass where the bound states are ultra-compact and indistinguishable from a Schwarzschild black hole. This limit can be associated with an entrapment limit discussed in \cite{Heidmann:2023thn}, where neutral configurations of self-gravitating charges  become sufficiently compact so that they generate a high-redshift that entraps their own electromagnetic field.  As such, they look like vacuum solutions right outside the charge locii.

\subsection{Ernst solution}
\label{sec:NRNGen}

The Ernst solution corresponding to $N$ Reissner-Nordström black holes has been derived in \cite{NoraBreton1998} as the static limit of $N$ Kerr-Newman black holes \cite{Ruiz:1995uh}. The metric and electromagnetic field is given by the generic ansatz \eqref{eq:StaticErnstMetric4d} in Weyl-Papapetrou coordinates and \eqref{eq:GenericMetricSpher} in spherical coordinates, and the fields $(Z,A,H,\nu)$ are sourced by $N$ rods on the $z$-axis as depicted in Fig.\ref{fig:NRNBHstructure}.

The solutions are given in terms of $3N$ parameters.  They can be chosen to be the $2N$ rod endpoints $\alpha_k$, and $N$ parameters,  denoted $\beta_k$,  that give indirectly rise to the charges at the black holes. One can take the extremal limit for one or multiple black holes in the chain by simply considering that the rod shrinks to a point:
\begin{equation}
    \text{The $i^\text{th}$ black hole is extremal} \quad \Leftrightarrow \quad \alpha_{2i-1} = \alpha_{2i}.
\end{equation}
If we take the extremal limit for each black hole, we end up with a bound state of $N$ extremal Reissner-Nordström black holes as in \cite{Heidmann:2023thn}, given by $2N$ parameters. Note that this does not lead to the linear Majumdar-Papapetrou solutions \cite{PhysRev.72.390},  or BPS multicenter solutions,  as the charges can have different signs (so BPS and anti-BPS).

The fields $(Z,A,H,\nu)$ are intricated functions of the spacetime coordinates $(\rho,z)$ and the parameters $(\alpha_k,\beta_k)$,
\begin{equation}
    Z = \frac{E_-}{\sqrt{E_+ E_- +F^2}}\,,\qquad e^{4\nu}= \frac{E_+ E_-+F^2}{K ^2\prod_{k=1}^{2N}r_k} \,,\qquad A=\frac{F}{E_-}\,,\qquad H=-\frac{I}{E_-}\,,
    \label{eq:FieldsNRN}
\end{equation}
where $E_\pm$, $F$, $I$, and $K$ are involved matrix determinants depending on $\rho$, $z$,  $\alpha_k$ and $\beta_k$ and detailed in the Appendix \ref{app:ErnstSolNRN},  in Eq.\eqref{eq:MatrixDet1} and \eqref{eq:MatrixDet2}. 

As outlined in Section \ref{sec:AxisData}, our analytic derivations will rely on the axis data, denoted as $(e(z),f(z))$, which uniquely define the solution and are given by:
\begin{equation}
e(z) = 1+ \sum_{i=1}^N \frac{e_i}{z-\beta_i},\quad f(z) = \sum_{i=1}^N \frac{f_i}{z-\beta_i},
\label{eq:AxisData2}
\end{equation}
where $f_i$ and $e_i$ are expressed in terms of $\alpha_i$ and $\beta_i$ as:
\begin{equation}
    f_i^2 \= \frac{\prod_{k=1}^{2N} (\beta_i-\alpha_k)}{\prod_{k\neq i} (\beta_i-\beta_k)^2}\,,\qquad e_i \= \left.\frac{\mathrm{d}}{\mathrm{d} z}\left(\frac{\prod_{k=1}^{2N} (z-\alpha_k)}{\prod_{k\neq i} (z-\beta_k)^2}\right)\right|_{z=\beta_i}-2 f_i \sum_{k \neq i} \frac{f_k}{\beta_i-\beta_k}.
    \label{eq:DefFiEi}
\end{equation}
The axis data also correspond to the values of the metric tensor and electromagnetic fields on the $z$-axis above the rods:
\begin{equation}
\begin{split}
Z(\rho=0,z\geq \alpha_{2N})^{-2} &\= e(z)+f(z)^2 \= \frac{\prod_{k=1}^{2N}(z-\alpha_k)}{\prod_{k=1}^{N}(z-\beta_k)^2},\\
A(\rho=0,z\geq \alpha_{2N}) &\= f(z),\qquad H(\rho=0,z\geq \alpha_{2N})=0.
\end{split}
\label{eq:FieldAxisData}
\end{equation}

Note that the signs of the $f_i$ are completely unrestricted. This will be crucial for constructing solutions where $A$ is nearly zero on the axis just above $z\gtrsim \alpha_{2N}$.

The gravitational and electromagnetic multipole moments can be derived by expanding the axis data at large $z$ \cite{Simon:1983kz,Sotiriou:2004ud}. For instance, the total mass, total dyonic charge, and dyonic dipole moment are respectively given by:\footnote{The total dyonic charge is defined as $Q = \sqrt{Q_m^2 + Q_e^2}$, where $Q_m$ and $Q_e$ represent the net magnetic and electric charges, respectively. These charges are determined using the conventions $Q_m = (8\pi)^{-1} \int F$ and $Q_e = (8\pi)^{-1} \int \star F$, where the integrals are taken over the two-sphere at the boundary. Similarly, the dyonic dipole moment is expressed as $\cJ = \sqrt{\cJ_m^2 + \cJ_e^2}$, with $\cJ_m$ and $\cJ_e$ corresponding to the magnetic and electric dipole moments, respectively.}
\begin{equation}
M = -\frac{1}{2}\, \sum_{i=1}^N e_i \= \sum_{i=1}^N\left(\frac{\alpha_{2i-1}+\alpha_{2i}}{2} -\beta_i \right),\qquad Q = \sum_{i=1}^N f_i,\qquad \cJ = M Q + \sum_{i=1}^N f_i \beta_i.
\label{eq:ConsChargeGen}
\end{equation}
It is important to note that $e_i$ and $f_i$ do not represent the individual masses and charges of the black holes.  Unfortunately,  these parameters,  as well as the $\beta_i$, are not directly related to physical observables but to mathematical structures behind the Sibgatullin method \cite{Manko_1993,Ruiz:1995uh}. The charges for instance are derivable only using the values of the fields at the rods, which are mainly obtainable in a case-by-case manner using \eqref{eq:FieldsNRN}.

\subsection{Spacetime structure}
\label{sec:SpacetimeStructure}

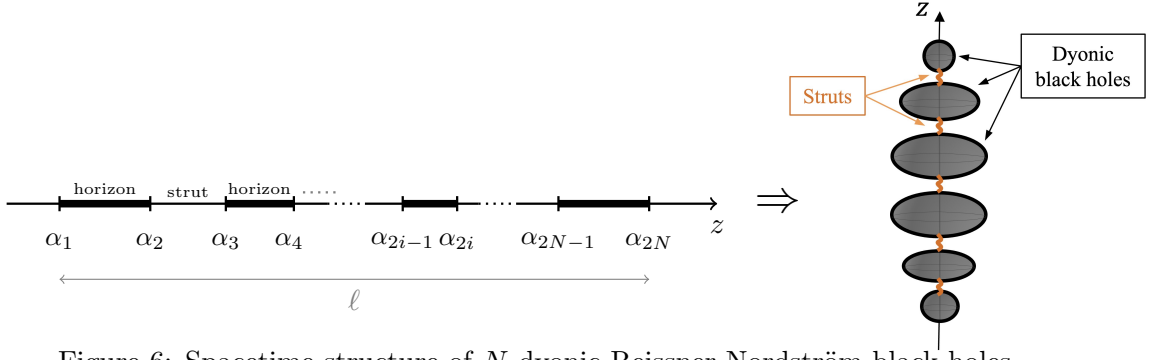
\begin{figure}
\centering
    \begin{tikzpicture}[dot/.style = {circle, fill, minimum size=#1,
              inner sep=0pt, outer sep=0pt},
dot/.default = 6pt  
                    ] 
\def\deb{-8} 
\def\inter{0.7} 
\def\ha{2.8} 
\def\zaxisline{5} 
\def\rodsize{1.2} 
\def\interrod{1} 
\def\fin{0.5} 
\def\posx{\deb+0.5+1.5*\rodsize}
\def\posy{\ha-1.1*\inter}




\draw (\fin+0.2,\ha-\zaxisline*\inter-0.3) node{$z$};

\draw[black,line width=0.3mm] (\deb-0.7,\ha-\zaxisline*\inter) -- (\deb+1.7*\rodsize+\interrod+0.5,\ha-\zaxisline*\inter);\draw[black,line width=0.3mm,dotted] (\deb+1.7*\rodsize+\interrod+0.5,\ha-\zaxisline*\inter) -- (\deb+1.7*\rodsize+\interrod+1,\ha-\zaxisline*\inter);

\draw[black,line width=0.3mm] (\deb+1.7*\rodsize+\interrod+1,\ha-\zaxisline*\inter) -- (\deb+1.7*\rodsize+\interrod+2.5,\ha-\zaxisline*\inter);
\draw[black,line width=0.3mm,dotted] (\deb+1.7*\rodsize+\interrod+2.5,\ha-\zaxisline*\inter) -- (\deb+1.7*\rodsize+\interrod+3,\ha-\zaxisline*\inter);
\draw[black,->, line width=0.3mm] (\deb+1.7*\rodsize+\interrod+3,\ha-\zaxisline*\inter) -- (\fin+0.2,\ha-\zaxisline*\inter);


\draw[black,line width=1mm] (\deb,\ha-\zaxisline*\inter) -- (\deb+\rodsize,\ha-\zaxisline*\inter);
\draw[line width=0.3mm] (\deb+\rodsize,\ha-\zaxisline*\inter+0.1) -- (\deb+\rodsize,\ha-\zaxisline*\inter-0.1);
\draw[line width=0.3mm] (\deb,\ha-\zaxisline*\inter+0.1) -- (\deb,\ha-\zaxisline*\inter-0.1);

\draw[black,line width=1mm] (\deb+\rodsize+\interrod,\ha-\zaxisline*\inter) -- (\deb+1.75*\rodsize+\interrod,\ha-\zaxisline*\inter);
\draw[line width=0.3mm] (\deb+\rodsize+\interrod,\ha-\zaxisline*\inter+0.1) -- (\deb+\rodsize+\interrod,\ha-\zaxisline*\inter-0.1);
\draw[line width=0.3mm] (\deb+1.75*\rodsize+\interrod,\ha-\zaxisline*\inter+0.1) -- (\deb+1.75*\rodsize+\interrod,\ha-\zaxisline*\inter-0.1);

\draw[black,line width=1mm] (\deb+1.7*\rodsize+\interrod+1.5,\ha-\zaxisline*\inter) -- (\deb+2.3*\rodsize+\interrod+1.5,\ha-\zaxisline*\inter);
\draw[line width=0.3mm] (\deb+1.7*\rodsize+\interrod+1.5,\ha-\zaxisline*\inter+0.1) -- (\deb+1.7*\rodsize+\interrod+1.5,\ha-\zaxisline*\inter-0.1);
\draw[line width=0.3mm] (\deb+2.3*\rodsize+\interrod+1.5,\ha-\zaxisline*\inter+0.1) -- (\deb+2.3*\rodsize+\interrod+1.5,\ha-\zaxisline*\inter-0.1);

\draw[black,line width=1mm] (\fin-0.7-\rodsize,\ha-\zaxisline*\inter) -- (\fin-0.7,\ha-\zaxisline*\inter);
\draw[line width=0.3mm] (\fin-0.7-\rodsize,\ha-\zaxisline*\inter+0.1) -- (\fin-0.7-\rodsize,\ha-\zaxisline*\inter-0.1);
\draw[line width=0.3mm] (\fin-0.7,\ha-\zaxisline*\inter+0.1) -- (\fin-0.7,\ha-\zaxisline*\inter-0.1);

\draw[gray,<->] (\deb,\ha-\zaxisline*\inter-1) -- (\fin-0.7,\ha-\zaxisline*\inter-1) node[midway,below,gray] { $\ell$};


\draw (\deb+0.5*\rodsize,\ha-\zaxisline*\inter+0.2) node{{\tiny horizon}};
\draw (\deb+\rodsize+0.5*\interrod,\ha-\zaxisline*\inter+0.15) node{{\tiny strut}};
\draw (\deb+1.37*\rodsize+\interrod,\ha-\zaxisline*\inter+0.2) node{{\tiny horizon}};
\draw[gray,line width=0.3mm,dotted] (\deb+1.85*\rodsize+\interrod,\ha-\zaxisline*\inter+0.15) -- (\deb+2.25*\rodsize+\interrod,\ha-\zaxisline*\inter+0.15);

\draw (\deb,\ha-\zaxisline*\inter-0.5) node{{\small $\alpha_1$}};
\draw (\deb+\rodsize,\ha-\zaxisline*\inter-0.5) node{{\small $\alpha_2$}};

\draw (\deb+\rodsize+\interrod,\ha-\zaxisline*\inter-0.5) node{{\small $\alpha_3$}};
\draw (\deb+1.7*\rodsize+\interrod,\ha-\zaxisline*\inter-0.5) node{{\small $\alpha_4$}};

\draw (\deb+1.7*\rodsize+\interrod+1.5,\ha-\zaxisline*\inter-0.5) node{{\small $\alpha_{2i-1}$}};
\draw (\deb+2.3*\rodsize+\interrod+1.5,\ha-\zaxisline*\inter-0.5) node{{\small $\alpha_{2i}$}};

\draw (\fin-0.7-\rodsize,\ha-\zaxisline*\inter-0.5) node{{\small $\alpha_{2N-1}$}};
\draw (\fin-0.7,\ha-\zaxisline*\inter-0.5) node{{\small $\alpha_{2N}$}};

\node[anchor=south,inner sep=-1cm] at (\fin+3.5,-1.7) {\includegraphics[width=.3\textwidth]{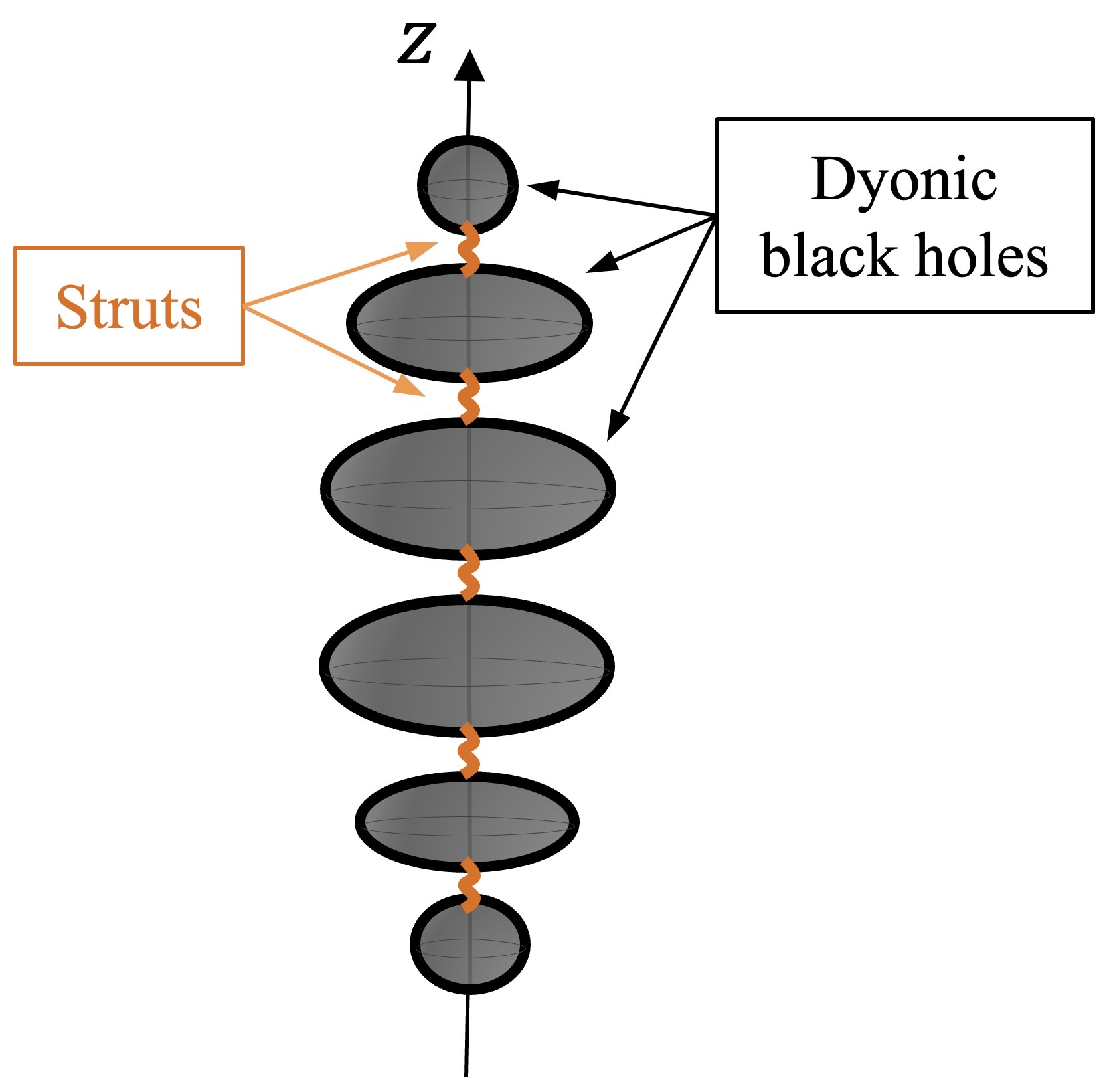}};

\draw (\fin+1,\ha-\zaxisline*\inter) node{{\LARGE $\Rightarrow$}};

\end{tikzpicture}
\caption{Spacetime structure of $N$ dyonic Reissner-Nordström black holes.}
\label{fig:NRNBHstructure}
\end{figure} 

The solution is regular for $\rho>0$ and approaches $\mathbb{R}^{1,3}$ asymptotically. On the symmetry axis at $\rho=0$, either the Killing vector $\partial_t$ or $\partial_\phi$ shrinks regularly. Three distinct structures are observed:
\begin{itemize}
    \item \underline{Regularity above and below the bound state:} 
    For $z>\alpha_{2N}$ and $z<\alpha_1$ on the $z$-axis, $\nu=0$, and the $\phi$-circle degenerates as the cylindrical angle:
    \begin{equation}
        ds_2^2 \propto d\rho^2+\rho^2 d\phi^2\,,
    \end{equation}
    which is a regular origin in $\IR^{2}$ under the assumption of $\phi$ being $2\pi$ periodic.
    \item \underline{A charged black hole at each rod:} For $z$ within the range $\alpha_{2i-1}\leq z\leq \alpha_{2i}$, the behavior of the functions $Z$ and $e^{4\nu}$ as $\rho\to 0$ indicates the locus of a black hole horizon.  Specifically, $Z$ scales as $\rho^{-1}$ and $e^{4\nu}$ scales as $\rho^2$ near $\rho=0$. As a result, the local metric near the rod resembles that of a horizon, characterized by:
    \begin{equation}
        ds_4^2 \sim \left(Z^2 e^{4\nu} \right)\bigl|_{\rho=0} \,\left[ - \kappa_i^2 \rho^2 dt^2 + d\rho^2 + dz^2+ \left(\frac{\rho^2}{e^{4\nu}} \right)\Biggl|_{\rho=0}\,d\phi^2 \right],
    \end{equation}
    where $\kappa_i$ is the surface gravity of the black hole. The temperature $T_i$ and area $\cA_i$ of the horizon, $\cA_i\equiv \int_{\phi=0}^{2\pi}\int_{z=\alpha_{2i-1}}^{\alpha_{2i}} \sqrt{g_{zz} g_{\phi\phi}}\bigl|_{\rho=0}$,  are expressed in terms of the field values as:
    \begin{equation}
        T_i \= \frac{\kappa_i}{2\pi} \= \frac{1}{2\pi} \left(\frac{1}{\rho Z^2 e^{2\nu}} \right)\Biggl|_{\rho=0}\,,\qquad \cA_i \= \frac{\alpha_{2i}-\alpha_{2i-1}}{T_i}\,.
        \label{eq:TemperatureAreaGen}
    \end{equation}
    Consequently, the Bekenstein-Hawking entropy $S_i$ of the black hole takes a simple form:
    \begin{equation}
        S_i \= \frac{\alpha_{2i}-\alpha_{2i-1}}{4G_4\,T_i},
        \label{eq:EntropyGen}
    \end{equation}
    where the only quantity requiring derivation from the intricate field values \eqref{eq:FieldsNRN} is $\rho Z^2 e^{2\nu}$ on the axis.

    Additionally, the black hole carries electromagnetic charges under the field strength $F$. The electric and magnetic charges, denoted as $Q_m^{(i)}$ and $Q_e^{(i)}$ respectively, are related by the dyonic parameter $\eta$ through electromagnetic duality \eqref{eq:MagDualErnst}:\footnote{We use the convention $Q_m^{(i)}=(8\pi)^{-1} \int F$, $Q_e^{(i)}=(8\pi)^{-1} \int \star F$, where the integral is along $z$ and $\phi$ at the rod.}
    \begin{equation}
        Q_m^{(i)} \= Q_i \, \sin \eta\,,\qquad Q_e^{(i)} \= Q_i\, \cos \eta ,
    \end{equation}
    where $Q_i$ is the dyonic charge at the black hole,  given in terms of the field values as
    \begin{equation}
    Q_i \= \sqrt{{Q_m^{(i)}}^2 + {Q_e^{(i)}}^2 } \=  \frac{1}{2} \,\left[ H(0,\alpha_{2i})-H(0,\alpha_{2i-1}) \right].
    \label{eq:DyonicChargeLocal}
    \end{equation}
Closed-form expressions of the local charges in terms of the parameters $(\alpha_i,\beta_i)$ cannot be obtained due to the complexity of the magnetic gauge potential, $H$ \eqref{eq:FieldsNRNApp}. Determining these charges must be done on a case-by-case basis when dealing with specific examples.
    \item \underline{A strut in between the rods:} 
    Between the black holes, within the range $\alpha_{2i}\leq z\leq \alpha_{2i+1}$ and at $\rho=0$, the fields have finite positive values, resulting in a region where the $\phi$-circle degenerates as the cylindrical angle. However, the base function $\nu$ becomes negative ($\nu < 0$), causing the local two-dimensional $(\rho,\phi)$ space to be described by:
    \begin{equation}
        ds_2^2 \propto d\rho^2 + \frac{\rho^2}{\gamma_i^2} \, d\phi^2 \,,\qquad \gamma_i\equi e^{2\nu}\bigl|_{\rho=0} <1\,.
    \end{equation}
    Consequently, the degenerating angle, $\phi/\gamma_i$, exhibits a periodicity greater than $2\pi$ ($2\pi/\gamma_i > 2\pi$). This segment characterizes an origin in $\IR^{2}$ with a conical excess. This structure is well understood and corresponds to a string with negative tension, known as a \emph{strut}, which prevents the collapse of the black holes \cite{Gibbons:1979nf,Costa:2000kf}. Struts are essential components that arise in four-dimensional theories of gravity where there are insufficient degrees of freedom to counteract gravitational attraction and maintain the staticity of multiple gravitational sources. However,  it is known that struts can be replaced by smooth topological spacetime bubbles in higher dimensions with compact dimensions \cite{Emparan:2001wk,Elvang:2002br,Bah:2021owp,Bah:2021rki,Heidmann:2021cms}. Therefore, struts can be viewed as four-dimensional effective descriptions of more intricate yet regular topological structures in higher dimensions.

    A strut corresponds to a conical singularity, for which an energy-momentum tensor can be associated, localized at $\rho=0$ and $\alpha_{2i}\leq z\leq \alpha_{2i+1}$. As established in \cite{Costa:2000kf,Bah:2021owp}, the strut tension $\cT_i$ and energy $E_i$ are given by:
    \begin{equation}
        \cT_i \= \frac{\gamma_i-1}{4 \gamma_i G_4} \,<\,0\,,\qquad E_i \= \frac{\gamma_i-1}{4 G_4}\left( \alpha_{2i+1}-\alpha_{2i}\right) \,<\,0\,. 
        \label{eq:StrutTensionEnergy}
    \end{equation}
    Hence, the strut exerts the necessary negative pressure to keep the black holes apart. As the black holes approach each other, $\gamma_i \to 0$, resulting in the divergence of the strut tension while the strut energy tends towards zero.
\end{itemize}

In summary, the solutions constructed in \cite{NoraBreton1998} correspond to arbitrary bound states of $N$ Reissner-Nordström black holes along a line. These black holes, carrying electromagnetic charges, are held apart by struts. The rod structure and geometry along the axis of symmetry are illustrated in Fig.\ref{fig:NRNBHstructure}.  In Weyl coordinates, the black holes are distributed along a segment of length $\ell$ at $\rho=0$,  while in spherical coordinates they are spread at the coordinate boundary,  $r=\ell$, as shown in Fig.\ref{fig:WeylSpherCoor}.

\subsection{Chain of microscopic black holes}
\label{sec:ClassMicroBH}

\begin{figure}
\centering
    \begin{tikzpicture}[dot/.style = {circle, fill, minimum size=#1,
              inner sep=0pt, outer sep=0pt},
dot/.default = 6pt  
                    ] 
\def\deb{-13} 
\def\inter{0.7} 
\def\ha{2.8} 
\def\zaxisline{5} 
\def\rodsize{1.2} 
\def\interrod{1} 
\def\fin{-1.5} 
\def\posx{\deb+0.5+1.5*\rodsize}
\def\posy{\ha-1.1*\inter}
\def\segment{2.5} 
\def\propa{0.63} 
\def\propb{0.5} 
\def\debB{-12.5}




\draw (\fin+0.2,\ha-\zaxisline*\inter-0.3) node{$z$};

\draw[black,line width=0.3mm] (\deb,\ha-\zaxisline*\inter)  -- (\fin-1.5*\segment,\ha-\zaxisline*\inter);
\draw[black,line width=0.3mm,dotted] (\fin-1.5*\segment,\ha-\zaxisline*\inter) -- (\fin-1.1*\segment,\ha-\zaxisline*\inter);
\draw[black,->, line width=0.3mm] (\fin-1.1*\segment,\ha-\zaxisline*\inter) -- (\fin+0.2,\ha-\zaxisline*\inter);


\draw[line width=0.3mm,byzantine] (\debB,\ha-\zaxisline*\inter+0.1) -- (\debB,\ha-\zaxisline*\inter-0.1);

\draw[line width=0.3mm,byzantine] (\debB+\segment,\ha-\zaxisline*\inter+0.1) -- (\debB+\segment,\ha-\zaxisline*\inter-0.1);

\draw[line width=0.3mm,byzantine] (\debB+2*\segment,\ha-\zaxisline*\inter+0.1) -- (\debB+2*\segment,\ha-\zaxisline*\inter-0.1);

\draw[line width=0.3mm,byzantine] (\fin-0.5-\segment+\propa*\segment-\propa*\propb*\segment,\ha-\zaxisline*\inter+0.1) -- (\fin-0.5-\segment+\propa*\segment-\propa*\propb*\segment,\ha-\zaxisline*\inter-0.1);


\draw[black,line width=1mm] (\debB+\propa*\propb*\segment,\ha-\zaxisline*\inter) -- (\debB+\propa*\propb*\segment+\segment-\propa*\segment,\ha-\zaxisline*\inter);
\draw[line width=0.3mm] (\debB+\propa*\propb*\segment,\ha-\zaxisline*\inter+0.1) -- (\debB+\propa*\propb*\segment,\ha-\zaxisline*\inter-0.1);
\draw[line width=0.3mm] (\debB+\propa*\propb*\segment+\segment-\propa*\segment,\ha-\zaxisline*\inter+0.1) -- (\debB+\propa*\propb*\segment+\segment-\propa*\segment,\ha-\zaxisline*\inter-0.1);

\draw[black,line width=1mm] (\debB+\propa*\propb*\segment+\segment,\ha-\zaxisline*\inter) -- (\debB+\propa*\propb*\segment+2*\segment-\propa*\segment,\ha-\zaxisline*\inter);
\draw[line width=0.3mm] (\debB+\propa*\propb*\segment+\segment,\ha-\zaxisline*\inter+0.1) -- (\debB+\propa*\propb*\segment+\segment,\ha-\zaxisline*\inter-0.1);
\draw[line width=0.3mm] (\debB+\propa*\propb*\segment+2*\segment-\propa*\segment,\ha-\zaxisline*\inter+0.1) -- (\debB+\propa*\propb*\segment+2*\segment-\propa*\segment,\ha-\zaxisline*\inter-0.1);

\draw[black,line width=1mm] (\debB+\propa*\propb*\segment+2*\segment,\ha-\zaxisline*\inter) -- (\debB+\propa*\propb*\segment+3*\segment-\propa*\segment,\ha-\zaxisline*\inter);
\draw[line width=0.3mm] (\debB+\propa*\propb*\segment+2*\segment,\ha-\zaxisline*\inter+0.1) -- (\debB+\propa*\propb*\segment+2*\segment,\ha-\zaxisline*\inter-0.1);
\draw[line width=0.3mm] (\debB+\propa*\propb*\segment+3*\segment-\propa*\segment,\ha-\zaxisline*\inter+0.1) -- (\debB+\propa*\propb*\segment+3*\segment-\propa*\segment,\ha-\zaxisline*\inter-0.1);

\draw[black,line width=1mm] (\fin-0.5-\segment+\propa*\segment,\ha-\zaxisline*\inter) -- (\fin-0.5,\ha-\zaxisline*\inter);
\draw[line width=0.3mm] (\fin-0.5-\segment+\propa*\segment,\ha-\zaxisline*\inter+0.1) -- (\fin-0.5-\segment+\propa*\segment,\ha-\zaxisline*\inter-0.1);
\draw[line width=0.3mm] (\fin-0.5,\ha-\zaxisline*\inter+0.1) -- (\fin-0.5,\ha-\zaxisline*\inter-0.1);

\draw[arsenic,<->] (\debB,\ha-\zaxisline*\inter+0.7) -- (\debB+\propa*\propb*\segment,\ha-\zaxisline*\inter+0.7) node[midway,above,gray] { {\footnotesize $ a \delta/2$}};

\draw[arsenic,<->] (\debB+\propa*\propb*\segment+\segment,\ha-\zaxisline*\inter+0.7) -- (\debB+\propa*\propb*\segment+2*\segment-\propa*\segment,\ha-\zaxisline*\inter+0.7) node[midway,above,gray] { {\footnotesize $ (1-a) \delta$}};

\draw[arsenic,<->] (\debB+\propa*\propb*\segment,\ha-\zaxisline*\inter+0.5) -- (\debB+\propa*\propb*\segment+\segment,\ha-\zaxisline*\inter+0.5) node[midway,above,gray] { $\delta$};

\draw[arsenic,<->] (\debB+\propa*\propb*\segment,\ha-\zaxisline*\inter-1) -- (\fin-0.5,\ha-\zaxisline*\inter-1) node[midway,below,gray] { $\ell$};


\draw (\debB,\ha-\zaxisline*\inter-0.5) node{{\color{byzantine}{\small $\beta_1$}}};
\draw (\debB+\segment,\ha-\zaxisline*\inter-0.5) node{{\color{byzantine}{\small $\beta_2$}}};
\draw (\debB+2*\segment,\ha-\zaxisline*\inter-0.5) node{{\color{byzantine}{\small $\beta_3$}}};
\draw (\fin-0.5-\segment+\propa*\segment-\propa*\propb*\segment,\ha-\zaxisline*\inter-0.5) node{{\color{byzantine}{\small $\beta_N$}}};

\draw (\debB+\propa*\propb*\segment,\ha-\zaxisline*\inter-0.5) node{{\small $\alpha_1$}};
\draw (\debB+\propa*\propb*\segment+\segment-\propa*\segment,\ha-\zaxisline*\inter-0.5) node{{\small $\alpha_2$}};

\draw (\debB+\propa*\propb*\segment+\segment,\ha-\zaxisline*\inter-0.5) node{{\small $\alpha_3$}};
\draw (\debB+\propa*\propb*\segment+2*\segment-\propa*\segment,\ha-\zaxisline*\inter-0.5) node{{\small $\alpha_4$}};

\draw (\debB+\propa*\propb*\segment+2*\segment,\ha-\zaxisline*\inter-0.5) node{{\small $\alpha_{5}$}};
\draw (\debB+\propa*\propb*\segment+3*\segment-\propa*\segment,\ha-\zaxisline*\inter-0.5) node{{\small $\alpha_{6}$}};

\draw (\fin-0.5-\segment+\propa*\segment,\ha-\zaxisline*\inter-0.5) node{{\small $\alpha_{2N-1}$}};
\draw (\fin-0.5,\ha-\zaxisline*\inter-0.5) node{{\small $\alpha_{2N}$}};

\end{tikzpicture}
\caption{Rod structure and $\beta$ parameters for a bound state of $N$ microscopic Reissner-Nordström black holes. The scale of the pattern is $\delta=\frac{\ell}{N(1-\frac{a}{N})}$.}
\label{fig:microscopicRN}
\end{figure}
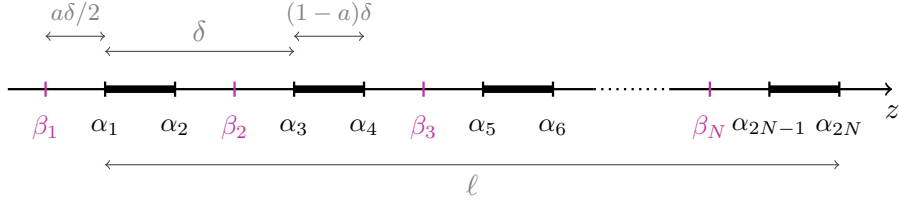

In this section, we specify to a subclass of bound states made of a large number of small black holes. For that purpose, we consider solutions with a rod configuration consisting of $N$ identical patterns of infinitesimal length $\delta$, so that the length of the configuration is $\ell \sim N \delta$. Each pattern contains one black hole and a value of $\beta_k$,  lying midway between the segments separating the black holes:
\begin{equation}
    \alpha_{2k-1} \= (k-1) \,\delta,\qquad \alpha_{2k} \= (k-a) \,\delta\,,\qquad \beta_k \= \left(k-1-\frac{a}{2}\right)\, \delta\,,\quad k=1,\ldots,N\,,
    \label{eq:alphabetaMicro}
\end{equation}
where $0 < a\leq 1$ is a parameter that fix the position of $\alpha_{2k}$ in each pattern as shown in Fig.\ref{fig:microscopicRN}. Thus the length configuration is given by
\begin{equation}
    \ell \= N \left(1-\frac{a}{N}\right) \,\delta.
    \label{eq:RelationEllDelta}
\end{equation}

The parameter $a$ can be associated to the extremality parameter of the black holes,  which gives the length of the black hole rods, $\alpha_{2k}-\alpha_{2k-1}= (1-a)\,\delta$.  Moreover, two interesting values have to be noted:
\begin{itemize}
    \item \underline{$a=1$ and the extremal limit:} The black holes are extremal as the sources become point-like.
    \item \underline{$a \to 0$ and the vacuum limit:} The black holes are merging into each other,  forming one single rod of length $\ell$ with zero net charge as $\beta_k\to \alpha_{2k-1}$.  Thus, the single rod is a Schwarzschild black hole.. 
\end{itemize}

The solutions define a family of electrovacuum geometries corresponding to a bound state of dyonic black holes, held apart by struts, as shown in Fig.\ref{fig:NRNBHstructure}.  The solutions are parametrized by three parameters: $N$ the number of black holes, $\delta$ the length of the microscopic elements, and $a$ which is associated to the extremality of the microscopic black holes. 

Moreover, there are other degrees of freedom in the solutions that do not materialize in terms of parameters but in terms of sign choices for the quantities $f_i$ in \eqref{eq:DefFiEi}. As we will see in a moment, these signs are associated to the sign of the black hole charges.

The total mass and dyonic charge of the bound states \eqref{eq:ConsChargeGen} are given by
\begin{equation}
    M \= \frac{N\delta}{2} \= \frac{\ell}{2\left(1-\frac{a}{N}\right)}\,,\qquad Q \= \sum_{i=1}^N f_i\,,
    \label{eq:ConservedChargesMicroRN}
\end{equation}
where the $f_i$ are given by
\begin{equation}
    f_i \= \pm \frac{\sqrt{a (2-a)}\,\delta}{2} \,\prod_{k\neq i}^N \sqrt{\left( 1+\frac{a}{2(k-i)}\right) \left( 1+\frac{2-a}{2(k-i)}\right)}\,,
    \label{eq:fparamNmicroRN}
\end{equation}
and the ``$\pm$'' reflects the sign freedom in the choice of $f_i$.

\subsubsection{Indistinguishability from the Schwarzschild black hole}
\label{sec:AxisDataMicroBH}

The metric tensor and electromagnetic fields of the solutions are generally given by \eqref{eq:StaticErnstMetric4d} with the corresponding fields from \eqref{eq:FieldsNRN}. However, these expressions pose challenges for analysis when dealing with a large number of black holes, as they involve determinants of matrices with a rank of order $2N$,  \eqref{eq:MatrixDet1} and \eqref{eq:MatrixDet2}.

Nevertheless, we can analyze the behavior of the solutions outside the rods by using the indistinguishability property of Ernst solutions discussed in Section \ref{sec:AxisData}. This consists in studying the solutions along their symmetry axis above the rod configuration, which encapsulates information about the solutions everywhere outside the sources (refer to Fig.\ref{fig:MatchingAxis}).

On the symmetry axis, the solutions are characterized by two axis data $(e(z),f(z))$, as introduced in \eqref{eq:AxisData2}. The gravitational and electromagnetic fields are then expressed in terms of this data in \eqref{eq:FieldAxisData}:
\begin{equation}
Z(\rho=0,z\geq \ell)^{-2} \= \frac{\prod_{k=1}^{2N}(z-\alpha_k)}{\prod_{k=1}^{N}(z-\beta_k)^2},\qquad A(\rho=0,z\geq \ell) \= \sum_{k=1}^N \frac{f_k}{z-\beta_k},
\label{eq:FieldAxisData2}
\end{equation}
where the norm of $f_k$ is given in \eqref{eq:fparamNmicroRN}.

In the limit of a large number of black holes ($N\gg 1$), the $\alpha_k$ and $\beta_k$ represent infinitesimal steps, allowing us to simplify the above expressions using Riemann sum approximations when $z$ is slightly above the rods. For detailed derivation, we refer the reader to Appendix \ref{App:MicroBH}.

\begin{itemize}
    \item \underline{Gravitational axis data:}

In Appendix \ref{App:RiemannSum}, we demonstrate the existence of an infinitesimal scale $\epsilon=\mathcal{O}(N^{-1})$ such that
\begin{equation}
    Z(0,z\gtrsim \ell(1+\epsilon))^{-2} = \frac{z-\ell}{z}\,\, \left( 1+\cO(N^{-1})\right).
\end{equation}
Consequently, the gravitational data on the symmetry axis directly above the chain of black holes becomes indistinguishable from the Schwarzschild axis data provided in \eqref{eq:SchwAxisData}. 

    \item \underline{Electromagnetic axis data:}

Although the solutions exhibit the same gravitational axis data as a Schwarzschild black hole,  the property discussed in Section \ref{sec:AxisData} applies only if the electric axis data, $f(z)=A(0,z)$, also match.  Schwarzschild being a vacuum solution,  it has $f(z)=0$. Therefore, the bound states of Reissner-Nordström black holes are indistinguishable from Schwarzschild only when their electric axis data are almost zero above the bound state.

Estimating $A(0,z)$ is more intricate than $Z(0,z)$ due to the expression of the $f_i$ parameters \eqref{eq:fparamNmicroRN}. In Appendix \ref{App:ApproxFi}, we demonstrate that, in the large $N$ limit,
\begin{equation}
\begin{split}
    f_i \,\sim\, \pm\,\cF\left(i-1+\frac{\sin(\frac{\pi}{2}\,a)}{\pi}\right)\,,\qquad  \cF(x) &\equi \frac{\delta \,\sin(\frac{\pi}{2}\,a)}{\pi}\,\sqrt{\frac{N-x}{x} }.
    \label{eq:ApproxFi}
\end{split}
\end{equation}

If all $f_i$ have the same sign, the solutions exhibit non-negligible electromagnetic fields, and their electromagnetic axis data above the black holes is given by
\begin{equation}
    A(0,z\gtrsim \ell(1+\epsilon)) \= \frac{2Q}{\ell}\left(1- \sqrt{\frac{z-\ell}{z}}  \right)+\cO(N^{-1}),
    \label{eq:A same sign}
\end{equation}
where $Q$ is the total dyonic charge, $Q=\ell\,\sin(\frac{\pi}{2}\,a)\,  (1+\mathcal{O}(N^{-1}))$. This is negligible only when $a\to 0$ which is not an interesting limit as the black holes are trivially merging into a single Schwarzschild rod in that limit.

To produce solutions with negligible electromagnetic fields above the rods, one must change the signs of the $f_i$ so that their contribution in $f(z)$ cancels out.  In Section \ref{sec:LocalCharges},  we will show that the signs of $f_i$ are associated to the signs of the local charges $Q_i$ at the black holes  \eqref{eq:DyonicChargeLocal}.  

Although the $f_i$ do not have constant norms, they are slowly varying except for the first few elements.\footnote{Indeed, one has $|f_{i+1}|-|f_{i}| = \cO(N^{-2})$ except for the few first elements where we have $|f_{i+1}|-|f_{i}| = \cO(N^{-\frac{1}{2}})$. } By alternately changing the signs of the $f_i$,  ``$+-+-\ldots$,'' the electromagnetic axis data can be made infinitesimal in the large $N$ limit, with the primary contribution coming from the first few terms.\footnote{Note that other sign patterns can be chosen,  as ``$+--++--\ldots$''. However,  we will consider only the simplest case in this paper.}  In Appendix \ref{App:RiemannSum}, we demonstrate that
\begin{equation}
    A(0,z\gtrsim \ell(1+\epsilon)) \= \frac{Q}{z} \left( 1+\cO(N^{-1})\right), \qquad Q \= \cO\left(\frac{M}{\sqrt{N}}\right),
    \label{eq:EMfieldAxisMicro}
\end{equation}
where $Q$ is the infinitesimal net charge of the bound state.  The total charge is parametrically smaller than the mass,  it tends to zero in the large $N$ limit,  and the electric axis data approaches its vacuum limit:
\begin{equation}
    A(0,z\gtrsim \ell(1+\epsilon)) \= \cO\left(\frac{1}{\sqrt{N}} \right)\,.
\end{equation}
Thus, we have shown that both the electric and gravitational axis data of the bound states understudy match the Schwarzschild axis data right above the sources $z\gtrsim \ell(1+\epsilon)$, where $\epsilon$ is of order $N^{-1}$.

The convergence of the electric axis data towards the vacuum value is slower compared to the gravitational axis data and requires a significantly large $N$ for the electromagnetic fields to be negligible above the configuration. However, this is a consequence of specializing to solutions with fixed internal parameters. By adjusting the parameters of one black hole in the chain, we could cancel the infinitesimal net charge and make the bound states much closer to a vacuum solution.  This will be made more precise in Section \ref{sec:AnalysisSmallN}.

\end{itemize}

To conclude, the analysis of the solutions on the symmetry axis above the sources has revealed the existence of an infinite subclass of solutions sharing the same axis data at leading order in $N$ as the  Schwarzschild black hole \eqref{eq:SchwAxisData},
\begin{equation}
    Z(0,z)^{-2} \sim \frac{z-2M}{z}, \qquad A(0,z) \sim 0\,,
\end{equation}
for $z$ slightly above the sources $z\gtrsim 2M(1+\mathcal{O}(N^{-1}))$.\footnote{We used \eqref{eq:ConservedChargesMicroRN} to substitute the configuration length $\ell$ with the ADM mass. }

Based on the argument presented in Section \ref{sec:AxisData}, these solutions are indistinguishable from the Schwarzschild spacetime within the range $2M (1+\mathcal{O}(N^{-1}))\lesssim r < \infty$, where $r$ denotes the spherical coordinates centered around the configuration \eqref{eq:SpherCoord}.  By indistinguishable, we mean that they are identical at leading order in large $N$.  Moreover,  due to the residual net charge of order $M/\sqrt{N}$,  we expect $1/\sqrt{N}$ corrections as we approach $r\sim 2M (1+\mathcal{O}(N^{-1}))$.  

The interested reader can find an explicit construction of $N=100$ Reissner-Nordström black holes with $a=1/2$ and its comparison to a Schwarzschild black hole with the same mass in Appendix \ref{App:OutOfAxis}.  We illustrate the result by analyzing the fields entering the metric and gauge field in the whole spacetime.  We show that they are indeed equal to the Schwarzschild values with infinitesimal corrections for $r\gtrsim 2M (1+\mathcal{O}(N^{-1}))$. \\

\begin{figure}[t]
\begin{center}
\includegraphics[width= 0.7 \textwidth]{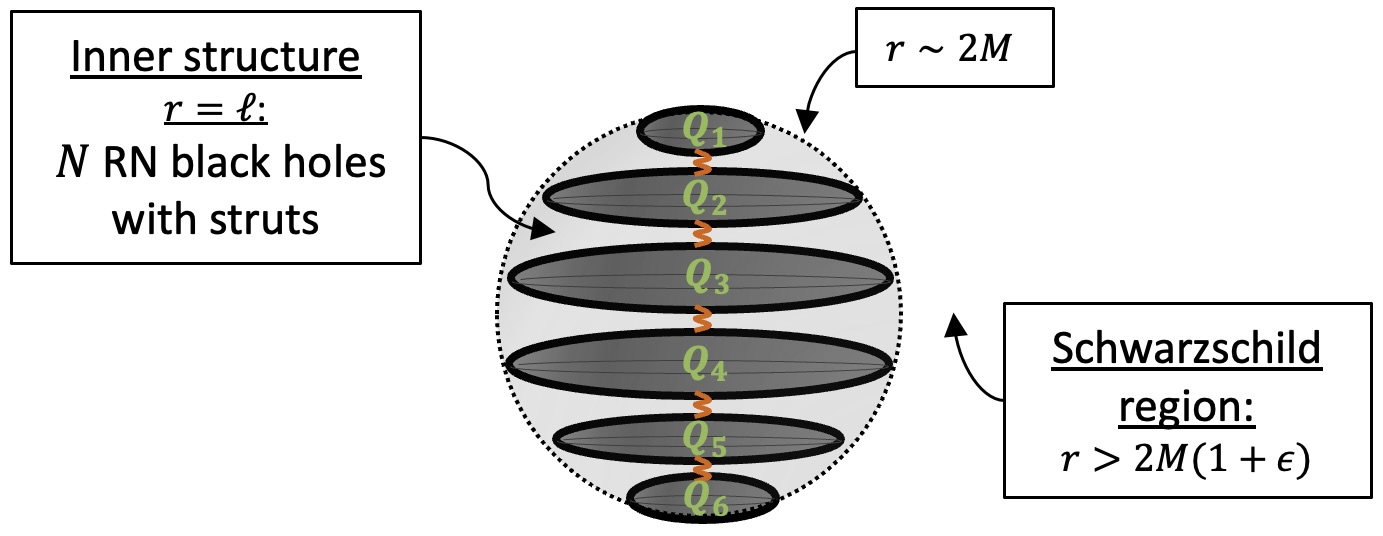}
\caption{Schematic description of the spacetime structure of $N$ microscopic Reissner-Nordström that are indistinguishable from a Schwarzschild black hole up to the horizon scale.}
\label{fig:SchwDissolNRN}
\end{center}
\end{figure}

Thus, our analysis demonstrates that the Schwarzschild black hole can effectively be ``resolved'' into chains of microscopic Reissner-Nordström black holes. This resolution implies that the solutions are indistinguishable from Schwarzschild from the asymptotics up to an infinitesimal distance away from the Schwarzschild horizon.  The latter is replaced by nontrivial structures of microscopic black holes held apart by struts. These novel structures, closely resembling the Schwarzschild metric, are depicted in Fig.\ref{fig:SchwDissolNRN}.

Remarkably, the Schwarzschild metric can be matched while maintaining the $a$ parameter entirely free, thereby defining an infinite family of solutions.  The fact that the solutions are indistinguishable when $a\to 0$ is not really surprising, as it is the limit where all black holes merge into one single uncharged rod.  This limit is like removing infinitesimal segments in favor of one big single Schwarzschild rod.  However,  the result is much more impressive when $a$ takes finite values when the separation between the black holes is the same order of their size.  More precisely,  for $a=1$,  the black hole are extremal point charges,  so we can form bound states arbitrarily close to extremality that generates a spacetime indistinguishable from the Schwarzschild metric.

While we gain a good understanding of the spacetime structure outside the sources solely from the axis data, we still lack information about the internal structure replacing the Schwarzschild horizon. This includes details such as the internal charges at the black holes, their entropy and temperature, and the energy of the struts holding them apart. These characteristics can only be derived by studying the metric tensor and electromagnetic fields using the intricate expressions \eqref{eq:FieldsNRN}, which will be the focus of the next section.

\section{Schwarzschild resolution into a large number of black holes}
\label{sec:AnalysisLargeN}

In the preceding section, we constructed bound states of Reissner-Nordström black holes on a line. These solutions closely resemble the Schwarzschild metric slightly away from its horizon, at $r>2M(1+\epsilon)$ where $\epsilon=\cO(N^{-1})$. This result highlights the possibility to turn on vast electromagnetic degrees of freedom near the Schwarzschild horizon, effectively replacing it with novel spacetime structures. 


In this section, our objective is to provide a more detailed description of theses internal structures:  at $\rho \sim 0$ and $0=\alpha_1 \lesssim z \lesssim \alpha_{2N} = \ell$.  In the spherical coordinates,  \eqref{eq:SpherCoord},  this corresponds to the region where the solutions differ significantly from the Schwarzschild geometry: $\ell\leq r< 2M(1+\cO(N^{-1}))$.  This region encompasses the near-horizon vicinity of the microscopic black holes and the regions near the struts as we move along $z$ (or equivalently $\theta$). Hence, our analysis of the microstructure will consists in:
\begin{itemize}
\item[•] \underline{Black hole properties:} This involves determining the internal charges of the microscopic Reissner-Nordström black holes \eqref{eq:DyonicChargeLocal}, their entropy \eqref{eq:EntropyGen}, and their temperature \eqref{eq:TemperatureAreaGen}.
\item[•] \underline{Strut properties:} This involves analyzing the tension and energy of the struts in the bound states \eqref{eq:StrutTensionEnergy}.
\item[•] \underline{Geometric characterization:} This involves studying how the internal topology differs from the Schwarzschild geometry, which is described by a spherical S$^2$ and a degenerating timelike Killing vector. To achieve this, we analyze the properties of the S$^2$ around the bound states up to where the black holes and struts are,  $r=\ell$.
\end{itemize}
Unlike the previous section, where we relied on axis data, accessing these characteristics requires the full expressions of the fields across the entire spacetime, as given by matrix determinants \eqref{eq:FieldsNRN},  \eqref{eq:MatrixDet1} and \eqref{eq:MatrixDet2}. However, obtaining analytic expressions from the intricate matrix determinants is challenging. Therefore, we will concentrate on specific examples to estimate these quantities in the large $N$ limit.

\subsection{Properties of the microscopic black holes}

We examine the solutions involving $N$ Reissner-Nordström black holes, as described in Section \ref{sec:ClassMicroBH}, where the $f_i$ alternate in sign.  The solutions are characterized by two parameters: $a$, representing the internal extremality parameter of the Reissner-Nordström black holes, and $N$,  the number of black holes.

\subsubsection{Charge distribution}
\label{sec:LocalCharges}

We start our analysis of the internal structure by deriving the dyonic charges of the Reissner-Nordström black holes. To compute these charges, we use the expression \eqref{eq:DyonicChargeLocal}, requiring the evaluation of the magnetic gauge potential $H$ at each black hole,  \eqref{eq:FieldsNRN} and \eqref{eq:MatrixDet2}.

Through analysis of various solutions with differing $a$ and $N$, we found:
\begin{itemize}
\item[•]  The charges alternate in sign, following the signs of the $f_i$. Consequently, all even-numbered black holes possess positive charges (resp.  negative), while all odd-numbered black holes have negative charges (resp.  positive). 

Thus, when the black holes are extremal ($a=1$), the bound state represents a chain of BPS and anti-BPS black holes separated by struts in four dimensions.

\item[•] The magnitudes of the charges, in the limit of large $N$, can be approximated by:
\begin{equation}
|Q_i| \,\sim\, 4M\, \sqrt{\frac{i-1+\sin(\frac{\pi a}{2})+\frac{1}{\pi}}{N}\left(1- \frac{i-\frac{1}{\pi}}{N} \right)}\,,  \qquad i=1,\ldots, N\,.
\label{eq:ApproxLocalCharges}
\end{equation}
\end{itemize}

\begin{figure}[t]
\begin{center}
\includegraphics[width= 0.7\textwidth]{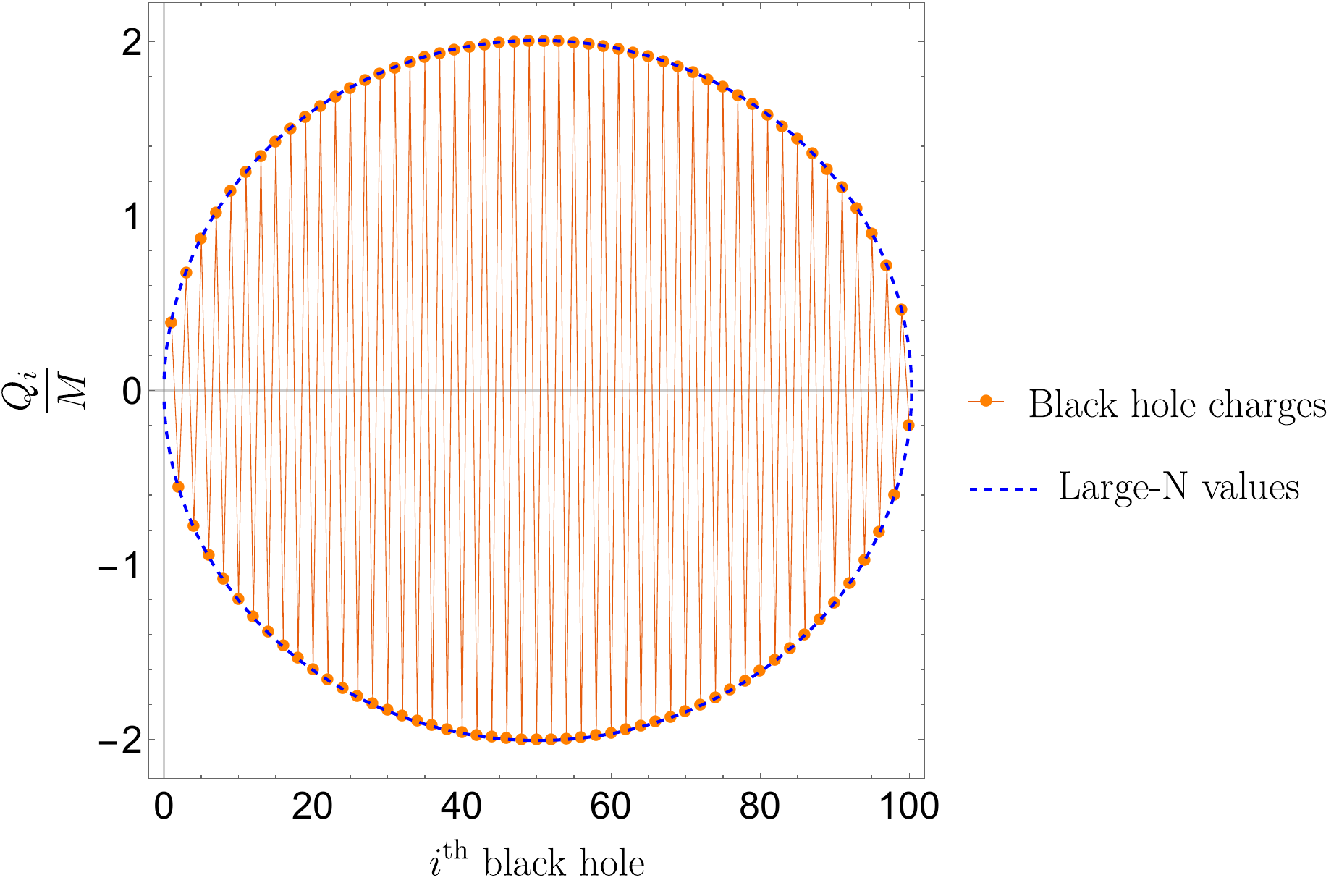}
\caption{The black hole charges \eqref{eq:DyonicChargeLocal} for a bound state made of $N=100$  black holes and $a=\frac{1}{2}$ \eqref{eq:alphabetaMicro}. The $y$ axis represents the charges normalized by the ADM mass. The $x$ axis denotes the black hole position within the bound state.  The red dots and lines are the numerical values, while the blue curve is the large-$N$ approximation \eqref{eq:ApproxLocalCharges}.}
\label{fig:LocalCharges}
\end{center}
\end{figure}

In Fig.\ref{fig:LocalCharges},  we plot the charges for a bound state with $N=100$ and $a=\frac{1}{2}$, along with the large $N$ approximation above.

Several notable implications arise from the expression of the local charges. Firstly, the local charges are significantly large compared to the microscopic scale of the bound state. Indeed, these charges naturally scale with $M$, the ADM mass, rather than $\delta$ \eqref{eq:ConservedChargesMicroRN}. One might question how microscopic black holes can carry macroscopic charges while still satisfying the extremality bound for each black hole. This can be explained by the tightly bound nature of the solutions. A substantial portion of the energy of the elements in the bound state is used as binding energy, with only a small fraction contributing to the total ADM mass. 

For the extremal limit ($a=1$), we expect the energy of each black hole to be on the order of their charges, resulting in a total energy of the order $N M$ when unbounded. However, when bounded and in the critical limit where they resemble the Schwarzschild black hole, the black holes are gravitationally squeezed up to a microscopic size.  A significant portion of their energy is used in binding energy, while their initial macroscopic charges remain. This characteristic is similar to the neutral bound state of two extremal black holes constructed in \cite{Heidmann:2023kry}, where the local charges were of the order $M^\frac{4}{3}$.

Secondly, the charges have minimal dependence on the extremality parameter $a$, which emerges at the $N^{-1/2}$ subleading order and affects only the first few charges. This suggests that the charge distribution derived here follows a universal profile that electromagnetic degrees of freedom must satisfy in order to replace the Schwarzschild horizon with nontrivial electromagnetic structure. This profile indicates distributing large charges with alternating signs over the Schwarzschild horizon,  and with a magnitude of order $\sim 2\sqrt{x(2M-x)}$, where $x$ denotes the position of the charge along the horizon of size $2M$.\footnote{By horizon size,  we mean the rod size $\ell=2M$ of the Schwarzschild black hole \eqref{eq:SchwMassEntropyTemp}.} Consequently, the charge distribution is contained in a circle where charges are smaller at the poles and larger towards the equator, as shown in Fig.\ref{fig:LocalCharges}.

Moreover, it is evident that the infinitesimal net charge remaining upon summing all charges \eqref{eq:EMfieldAxisMicro} is a consequence of having fixed beforehand all parameters of the black holes in the class of solutions considered. Slight modifications in one charge, by an amount on the order of $M/\sqrt{N}$, would cancel out the total charge without affecting the main physics.

Lastly, it is noteworthy that substantial charges spread across the horizon of a Schwarzschild black hole can be activated without inducing significant electromagnetic effects on a large scale. Indeed, the solutions are indistinguishable from a Schwarzschild black hole for $r>2M(1+\cO(N^{-1}))$. This implies that the electromagnetic flux, while intense at $r=\ell \sim 2M$ to generate the charges, is almost vanishing just above $r=2M$. This phenomenon, termed ``electromagnetic entrapment,'' has been studied in \cite{Heidmann:2023thn}. It illustrates the capacity of ultra-compact, neutral, and highly-redshifted geometries to entrap their own electromagnetic flux, resulting in almost zero electromagnetic field outside the high-redshift region, regardless of the internal charge distribution. The current construction further exemplifies this phenomenon through a direct application to the Schwarzschild black hole.

\subsubsection{Entropy and temperature}
\label{sec:EntropyTemperature}

The entropy  $S_i$,  and temperature  $T_i$, of each black hole are determined by the generic formulas \eqref{eq:TemperatureAreaGen} and \eqref{eq:EntropyGen}. Given that the microscopic black holes are fixed beforehand, we do not anticipate them to be in thermal equilibrium, leading to different temperatures among the black holes. However, in the large $N$ limit, the system's symmetry imposes that the black holes have temperatures very close to the average temperature, $T$, expressed as:
\begin{equation}
T \equi \braket{\,T_i\,} \= \frac{1}{N} \sum_{i=1}^N T_i\,.
\end{equation}
Specifically, we have found that all black holes exhibit a temperature $T_i=T$ with corrections of order $N^{-1}$, except for the first few and last few black holes in the configuration, which deviate by a few percent. This implies that the black holes are nearly in thermal equilibrium, and slight adjustments to their positions could result in a fixed temperature for each black hole.

Additionally, we introduce the entropy of the bound state, given by the sum of the black hole entropies:\footnote{For the second equality,  we have simplified \eqref{eq:EntropyGen} by using that $T_i = T(1+\cO(N^{-1}))$ for most of the black holes in the bound state.}
\begin{equation}
S \= \sum_{i=1}^N S_i \= \sum_{i=1}^N \frac{(1-a) \delta}{4 T_i} \,\sim\, \frac{1-a}{T}\,\frac{M}{2}.
\end{equation}
This suggests that, like the Schwarzschild black hole, the entropy of the bound state increases linearly with the mass at a fixed temperature.

\begin{figure}%
    \centering
    \subfloat[\centering Entropy of the bound state.]{{\includegraphics[width=0.45\textwidth]{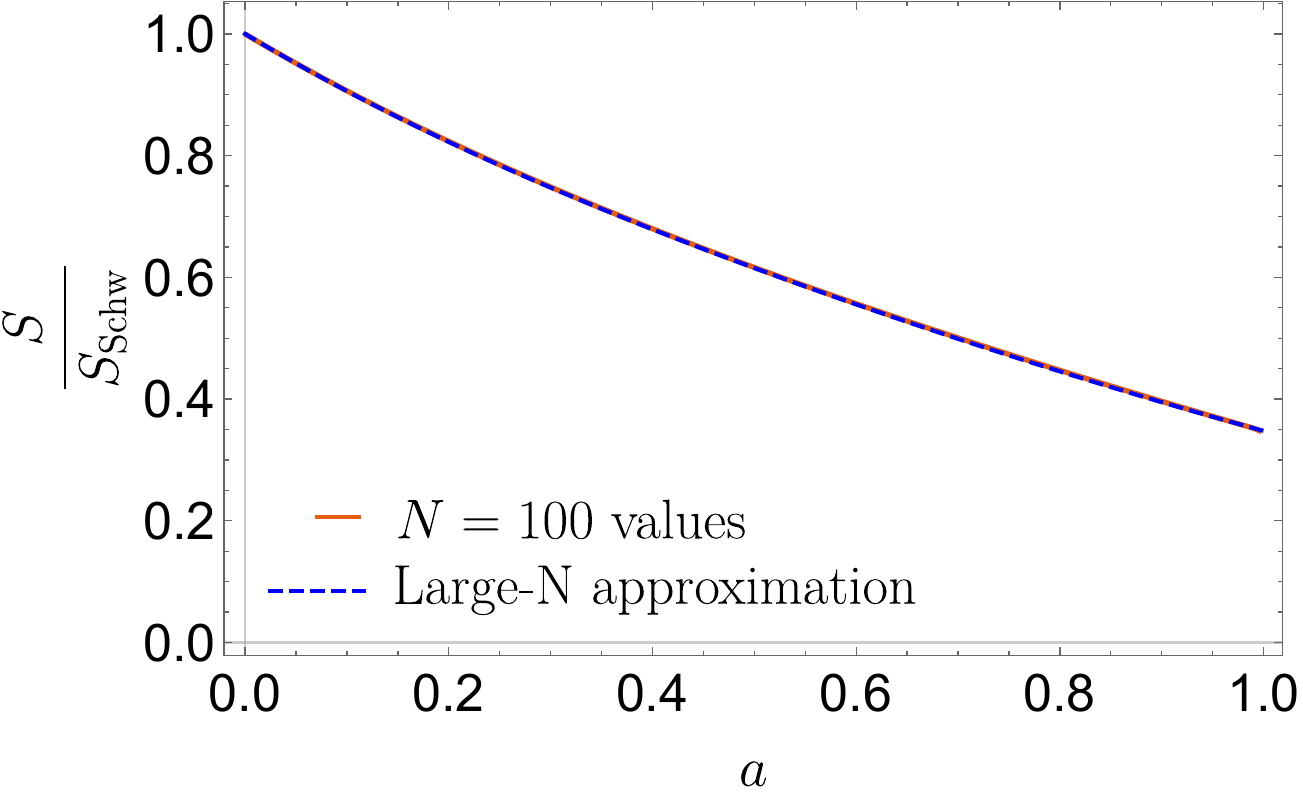} }}%
    \qquad
    \subfloat[\centering Average temperature of the black holes.]{{\includegraphics[width=0.45\textwidth]{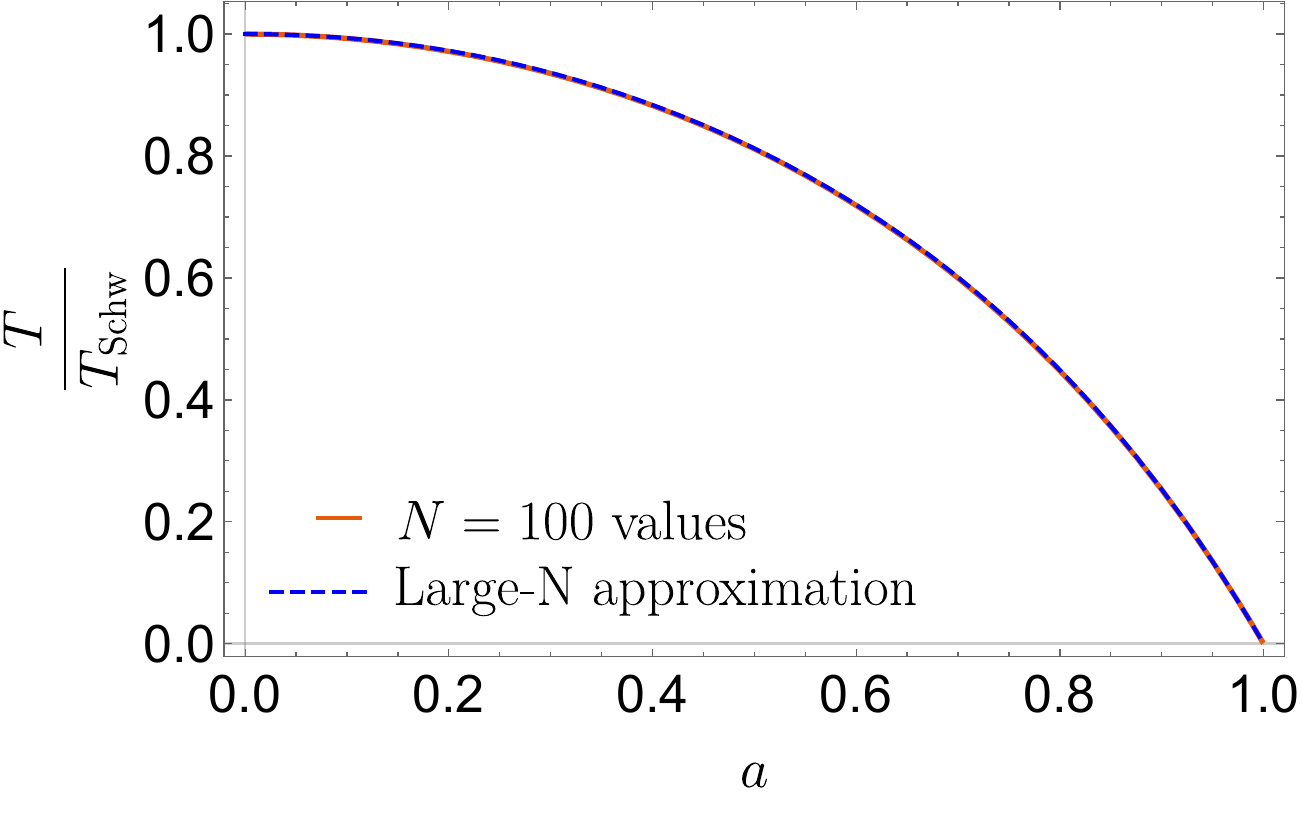} }}%
    \caption{Entropy and average temperature of the bound states as a function of $a$ and normalized by the Schwarzschild values \eqref{eq:SchwMassEntropyTemp}.  The plots show the values for bound states with $N=100$ black holes and the approximated values \eqref{eq:TempEntropApprox}.}%
    \label{fig:EntropyTemp}%
\end{figure}

In Fig.\ref{fig:EntropyTemp}, we plot the entropy $S$ and average temperature $T$ of the bound state for $N=100$ black holes as a function of the extremality parameter $a$. Unlike the local charges in the previous section, we were unable to find a satisfactory approximation using $\sin(\frac{\pi a}{2})$ as we would have expected. Instead, we found an approximate fit at large $N$, also plotted in Fig.\ref{fig:EntropyTemp}:\footnote{It is worth noting that the coefficient ``$1+a+\frac{3}{8} a^2+\frac{1}{2} a^4$" is simply a polynomial fit and other polynomials matching the values could have been found.}
\begin{equation}
T \,\sim\, (1-a)\left(1+a+\frac{3}{8} a^2+\frac{1}{2} a^4 \right)\,T_\text{Schw}\,,\qquad S \, \sim\, \frac{S_\text{Schw}}{1+a+\frac{3}{8} a^2+\frac{1}{2} a^4}\,,
\label{eq:TempEntropApprox}
\end{equation}
where $T_{\text{Schw}}=(8\pi M)^{-1}$ and $S_{\text{Schw}}=4\pi M^2$ are the temperature and entropy of a Schwarzschild black hole with the same mass.

Firstly, the extremality parameter significantly influences the temperature of the bound state as expected. The temperature is zero when the Reissner-Nordström black holes are extremal ($a=1$), while it matches the Schwarzschild temperature when the black holes merge into each other to form a single Schwarzschild rod ($a=0$). However, it is noteworthy that, for all values of $a$ from $0$ to $1$,  the entropy remains a fraction of the Schwarzschild entropy, ranging from $S_{\text{Schw}}$ when they merge to $\frac{8}{23}\,S_{\text{Schw}}\approx 0.35\, S_{\text{Schw}}$ when they are extremal. Thus, the bound states develop an \emph{effective temperature} from the relation between their entropy and mass:
\begin{equation}
2M \,\=\,  T_\text{eff}\, S \,,\qquad T_\text{eff} \sim \left(1+a+\frac{3}{8} a^2+\frac{1}{2} a^4\right) \, T_\text{Schw}\,.
\label{eq:EffectiveTemp}
\end{equation}
Consequently, all bound states exhibit a similar mass/entropy relation to the Schwarzschild black hole, even at extremality when the temperature is zero. The effective temperature ranges from the Schwarzschild temperature when the Reissner-Nordström black holes merge, to approximately $2.9 \,T_{\text{Schw}}$ at extremality. Future research may explore whether this effective temperature can be associated with radiation from the bound state through (charged) particle pair creation, akin to the Hawking process \cite{Hawking:1974rv,Hawking:1975vcx}.\\

This concludes the analysis of the microscopic black holes that have replaced the horizon of the Schwarzschild black hole. First, we observed that the charges at the black holes follow an apparently universal distribution, mainly independent of the extremality parameter $a$, and necessitating the charge distribution to take the form $\sim 2 \sqrt{x(2M-x)}$, where $x$ is the position of the black holes in the bound state of size $\sim 2M$. Second, we demonstrated that despite having a temperature that differs significantly from the Schwarzschild temperature, especially at extremality when $a=1$ and $T=0$, the bound state exhibits an effective temperature, entropy, and mass/entropy relation closely resembling the Schwarzschild thermodynamic properties.

\subsection{Strut properties}
\label{sec:StrutProp}

The internal structure is described not only by  Reissner-Nordström black holes but also by struts. Struts are well-understood physical conical singularities that naturally emerge in four dimensions when dealing with black hole bound states \cite{Costa:2000kf}. They correspond to strings with negative tension and energy, necessary to counterbalance the lack of repulsion between the black holes and prevent the bound state from gravitational collapse \cite{Gibbons:1979nf,Costa:2000kf,Bah:2021owp}. As argued in \cite{Elvang:2002br,Bah:2021owp}, struts are features of four-dimensional classical theories of gravity and can be classically resolved into smooth topological bubbles, a point we will discuss in the Section \ref{sec:Conclusion}.

\begin{figure}%
    \centering
    \subfloat[\centering Strut tensions for $a=10^{-3}$.]{{\includegraphics[width=0.45\textwidth]{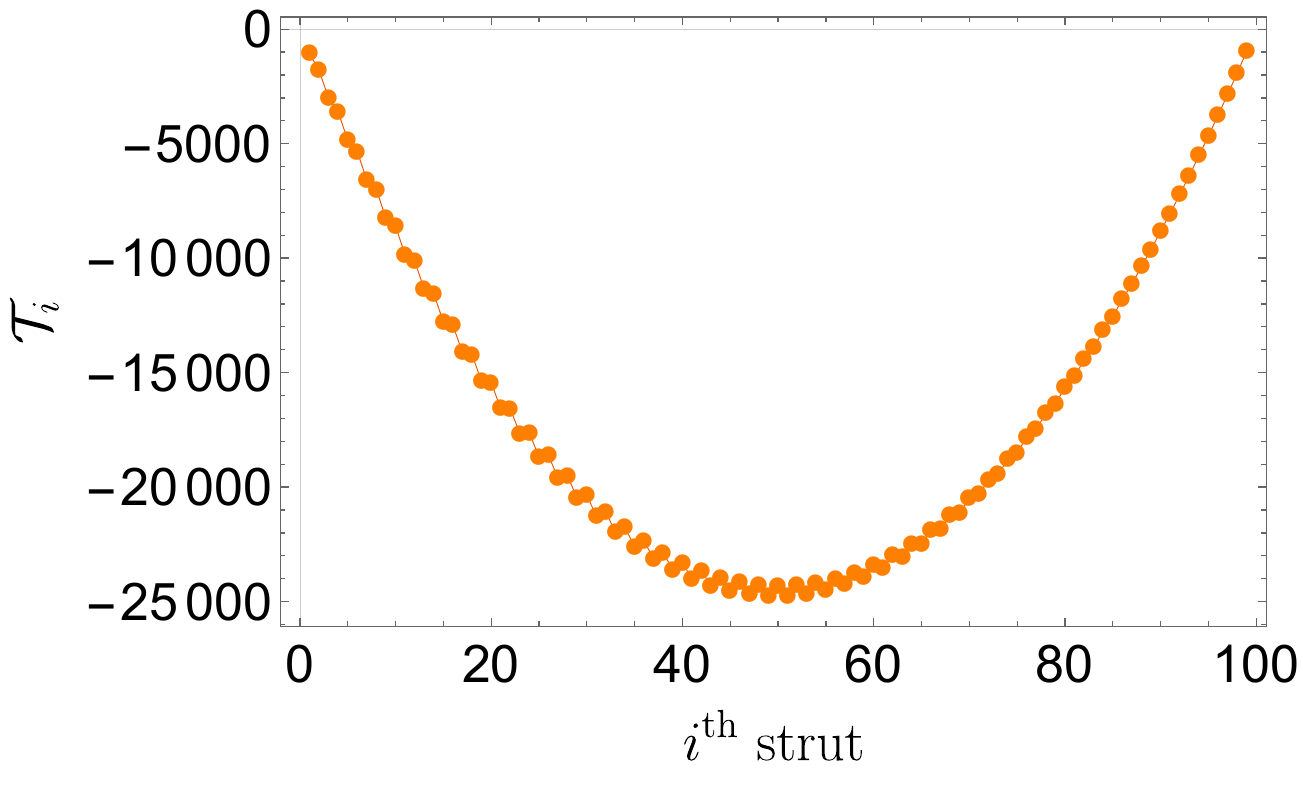} }}%
    \qquad
    \subfloat[\centering Strut tensions for $a=1-10^{-3}$.]{{\includegraphics[width=0.45\textwidth]{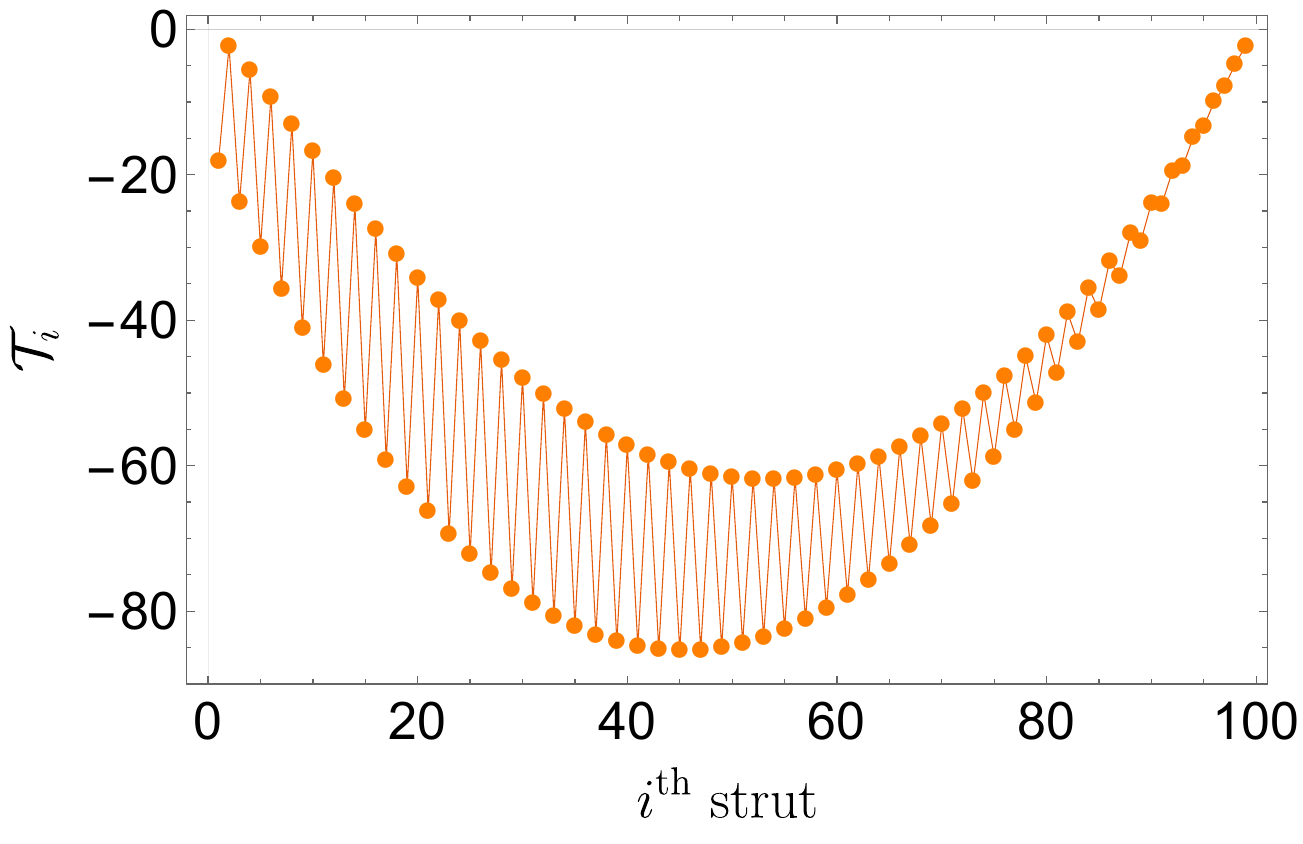} }}%
    \caption{Tensions of the struts \eqref{eq:StrutTensionEnergy} in a bound state with $N=100$ black holes and for two illustrative value of $a$ (in units $G_4=1$).  }%
    \label{fig:StrutTension}%
\end{figure}

Despite being singular, struts carry physical features of the bound state. In this section, we analyze their characteristics in terms of energy and tension and how they evolve as we change the internal extremality parameter $a$.  \\

In Fig.\ref{fig:StrutTension}, we present the strut tensions, $\cT_i$ \eqref{eq:StrutTensionEnergy}, of the $N-1$ struts for two bound states made of $N=100$ Reissner-Nordström black holes with $a=10^{-3}$  (close to merging) and $a=1-10^{-3}$ (close to extremality).

First, the strut tension is indeed negative and reaches its extremum in the middle of the bound state where gravitational contraction is most intense. Moreover, the strut tension oscillates slightly, especially close to extremality,  following the pattern of alternative signs of the charges.
 
\begin{figure}%
    \centering
    \subfloat[\centering Maximal strut tension in the bound state.]{{\includegraphics[width=0.45\textwidth]{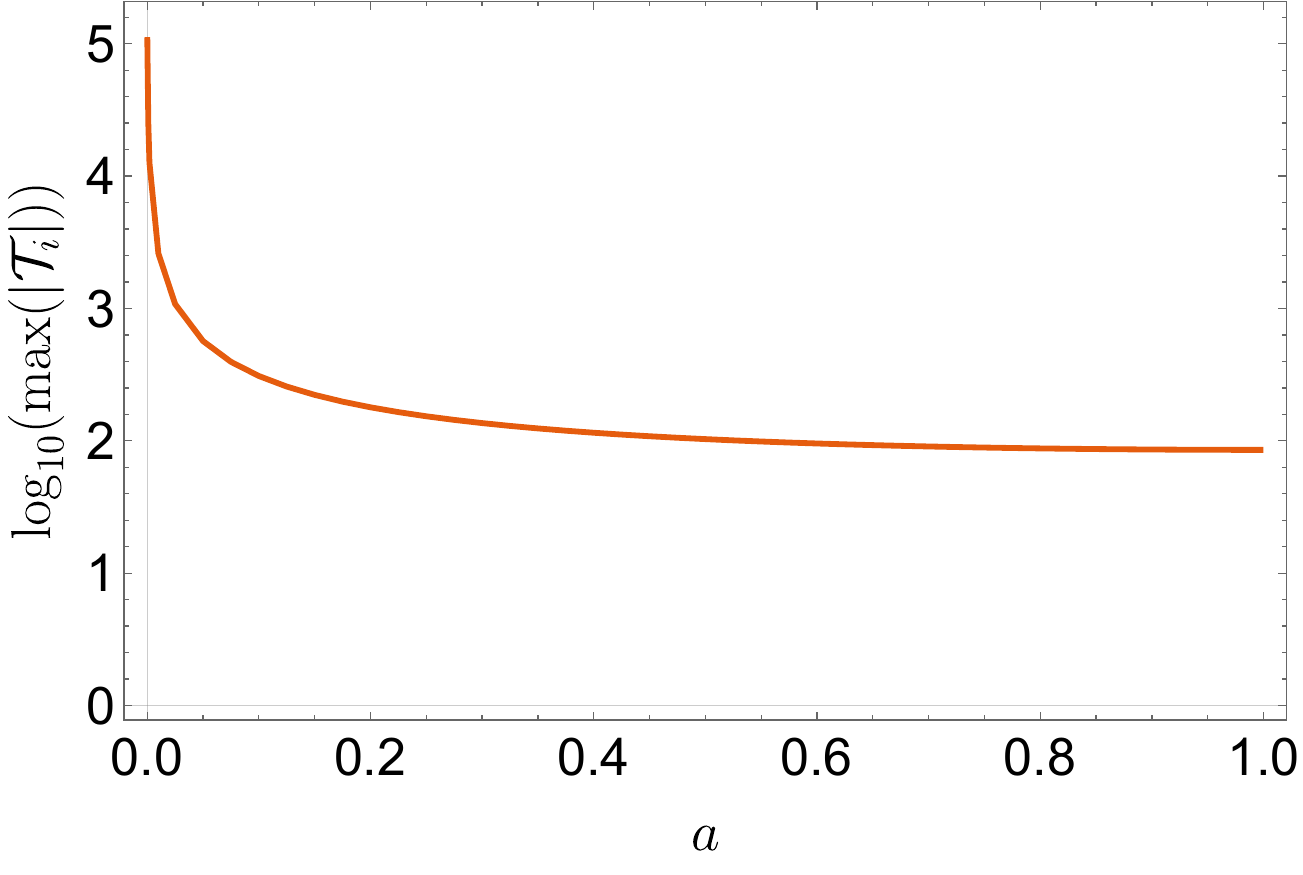} }}%
    \qquad
    \subfloat[\centering Total strut energy in the bound state.]{{\includegraphics[width=0.45\textwidth]{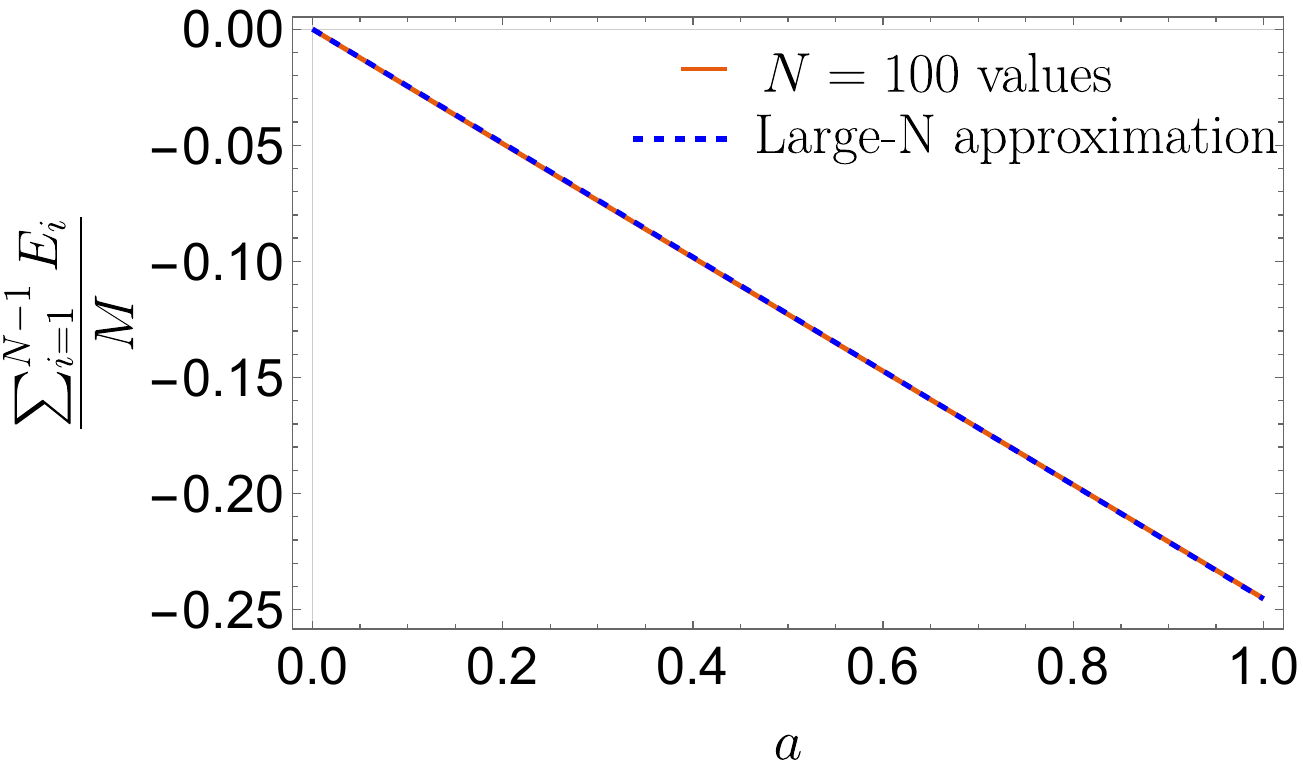} }}%
    \caption{Maximal strut tension and total strut energy (normalized by the ADM mass) in the bound state as a function of $a$. }%
    \label{fig:StrutEnergyTension}%
\end{figure}

 Furthermore, as the black holes approach their merging limit $a\to 0$, the strut tensions generally increase. The Fig.\ref{fig:StrutEnergyTension}(a) show a logarithmic plot of the maximal tension in the bound state as a function of $a$. As $a\to 0$, the black holes are on the verge of merging, necessitating the strut tension to diverge to withstand the intense attraction between them. However, for $a\gtrsim 0.2$, the strut tensions do not undergo significant changes, and the maximal tension is of order (in units $G_4=1$):
 \begin{equation}
 \text{max}(|\cT_i|) \= \cO(N)\,, \qquad 0.2 \lesssim a \leq 1\,.
 \label{eq:MaxTensionApprox}
 \end{equation}
This is because the separation between the black holes remains a fraction of their overall size as long as $a$ is not close to $0$. Consequently, strut tensions do not need to be extreme and simply increase linearly with the number of black holes in the bound state.
 
Finally, we derive the energy of the struts, $E_i$ \eqref{eq:StrutTensionEnergy}. We find that the struts along the bound state have the same energy with $N^{-1}$ corrections:
 \begin{equation}
 E_i \,\sim\, - \frac{a\,\delta}{4}\,.
 \end{equation}
In Fig.\ref{fig:StrutEnergyTension}(b), we plot the total energy of the struts $E_\text{struts}\equiv \sum_{i=1}^{N-1} E_i$. We find that the energy is mainly given by the following large $N$ approximation:
 \begin{equation}
 E_\text{struts} \,\sim\,  - \frac{a\,M}{2} \left(1- \frac{1.84}{N} \right) .
 \end{equation}
Thus, the negative energy carried by the strut reaches a maximum magnitude of $M/2$ when the black holes are extremal, and approaches zero when the black holes are almost merging.

Therefore, despite the bound states being in their ultra-compact limit and indistinguishable from a Schwarzschild black hole, the characteristics of the struts to prevent total collapse are not as extreme as one might have expected. Their negative energy scales with the ADM mass, while their negative tensions are not extremely high when the black holes are significantly far from their merging regime $a\gtrsim 0.2$. These are good indicators that the struts can be resolved in higher dimensions by topological bubbles that will provide the necessary pressure to replace the strut tension and energy without singularities. Examples of such resolution can be found in \cite{Elvang:2002br,Bah:2021owp,Bah:2021rki,Bah:2022yji,Astorino:2022fge,Heidmann:2023thn,Heidmann:2023kry}.

\subsection{Deformation of the two-sphere}
\label{sec:S2Profile}

In this section, our objective is to describe the geometry of the bound states and compare it to that of the Schwarzschild black hole.  Thus, we analyze the two-sphere for $r\geq \ell \sim 2M$. In the spherical coordinates \eqref{eq:SpherCoord},  we remind that the metric is given by 
\begin{equation}
\begin{split}
ds_4^2 = - \frac{dt^2}{Z^2}+ Z^2\left(1-\frac{\ell}{r}\right) \left[G \left( \frac{dr^2}{1-\frac{\ell}{r}}+r^2 d\theta^2\right) +r^2 \sin^2\theta \, d\phi^2 \right],
\end{split}
\end{equation}
where $G\equiv e^{4\nu}\,\left( 1- \frac{\ell\cos^2 \frac{\theta}{2}}{r}\right)\left( 1- \frac{\ell\sin^2 \frac{\theta}{2}}{r}\right)\left(1-\frac{\ell}{r}\right)^{-1}$.  The two-sphere is described by the line element:
\begin{equation}
ds_{S^2}^2 \= r^2 \,Z^2\left(1-\frac{\ell}{r}\right)  \left[G \,d\theta^2 + \sin^2 \theta \,d\phi^2 \right]\,.
\label{eq:TwoSphereGen}
\end{equation}

For a Schwarzschild black hole,  $Z^{-2}=1-\frac{\ell}{r}$ and $G=1$,  resulting in a round two-sphere described by $ds_{S^2}^2 = r^2 \left( d\theta^2 + \sin^2 \theta \,d\phi^2\right)$, with a radius $r$ ranging from infinity to $\ell = 2M$ at the horizon.

Since the bound states of $N$ Reissner-Nordström black holes are indistinguishable from a Schwarzschild black hole from the asymptotic up to $r\gtrsim 2M(1+\cO(N^{-1}))$, we expect the two-sphere to be similarly round in this range of values, and significantly deformed below, up to the location of the black holes and struts at $r=\ell$.

\begin{figure}%
    \centering
    \subfloat[\centering Radius of the two-sphere, normalized by $2M$,  as a function of $r$.]{{\includegraphics[width=0.45\textwidth]{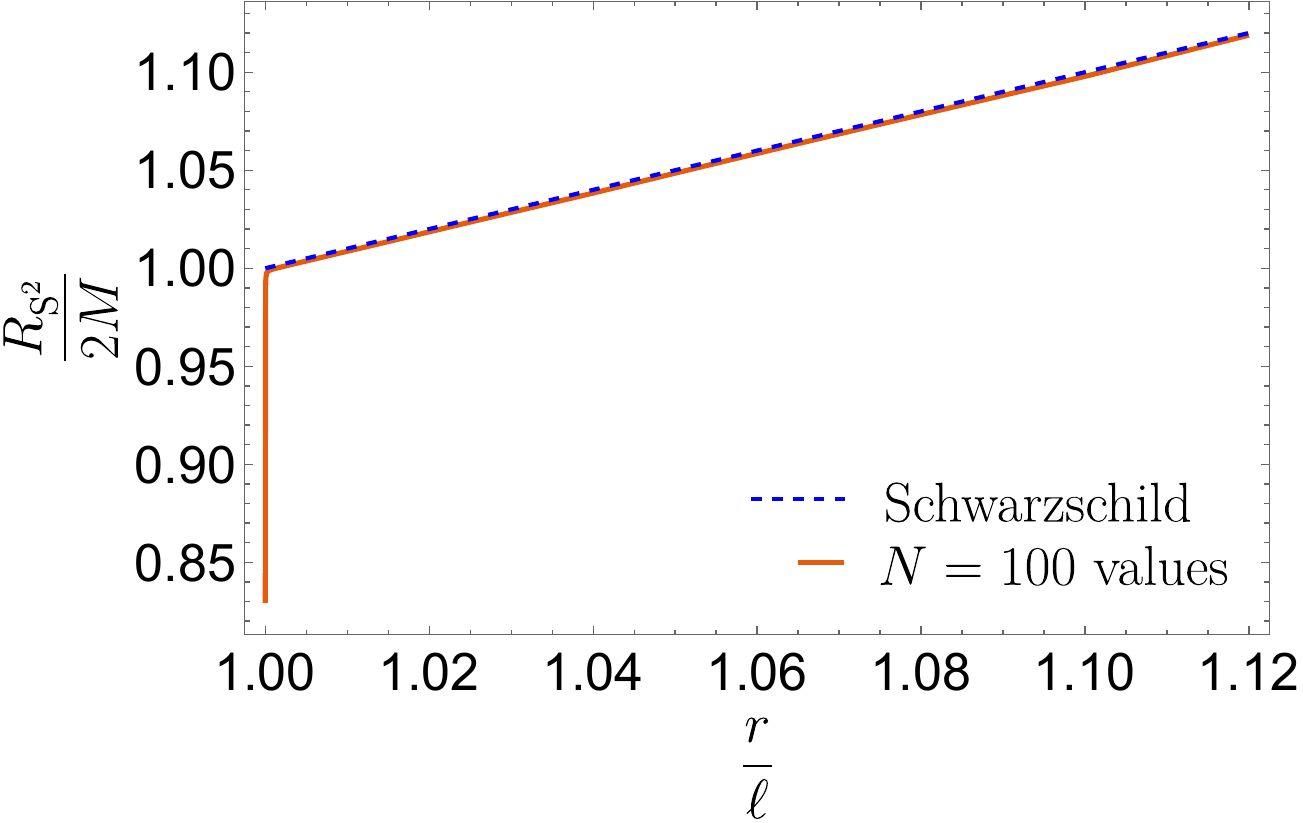} }}%
    \qquad
    \subfloat[\centering Axisymmetry factor of the two-sphere as a function of $r$.]{{\includegraphics[width=0.45\textwidth]{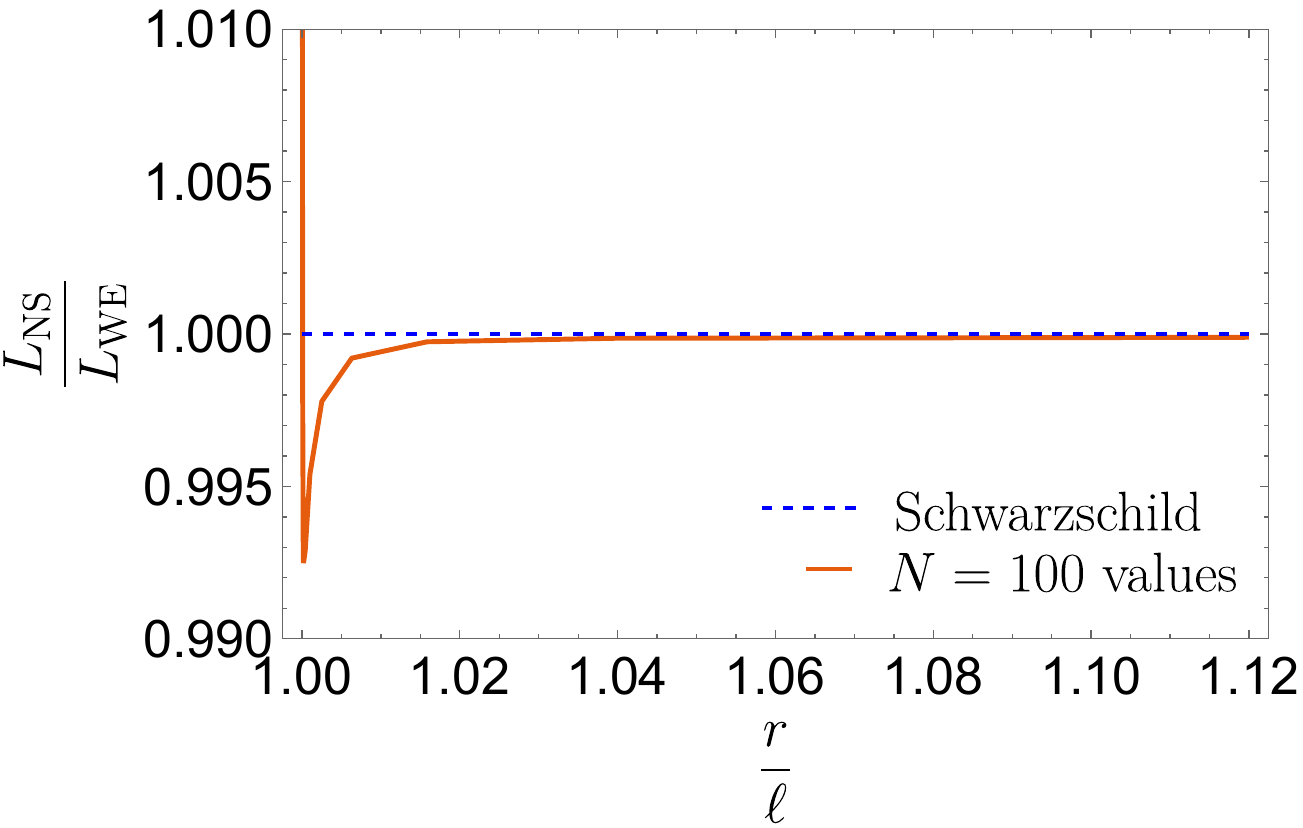} }}%
    \caption{Global characteristics of the two sphere as a function of the radial coordinate $r\geq \ell$.   }%
    \label{fig:SphereProp}%
\end{figure}

In Fig.\ref{fig:SphereProp}, we provide the global characteristics of the two-sphere for $N=100$ black holes and $a=\frac{1}{2}$. We introduce the radius of the two-sphere, denoted as $R_{S^2}$, its meridian length $L_{NS}$, and its equator length $L_{WE}$ as follows:
\begin{equation}
R_{S^2}(r)^2 \equi \frac{1}{4 \pi}\,  \int \sqrt{ g_{\theta \theta} g_{\phi \phi}}  \,d\theta d\phi \,,\quad L_{NS}(r) \equi \int_{\theta=0}^\pi  \sqrt{ g_{\theta \theta}}\,d\theta \,,\quad L_{WE}(r) \equi \pi \sqrt{g_{\phi \phi}}|_{\theta=\frac{\pi}{2}}\,.
\label{eq:DefS2RadAxisym}
\end{equation}
Moreover,  we have defined the axisymmetry factor, $L_{NS}/L_{WE}$, which indicates how much the two-sphere is stretched along its poles by comparing its meridian and equator lengths.

The Figure \ref{fig:SphereProp} notably illustrates that despite the bound state beginning to deviate from  Schwarzschild around $r\sim 2M(1+\cO(N^{-1}))$, its global characteristics remain closely aligned. Specifically, the radius of the S$^2$ matches that of Schwarzschild up to $r\sim \ell (1+10^{-4})$ for $N=100$ and $a=1/2$.  Below this distance,  it abruptly drops to approximately $0.8$,  which aligns with the value of the bound state entropy given in Fig.\ref{fig:EntropyTemp}(a). Additionally, axisymmetry begins to manifest at the same range of radius $r$.

Therefore, despite significant local deviations with the Schwarzschild black hole at $r\sim 2M(1+\cO(N^{-1}))$, these discrepancies have minimal impact on global quantities such as the overall sphere radius or axisymmetry factor.  \\

To provide a more detailed description of the two-sphere, we derive the two-dimensional surface corresponding to the sphere and described by the line element \eqref{eq:TwoSphereGen}.  We follow the procedure outlined in \cite{Costa:2000kf}, which involves changing the angular coordinate to:
\begin{equation}
ds_{S^2}^2 \= R(r,\widetilde{\theta})^2 \left[ d\widetilde{\theta}^2 + \sin^2 \widetilde{\theta}\, d\phi^2\right]\,.
\end{equation}
Subsequently, we plot the two-dimensional surface of radius $R(r,\widetilde{\theta})$, parameterized by $0\leq \phi<2\pi$ and $0\leq \widetilde{\theta}\leq \pi$, to represent the geometry of the two-sphere at the radial coordinate $r$.

\begin{figure}[t]
\begin{center}
\includegraphics[width= 0.9 \textwidth]{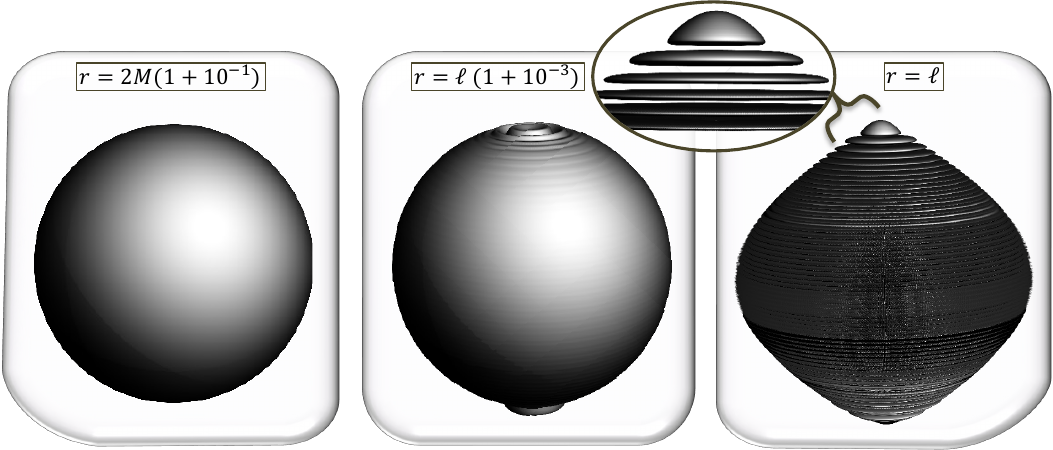}
\caption{The two-sphere of a bound state with $N=100$ Reissner-Nordström black hole and $a=\frac{1}{2}$.  The two-sphere is given for three values of $r$ from left to right: at $r\gtrsim 2M(1+\epsilon)$ where the bound state is indistinguishable from a Schwarzschild black hole (round sphere),  at $r\sim 2M(1+\epsilon)$ where the geometry starts to deviate from Schwarzschild and at the sources $r=\ell$  corresponding to $N$ black holes held apart by struts. }
\label{fig:TwoSphere100}
\end{center}
\end{figure}

The first two plots in Fig.\ref{fig:TwoSphere100} illustrate the two-sphere at radii very close to $2M$. The first plot corresponds to $r=2M(1+10^{-1})$, within a range where the bound state closely resembles a Schwarzschild black hole.  As expected,  the plot indicates a spherical shape with minor corrections. 

The second plot, for $r=2M(1+10^{-3})$, falls below the range where significant $N^{-\frac{1}{2}}$ deviations should occur. Notably, it reveals noticeable local deviations, manifesting as oscillations atop the spherical shape. These oscillations are imprints of the internal black holes which start to manifest at this scale.  Moreover,  having oscillations explains why the macroscopic features of the sphere, such as its radius and axisymmetry, remain nearly identical to those of a round sphere.

The final plot depicts the two-dimensional space at the coordinate boundary $r=\ell$, where the microscopic black holes and struts are located. This region is not topologically S$^2$ due to the presence of struts, which cause the $\phi$ circle to degenerate across entire ranges of $\theta$. However, each black hole has a S$^2$ horizon.  Therefore, following the same methodology outlined in \cite{Costa:2000kf}, we derive the two-sphere at each black hole, providing the shape of their horizons. Subsequently, we stack them together.  The distances that we use to separate the two-spheres are the physical distances between the black holes (the length of the struts),  given by $\int_{z=\alpha_{2i}}^{\alpha_{2i+1}} \sqrt{g_{zz}}$.

This plot shows the microscopic structure that has replaced the Schwarzschild horizon. The black holes experience significant compression within the intense gravitational environment surrounding the bound states. With the exception of the black holes located at the extremities,  their horizons are significantly flattened at their poles and extended along the $\phi$ direction, resulting in an elongated direction. 

 Remarkably,  when all black holes are put together,  these elongated directions follow the shape of a sphere with lengths approximately of $\sim 2M\sin \theta$,  where $\theta$ corresponds to the angular position of the black hole within the bound state.  We believe that this striking sphere-like shape,  which is necessary for having an area that scales as $(2M)^2$,  arises from an interplay between strong gravitational contraction at the equator and the specific charge distribution of the black holes derived earlier. 
 
 Moreover, despite the pronounced differences in shape, the black holes have very close horizon areas.  Furthermore,  the presence of the struts within the bound state is not visually apparent; they exist as segments situated between the black holes where the two-dimensional space degenerates as the cylindrical coordinate degeneracy.

\section{Schwarzschild resolution into a small number of black holes}
\label{sec:AnalysisSmallN}

In the preceding section, we provided a detailed analysis of the class of solutions introduced in Section \ref{sec:ClassMicroBH}. These solutions correspond to bound states of $N \gg 1$ microscopic Reissner-Nordström black holes separated by struts. 
They offer the advantage of providing a clear technique for generating solutions that highlight important internal parameters. A crucial parameter in this context is the extremality parameter $a$, which allows for tuning the temperature from zero (where the black holes are extremal) to the Schwarzschild temperature (where the black holes merge into a single black hole), while maintaining indistinguishability from Schwarzschild and an entropy that remains a fraction of the Schwarzschild entropy.

However, a significant drawback is that the solutions retain a negligible net charge of the order $M/\sqrt{N}$, resulting in different conserved charges compared to the Schwarzschild black hole. Moreover,  this charge induced a slow convergence towards Schwarzschild as $N$ becomes large. In this section, we modify slightly the class of solutions to construct bound states with zero net charge, while remaining indistinguishable from Schwarzschild in a range $r\gtrsim 2M(1+\epsilon)$. The absence of residual charge enhances the efficiency of achieving indistinguishability from Schwarzschild, enabling consideration of bound states with a smaller number of black holes.


In the examples analyzed in this section, we primarily focus on bound states composed of $N=10$ black holes. We first elaborate on the class of solutions, demonstrating their indistinguishability from a Schwarzschild black hole. Finally, we explore the internal structure of the bound states, mirroring the approach taken for the previous solutions.

\subsection{Ernst solution}
\label{sec:BSRN2}

The solutions under consideration also belong to a subset of the Ernst solutions corresponding to $N$ collinear Reissner-Nordström black holes \cite{NoraBreton1998}. These solutions are described by the four-dimensional metric and electromagnetic fields, as expressed in \eqref{eq:StaticErnstMetric4d} in Weyl-Papapetrou coordinates and \eqref{eq:GenericMetricSpher} in spherical coordinates.  Further details regarding these fields can be found in Section \ref{sec:NRNGen} and Appendix \ref{app:ErnstSolNRN}.

In Section \ref{sec:ClassMicroBH}, we constructed a family of solutions by fixing the positions of all black holes,  $\alpha_k$, along with the parameters $\beta_k$,  in microscopic patterns of length $\delta$. Consequently, the parameters $f_k$ \eqref{eq:fparamNmicroRN}, which contribute to the electromagnetic axis data, were fully determined except for their signs. The resulting solutions carried a residual net charge of order $M/\sqrt{N}$.

In this section, we adopt a different approach and construct a class of solutions,  similar to the previous one,  but characterized by zero net charge and possessing a less predictable internal structure.  The generic solutions of $N$ collinear Reissner-Nordström black holes \cite{NoraBreton1998} are given by $3N$ parameters.  Instead of fixing all the black hole positions and parameters $\beta_k$,  we will fix the $f_k$, and have a greater control on the electromagnetic properties of the solutions.  \\

First, we maintain the same values for the parameters $\beta_k$ and for $N-1$ edges of black hole rods:
\begin{equation}
   \beta_1 \= - \frac{a}{2}\,\delta \,,\qquad \beta_k \= \left(k-1-\frac{a}{2}\right)\, \delta\,,\qquad   \alpha_{2k-1} \= (k-1) \,\delta,\qquad k=2,\ldots,N\,.
   \label{eq:NewParam1}
\end{equation}
Thus,  $\delta$ represents similarly the microscopic scale of the bound state,  and $a$ adjusts the distance between the $\beta_k$ and $\alpha_{2k-1}$. Then,  we fix the $f_k$ in pairs of alternative sign:
\begin{equation}
f_{2i-1} \= -f_{2i} \= \cF\left(2i-\frac{3}{2}+\frac{\sin(\frac{\pi}{2}\,a)}{\pi}\right)\,,\quad  \cF(x) \equi \frac{\delta \,\sin(\frac{\pi}{2}\,a)}{\pi}\,\sqrt{\frac{N-x}{x} },\quad i=1,\ldots ,\frac{N}{2}.
\label{eq:choiceFi}
\end{equation}
Therefore, the remaining $N$ rod endpoints,  $\alpha_{2k}$, are determined to fulfill the constraints on the norm of the $f_i$ \eqref{eq:DefFiEi}:
\begin{equation}
    f_i^2 \= \frac{\prod_{k=1}^{2N} (\beta_i-\alpha_k)}{\prod_{k\neq i} (\beta_i-\beta_k)^2}\,,
    \label{eq:FiConstraint}
\end{equation}
and $\alpha_1$ is determined so that the ADM mass of the bound state equals half the rod configuration, $\ell = 2M = \alpha_{2N} - \alpha_1$, like the Schwarzschild solution \eqref{eq:SchwMassEntropyTemp}:
\begin{equation}
M\=\sum_{i=1}^N\left(\frac{\alpha_{2i-1}+\alpha_{2i}}{2} -\beta_i \right) \= \frac{\alpha_{2N}-\alpha_1}{2}\,.
\label{eq:MassConstraint}
\end{equation}

It is worth noting that these constraints cannot be solved analytically and must be addressed on a case-by-case basis. However, fixing the $f_i$ rather than the $\alpha_{2i}$ provides better control over the electromagnetic properties of the bound state.

Primarily, the choice \eqref{eq:choiceFi} ensures that the total charge, $Q = \sum f_i$ \eqref{eq:ConsChargeGen}, is zero, rendering the bound state with the exact same conserved charge as the Schwarzschild black hole. Additionally, it can be demonstrated that all electromagnetic multipole moments scale proportionally to $M^n/N$.  Consequently, we anticipate the bound states to be in their entrapment limit, as defined in \cite{Heidmann:2023thn}, and to resemble the Schwarzschild black hole.

Moreover, we deliberately selected the $f_i$ \eqref{eq:choiceFi} to be near the values obtained from the previous  solutions \eqref{eq:ApproxFi}. This ensures that the values of $\alpha_1$ and $\alpha_{2k}$,  obtained after solving the constraints \eqref{eq:FiConstraint} and \eqref{eq:MassConstraint},  are equal to those of the previous class of solutions with $N^{-1}$ corrections:
\begin{equation}
\alpha_1 \sim 0 \,,\qquad \alpha_{2k} \sim (k-a)\delta\,, \qquad k=1,\ldots,N\,.
\label{eq:NewParam2}
\end{equation}

Therefore, the class of solutions presented here can be viewed as a refinement of the previous solutions, ensuring:
\begin{itemize}
\item The charges and positions of the microscopic black holes are slightly modified to ensure that the bound state is neutral and that all higher electromagnetic multipoles vanish at large $N$ as $M^n/N$.
\item The size of the rod configuration strictly corresponds to twice the ADM mass,  $\ell=2M$,  as a Schwarzschild solution.  Moreover,   the configuration still splits in microscopic patterns of length $\delta$ containing one microscopic Reissner-Nordström black hole so that
\begin{equation}
\delta \,\sim\, \frac{\ell}{N} \= \frac{2M}{N}\,.
\end{equation}
\end{itemize}

\begin{figure}
\centering
    \begin{tikzpicture}[dot/.style = {circle, fill, minimum size=#1,
              inner sep=0pt, outer sep=0pt},
dot/.default = 6pt  
                    ] 
\def\deb{-13} 
\def\inter{0.7} 
\def\ha{2.8} 
\def\zaxisline{5} 
\def\rodsize{1.2} 
\def\interrod{1} 
\def\fin{-1.5} 
\def\posx{\deb+0.5+1.5*\rodsize}
\def\posy{\ha-1.1*\inter}
\def\segment{2.5} 
\def\propa{0.63} 
\def\propb{0.5} 
\def\debB{-12.5}




\draw (\fin+0.2,\ha-\zaxisline*\inter-0.3) node{$z$};

\draw[black,line width=0.3mm] (\deb,\ha-\zaxisline*\inter)  -- (\fin-1.5*\segment,\ha-\zaxisline*\inter);
\draw[black,line width=0.3mm,dotted] (\fin-1.5*\segment,\ha-\zaxisline*\inter) -- (\fin-1.1*\segment,\ha-\zaxisline*\inter);
\draw[black,->, line width=0.3mm] (\fin-1.1*\segment,\ha-\zaxisline*\inter) -- (\fin+0.2,\ha-\zaxisline*\inter);


\draw[line width=0.3mm,byzantine] (\debB,\ha-\zaxisline*\inter+0.1) -- (\debB,\ha-\zaxisline*\inter-0.1);

\draw[line width=0.3mm,byzantine] (\debB+\segment,\ha-\zaxisline*\inter+0.1) -- (\debB+\segment,\ha-\zaxisline*\inter-0.1);

\draw[line width=0.3mm,byzantine] (\debB+2*\segment,\ha-\zaxisline*\inter+0.1) -- (\debB+2*\segment,\ha-\zaxisline*\inter-0.1);

\draw[line width=0.3mm,byzantine] (\fin-0.5-\segment+\propa*\segment-\propa*\propb*\segment,\ha-\zaxisline*\inter+0.1) -- (\fin-0.5-\segment+\propa*\segment-\propa*\propb*\segment,\ha-\zaxisline*\inter-0.1);


\draw[black,line width=1mm] (\debB+\propa*\propb*\segment,\ha-\zaxisline*\inter) -- (\debB+\propa*\propb*\segment+\segment-\propa*\segment,\ha-\zaxisline*\inter);
\draw[line width=0.3mm] (\debB+\propa*\propb*\segment,\ha-\zaxisline*\inter+0.1) -- (\debB+\propa*\propb*\segment,\ha-\zaxisline*\inter-0.1);
\draw[line width=0.3mm] (\debB+\propa*\propb*\segment+\segment-\propa*\segment,\ha-\zaxisline*\inter+0.1) -- (\debB+\propa*\propb*\segment+\segment-\propa*\segment,\ha-\zaxisline*\inter-0.1);

\draw[black,line width=1mm] (\debB+\propa*\propb*\segment+\segment,\ha-\zaxisline*\inter) -- (\debB+\propa*\propb*\segment+2*\segment-\propa*\segment,\ha-\zaxisline*\inter);
\draw[line width=0.3mm] (\debB+\propa*\propb*\segment+\segment,\ha-\zaxisline*\inter+0.1) -- (\debB+\propa*\propb*\segment+\segment,\ha-\zaxisline*\inter-0.1);
\draw[line width=0.3mm] (\debB+\propa*\propb*\segment+2*\segment-\propa*\segment,\ha-\zaxisline*\inter+0.1) -- (\debB+\propa*\propb*\segment+2*\segment-\propa*\segment,\ha-\zaxisline*\inter-0.1);

\draw[black,line width=1mm] (\debB+\propa*\propb*\segment+2*\segment,\ha-\zaxisline*\inter) -- (\debB+\propa*\propb*\segment+3*\segment-\propa*\segment,\ha-\zaxisline*\inter);
\draw[line width=0.3mm] (\debB+\propa*\propb*\segment+2*\segment,\ha-\zaxisline*\inter+0.1) -- (\debB+\propa*\propb*\segment+2*\segment,\ha-\zaxisline*\inter-0.1);
\draw[line width=0.3mm] (\debB+\propa*\propb*\segment+3*\segment-\propa*\segment,\ha-\zaxisline*\inter+0.1) -- (\debB+\propa*\propb*\segment+3*\segment-\propa*\segment,\ha-\zaxisline*\inter-0.1);

\draw[black,line width=1mm] (\fin-0.5-\segment+\propa*\segment,\ha-\zaxisline*\inter) -- (\fin-0.5,\ha-\zaxisline*\inter);
\draw[line width=0.3mm] (\fin-0.5-\segment+\propa*\segment,\ha-\zaxisline*\inter+0.1) -- (\fin-0.5-\segment+\propa*\segment,\ha-\zaxisline*\inter-0.1);
\draw[line width=0.3mm] (\fin-0.5,\ha-\zaxisline*\inter+0.1) -- (\fin-0.5,\ha-\zaxisline*\inter-0.1);

\draw[arsenic,<->] (\debB+\segment,\ha-\zaxisline*\inter+0.7) -- (\debB+\propa*\propb*\segment+\segment,\ha-\zaxisline*\inter+0.7) node[midway,above,gray] { {\footnotesize $ a \delta/2$}};

\draw[arsenic,<->] (\debB+\propa*\propb*\segment+2*\segment,\ha-\zaxisline*\inter+0.7) -- (\debB+\propa*\propb*\segment+3*\segment-\propa*\segment,\ha-\zaxisline*\inter+0.7) node[midway,above,gray] { {\footnotesize $ \sim (1-a) \delta$}};

\draw[arsenic,<->] (\debB,\ha-\zaxisline*\inter+0.5) -- (\debB+\segment,\ha-\zaxisline*\inter+0.5) node[midway,above,gray] { $\delta$};

\draw[arsenic,<->] (\debB+\propa*\propb*\segment,\ha-\zaxisline*\inter-1) -- (\fin-0.5,\ha-\zaxisline*\inter-1) node[midway,below,gray] { $\ell=2M$};


\draw (\debB,\ha-\zaxisline*\inter-0.5) node{{\color{byzantine}{\small $\beta_1$}}};
\draw (\debB+\segment,\ha-\zaxisline*\inter-0.5) node{{\color{byzantine}{\small $\beta_2$}}};
\draw (\debB+2*\segment,\ha-\zaxisline*\inter-0.5) node{{\color{byzantine}{\small $\beta_3$}}};
\draw (\fin-0.5-\segment+\propa*\segment-\propa*\propb*\segment,\ha-\zaxisline*\inter-0.5) node{{\color{byzantine}{\small $\beta_N$}}};

\draw (\debB+\propa*\propb*\segment,\ha-\zaxisline*\inter-0.5) node{{\small $\alpha_1$}};
\draw (\debB+\propa*\propb*\segment+\segment-\propa*\segment,\ha-\zaxisline*\inter-0.5) node{{\small $\alpha_2$}};

\draw (\debB+\propa*\propb*\segment+\segment,\ha-\zaxisline*\inter-0.5) node{{\small $\alpha_3$}};
\draw (\debB+\propa*\propb*\segment+2*\segment-\propa*\segment,\ha-\zaxisline*\inter-0.5) node{{\small $\alpha_4$}};

\draw (\debB+\propa*\propb*\segment+2*\segment,\ha-\zaxisline*\inter-0.5) node{{\small $\alpha_{5}$}};
\draw (\debB+\propa*\propb*\segment+3*\segment-\propa*\segment,\ha-\zaxisline*\inter-0.5) node{{\small $\alpha_{6}$}};

\draw (\fin-0.5-\segment+\propa*\segment,\ha-\zaxisline*\inter-0.5) node{{\small $\alpha_{2N-1}$}};
\draw (\fin-0.5,\ha-\zaxisline*\inter-0.5) node{{\small $\alpha_{2N}$}};

\end{tikzpicture}
\caption{Rod structure and $\beta$ parameters for a bound state of $N$ microscopic Reissner-Nordström black holes. The scale of the pattern is $\delta\sim \frac{\ell}{N}= \frac{2M}{N}$.}
\label{fig:microscopicRN2}
\end{figure}
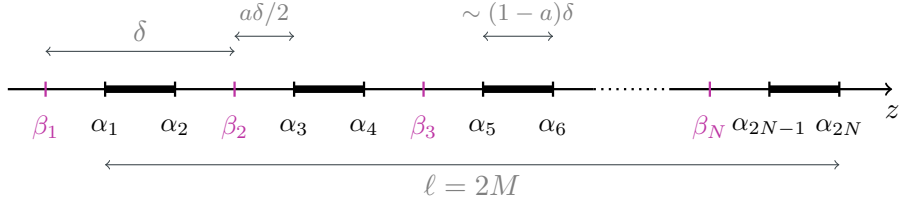

These two properties enhance the comparability of the new class of bound states with the Schwarzschild black hole. Consequently, solutions with smaller $N$ can be constructed and analyzed. In Fig.\ref{fig:microscopicRN2}, we depict the rod structure and parameters of the new class of solutions.

\subsection{Indistinguishability from the Schwarzschild black hole}
\label{sec:IndistinguishMicro2}

In this section, we demonstrate that the solutions closely resemble a Schwarzschild black hole for $r\gtrsim 2M(1+\epsilon)$, where $\epsilon=\cO(N^{-1})$. To achieve this, we follow a similar approach to that outlined in Section \ref{sec:AxisDataMicroBH}. First, we show that the solutions match the Schwarzschild values along the symmetry axis just above the rod configuration $z\gtrsim 2M(1+\epsilon)$.  As elaborated in Section \ref{sec:AxisData},  this correspondence implies that the indistinguishability extends throughout the entire spacetime within the radial range specified above.  We illustrate this result with a concrete example.

On the symmetry axis, beyond the rod configuration at $z\geq \ell=2M$, the gravitational and electric fields express in terms of the axis data \eqref{eq:FieldAxisData}:
\begin{equation}
Z(\rho=0,z\geq 2M)^{-2} \= \frac{\prod_{k=1}^{2N}(z-\alpha_k)}{\prod_{k=1}^{N}(z-\beta_k)^2},\qquad A(\rho=0,z\geq 2M) \= \sum_{k=1}^N \frac{f_k}{z-\beta_k}.
\label{eq:FieldAxisData3}
\end{equation}

Given that $\alpha_k$ and $\beta_k$ represent small steps of order $\delta$ \eqref{eq:NewParam1} and \eqref{eq:NewParam2}, we can approximate the gravitational field using a Riemann sum approximation, as done in Section \ref{sec:AxisDataMicroBH}, for $z\gtrsim \alpha_{2N} + \cO(\delta) \sim 2M (1+\cO(N^{-1}))$. Similarly, we found that
\begin{equation}
Z(0,z\gtrsim 2M(1+\epsilon))^{-2} = \frac{z-2M}{z} \left( 1+\cO(N^{-1})\right), 
\end{equation}
where $\epsilon=\cO(N^{-1})$. Thus, we match the values of a Schwarzschild black hole as expected.

As for the electric potential, the $f_k$ terms cancel each other in pairs, while the $\beta_k$ are separated by an infinitesimal distance $\delta \sim 2M/N$, ensuring that the electric potential also vanishes at large $N$ just above the configuration. Using the Riemann sum approximation (refer to Appendix \ref{App:RiemannSum}), we find that, for $z\gtrsim 2M (1+\cO(N^{-1})) $,
\begin{equation}
A(0,z) \sim \frac{\sin (\frac{\pi a}{2})}{N} \,\frac{M^2}{z^2} \,\sqrt{\frac{z}{z-2M}}\,,
\label{eq:AxisDataSol2}
\end{equation}
which is,  in the worst case, of order $\cO(N^{-1})$ when $z\gtrsim 2M (1+\cO(N^{-1}))$. Moreover, this demonstrates that the solution is indeed neutral.

Hence, the gravitational and electric fields are identical at leading order in large $N$ to the Schwarzschild fields on the symmetry axis for $z\gtrsim 2M (1+\cO(N^{-1}))$. As argued in section \ref{sec:AxisData}, this is sufficient to establish that the solutions are indistinguishable from a Schwarzschild black hole throughout the entire spacetime up to the locations of the sources,  $r\gtrsim 2M (1+\cO(N^{-1}))$.  \\

\begin{figure}[t]
\begin{center}
\includegraphics[width=  \textwidth]{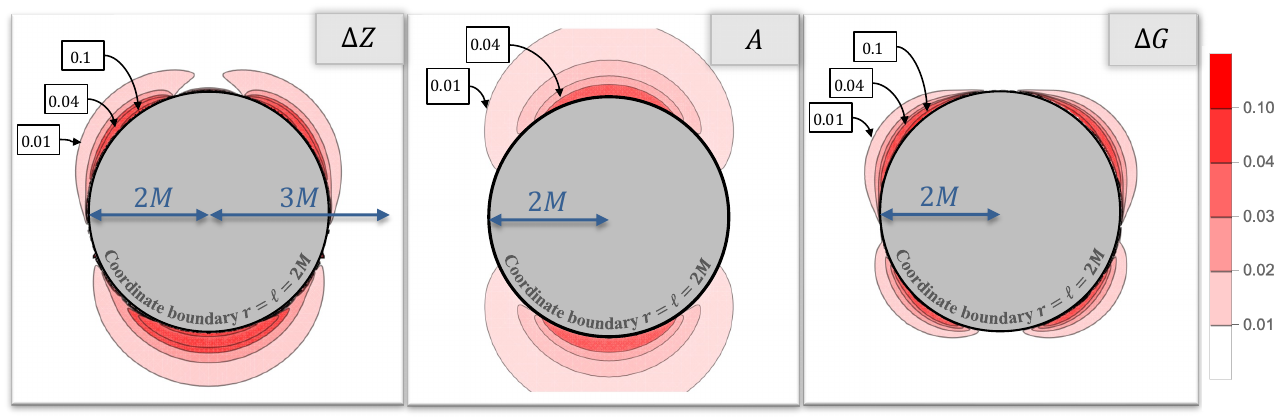}
\caption{Comparison of the gravitational field,  $Z$,  electric potential,  $A$,  and base function, $G$ with the Schwarzschild values for a bound state made of $N=10$ black holes and $a=1/2$.  The contour plots are given in the polar $(r,\theta)$ plane from the coordinate boundary $r=\ell=2M$ where the microscopic black holes are to $r=3M$.   From left to right: the deviation of $Z$,  $\Delta Z$ \eqref{eq:DeviationFunction},  $A$,  and the deviation of $G$,  $\Delta G$ \eqref{eq:DeviationFunction}.}
\label{fig:OutOfAxis10BH}
\end{center}
\end{figure}

To illustrate this assertion, we compare the fields entering the metric and gauge field with the Schwarzschild fields.  We choose a bound state with $N=10$ black holes and $a=\frac{1}{2}$.  In Fig.\ref{fig:OutOfAxis10BH}, we plot the following functions in the $(r,\theta)$ plane,
\begin{equation}
\Delta Z(r,\theta) \equi \left|\frac{Z^{-2}-\left(1-\frac{2M}{r}\right)}{1-\frac{2M}{r}} \right|\,,\qquad \Delta G(r,\theta) \equi \left| G-1\right|\,,
\label{eq:DeviationFunction}
\end{equation}
which measure the deviations of the main functions from the Schwarzschild values.

The plots clearly demonstrate that the bound state is indeed indistinguishable from a Schwarzschild black hole up to infinitesimal distance to the Schwarzschild horizon, as expected.  It is noteworthy that the precision is even higher than anticipated, as we are already achieving a very good match even with just $N=10$ black holes.

Furthermore, we provide a plot of the electric gauge potential in Fig.\ref{fig:OutOfAxis10BH}. The potential exhibits a dipole structure as expected from \eqref{eq:AxisDataSol2}, and we find $A(r,\theta) < \frac{\sin (\frac{\pi a}{2})}{4N}$ for $r\gtrsim 2M (1+\cO(N^{-1}))$. This is quite remarkable considering that at the coordinate boundary $r=2M$, we have Reissner-Nordström black holes distributed over the surface with charges of order the ADM mass. This illustrates the concept of electromagnetic entrapment, introduced in \cite{Heidmann:2023thn}, allowing neutral ultra-compact charge configurations to entrap their electromagnetic flux, thus resembling vacuum solutions slightly away from the sources.

Note that we did not analyze the magnetic gauge potential $H$ in our discussion. This is because its characteristics are intricately linked to the electric potential through electromagnetic duality \eqref{eq:MagDualErnst}. Consequently, all the properties found for the electric flux and charges are directly applicable to the magnetic flux.

\subsection{Internal structure}

We have demonstrated that the spacetime produced by the bound states introduced in Section \ref{sec:BSRN2} are indistinguishable from a Schwarzschild black hole from the asymptotic region up to an infinitesimal scale away from its horizon. Consequently, they resolve the Schwarzschild horizon into a novel structure supported by intense electromagnetic flux.  This flux is generated by dyonic black holes that can be made arbitrarily close to extremality, along with struts. In this section, our focus is directed towards describing this structure.
 
The solutions are determined by two internal parameters: $N$,  the number of black holes, thus defining the size of the microscopic patterns $\delta \sim 2M/N$; and $a$, the extremality parameter of the  black holes. It is worth noting that $a$ no longer directly controls the size of the black hole rods, as the $\alpha_{2k}$ are fixed by constraints on the $f_k$ \eqref{eq:FiConstraint}. Consequently, the rod lengths are only approximated by $\alpha_{2k}-\alpha_{2k-1}\sim a \delta$, with $N^{-1}$ corrections, so that $a$ only provides an average extremality parameter.

These solutions are refinements of the bound states extensively studied in Section \ref{sec:AnalysisLargeN}.  Therefore, we mainly refer to the comprehensive analysis conducted in that section. Here, we provide a concise summary, emphasizing the differences, and focus on the novel aspect: the ability of the new solutions to be Schwarzschild-like with a relatively small number of black holes.

\begin{itemize}
\item[•] \underline{Internal charges:}

The black hole charges exhibit a very similar profile to those studied in Section \ref{sec:LocalCharges}, with slightly more sensitivity to $N^{-1}$ corrections to ensure neutrality, $\sum_{i=1}^N Q_i=0$.   The charge signs alternate from a black hole to another, and once again, their magnitude is independent of the extremality parameter $a$ at leading order in large $N$.  More precisely, we found:
 \begin{equation}
|Q_i| \,\sim\, 4M\, \left[\frac{i-1+\sin(\frac{\pi a}{2})+\frac{1}{\pi}}{N}\left(1- \frac{i-\frac{1}{\pi}}{N} \right)\right]^{\frac{1}{2}+\frac{5 a-2}{2N}}\,, \qquad i=1,\ldots, N\,.
\label{eq:ApproxLocalCharges2}
\end{equation}
The distinction with the charge distribution \eqref{eq:ApproxLocalCharges} lies only in $N^{-1}$ corrections in the power. This further reinforces the notion that this charge distribution serves as a universal replacement for the Schwarzschild horizon.

\item[•] \underline{Temperature and entropy:}

Similarly, the temperatures and entropies of the microscopic black holes closely match those discussed in Section \ref{sec:EntropyTemperature} at leading order in large $N$.  The average temperature, $T$,  and the entropy of the bound state, $S$,  can be expressed in terms of the ADM mass and extremality parameter as in \eqref{eq:TempEntropApprox}:
\begin{equation}
T \,\sim\, (1-a)\left(1+a+\frac{3}{8} a^2+\frac{1}{2} a^4 \right)\,T_\text{Schw}\,,\qquad S \, \sim\, \frac{S_\text{Schw}}{1+a+\frac{3}{8} a^2+\frac{1}{2} a^4}\,,
\label{eq:TempEntropApprox2}
\end{equation}
where $T_{\text{Schw}}=(8\pi M)^{-1}$ and $S_{\text{Schw}}=4\pi M^2$ are the temperature and entropy of a Schwarzschild black hole.  Hence, the entropy of the bound states always remains a fraction of the Schwarzschild entropy.  Moreover,  this shows that the bound states develop an effective temperature that relates the ADM mass to the entropy as in \eqref{eq:EffectiveTemp}, $2M= T_\text{eff}\, S$ with $T_\text{eff} \sim \left(1+a+\frac{3}{8} a^2+\frac{1}{2} a^4\right) \, T_\text{Schw}$. 

\item[•] \underline{Strut properties:}

The properties of the struts also match those of the bound states analyzed in section \ref{sec:StrutProp}.  The strut tensions are relatively mild when $a\gtrsim 0.2$, meaning when the black holes are not too close to merging. The maximum tension and and the total strut energy are given by
\begin{equation}
\text{max}(|\cT_i|) \= \cO(N) \,,\qquad E_\text{struts} \,\sim\,  - \frac{a\,M}{2} \left(1- \frac{1.84}{N} \right) .
\end{equation}
Thus, despite forming an ultra-compact geometry with intense redshift, the struts do not need to be extreme to prevent the bound state from gravitational collapse. This becomes even more noteworthy with the present solutions, which can approach the Schwarzschild limit with a relatively small number of black holes, $N\sim 10$. 

\item[•] \underline{Deformation of the two-sphere:}

\begin{figure}%
    \centering
    \subfloat[\centering Radius of the two-sphere, normalized by $2M$,  as a function of $r$.]{{\includegraphics[width=0.45\textwidth]{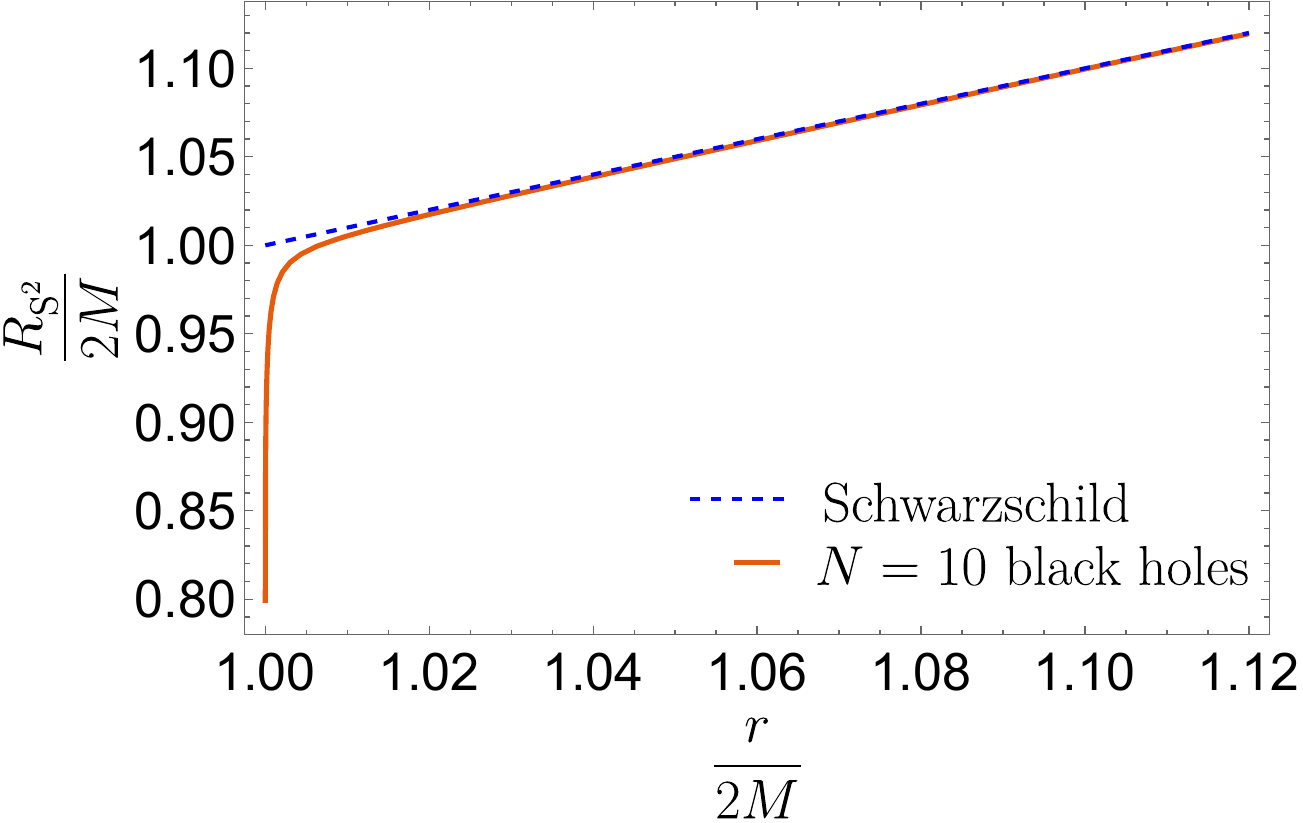} }}%
    \qquad
    \subfloat[\centering Axisymmetry factor of the two-sphere as a function of $r$.]{{\includegraphics[width=0.45\textwidth]{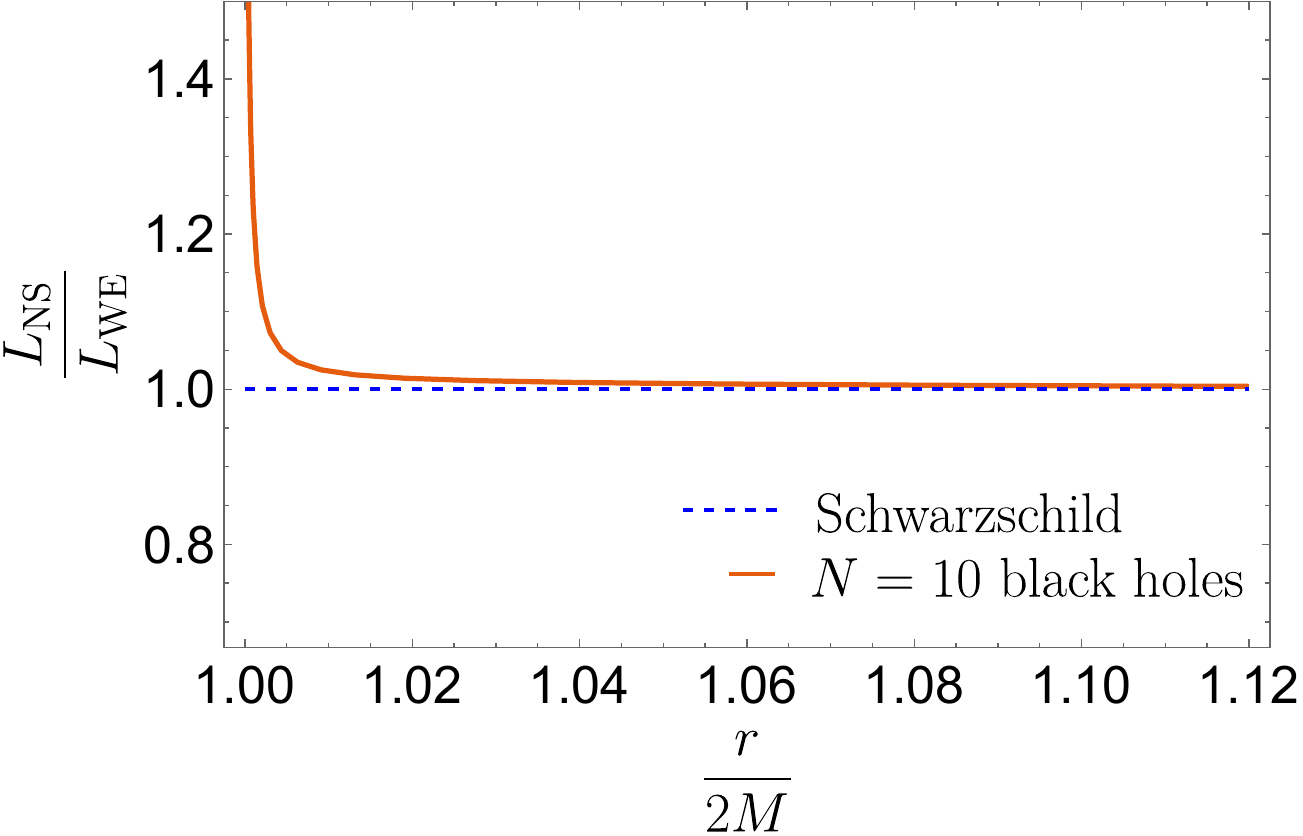} }}%
    \caption{Global characteristics of the two sphere as a function of the radial coordinate $r\geq \ell=2M$ for a bound state of $N=10$ black holes with $a=\frac{1}{2}$.   }%
    \label{fig:SphereProp2}%
\end{figure}

We describe the geometry of the solutions by analyzing the two-sphere around the bound state as in Section \ref{sec:S2Profile},  and we focus on bound states with a small number of black holes.

Firstly, Figure \ref{fig:SphereProp2} provides a description of the global characteristics of the two-sphere in terms of radius and axisymmetry factor, $R_{S^2}$ and $L_\text{NS}/L_\text{WE}$ \eqref{eq:DefS2RadAxisym}. These plots are obtained for bound states with $N=10$ black holes with  $a=\frac{1}{2}$. 

The plots demonstrate that the macroscopic properties of the two-sphere closely match the Schwarzschild expectation, even more so than anticipated. In fact, the radius and axisymmetry factor deviate from the Schwarzschild values by only 1\% below $r<2M(1+10^{-2})$. This close resemblance can be attributed to two main factors. First, as observed in Section \ref{sec:S2Profile}, the deformations induced by the black holes result in local oscillations along the two-sphere, which do not significantly impact macroscopic quantities such as overall lengths and area.  Second, the present solutions are refinements with stronger similarities to the Schwarzschild solutions. In these solutions, the number of black holes is not the sole factor determining the resolution scale.  Other factors, such as the net neutral charge distribution and positions of the black holes, also likely influence this scale.

\begin{figure}[t]
\begin{center}
\includegraphics[width=  0.9 \textwidth]{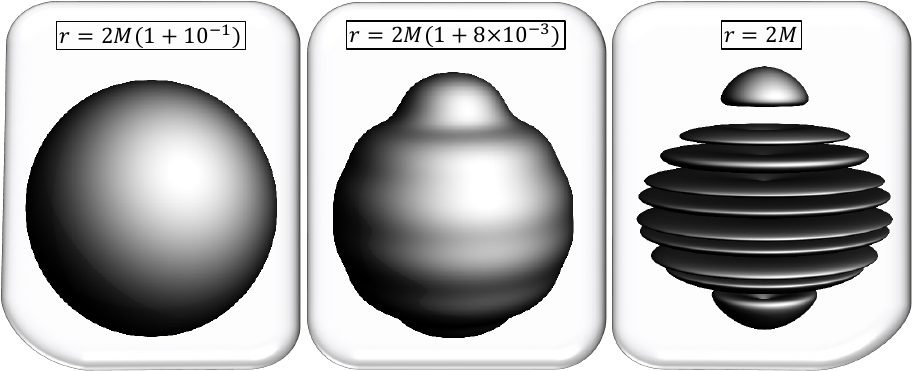}
\caption{The two-sphere of a bound state with $N=10$ Reissner-Nordström black hole and $a=\frac{1}{2}$.  The two-sphere is given for three values of $r$ from left to right: at $r\gtrsim 2M(1+\epsilon)$ where the bound state is indistinguishable from a Schwarzschild black hole (round sphere),  at $r\sim 2M(1+\epsilon)$ where the geometry starts to deviate from Schwarzschild and at the sources $r=\ell=2M$  corresponding to $N$ black holes held apart by struts. }
\label{fig:TwoSphere10}
\end{center}
\end{figure}

Figure \ref{fig:TwoSphere10} provides a local description of the two-sphere for three illustrative radii $r$. The two-sphere is described as the two-dimensional surface derived from the line element of the $(\theta,\phi)$ space at these radii, as detailed in Section \ref{sec:S2Profile}. The first plot on the left, at $r=2M(1+N^{-1})$, illustrates that the two-sphere is spherical at this radius, identical to the Schwarzschild two-sphere. The second plot, at $r=2M(1+8\times 10^{-3})$, depicts the region where the internal structure of the bound state begins to manifest, significantly deforming the S$^2$.

The last plot shows the structure of the bound state at the coordinate boundary $r=\ell=2M$, where the black holes and struts are located. This reveals the profiles of the ten horizons, while the empty regions between them correspond to the struts. Despite the high degree of deformation at the black hole horizons, as a whole, they closely resemble the shape of a sphere, even with few black holes, ensuring that the total area still scales as the Schwarzschild horizon area.

\end{itemize}

This concludes our analysis of the second class of solutions introduced in Section \ref{sec:BSRN2}. We have demonstrated that the solutions are refined versions of the bound states studied in Section \ref{sec:AnalysisLargeN}, such that they have very similar properties with small $N^{-1}$ corrections to cancel the infinitesimal net residual charge.  Consequently, they do not necessarily require a large number of black holes to be indistinguishable from the Schwarzschild black hole up to a small distance above its horizon. As a result, they ``resolve'' the Schwarzschild horizon into a novel structure generated by electromagnetic degrees of freedom. These electromagnetic degrees of freedom account for a fraction of the Schwarzschild entropy and manifest as Reissner-Nordström black holes held apart by struts in four dimensions.

\section{Discussion}
\label{sec:Conclusion}

In this section,  we provide a summary of the results and some directions for future research.

\subsection{Summary}

We have demonstrated that the Ernst formalism enables the construction and analysis of neutral bound states of Reissner-Nordström black holes in four dimensions, which are indistinguishable from the Schwarzschild geometry from the asymptotic region up to an infinitesimal scale above the Schwarzschild horizon. Below this scale, the latter is replaced by a novel structure supported by intense electromagnetic flux generated by a chain of microscopic Reissner-Nordström black holes held apart by struts. Through a comprehensive analysis of this novel structure, we have identified the following properties:
\begin{itemize}
\item[•] The extremality properties of the internal black holes can be arbitrarily adjusted, allowing for instance the resolution of the Schwarzschild black hole into a neutral bound state of extremal black holes.
\item[•] The crucial condition for resolving Schwarzschild lies in the charge distribution of the internal black holes. We have demonstrated that all bound states adhere to a fairly universal charge distribution.\footnote{This distribution applies to bound states with alternating signs of charges only (e.g., "$+-+-\ldots$"). We believe that another universal distribution could be found for a different pattern of signs, such as "$+--++--+\ldots$". } This distribution requires that the charge at each internal black hole has a magnitude of $|Q_i| \sim 2 \sqrt{x_i(2M-x_i)}$, where $0<x_i<2M$ is the position of the $i^\text{th}$ black hole replacing the Schwarzschild horizon of rod size $2M$.
\item[•] The entropy of the bound states always represents a fraction of the Schwarzschild entropy, with a minimum of $S\sim 0.35 S_\text{Schw}$ when all black holes are extremal.  Thus,  the degrees of freedom within the microscopic charged black holes accounts for a significant portion of the degrees of freedom in the Schwarzschild black hole.
\item[•] The temperature of the bound state differs significantly from the Schwarzschild temperature and is always smaller. For example, the temperature is strictly zero when the microscopic black holes are extremal. However, the bound states exhibit an effective temperature determined by the relation between its entropy and mass, $2M = T_\text{eff} S$, where $T_\text{eff}$ scales between $T_\text{Schw}$ and $2.9\,T_\text{Schw}$, even at extremality.
\item[•] The horizons of the internal black holes are highly deformed due to the intense gravitational environment near the bound states. The black holes in the middle of the bound states are completely flattened at their poles, with one extended direction. Remarkably, this extended direction has a length such that the horizons, when combined, maintain the shape of a sphere, as shown in Fig.\ref{fig:TwoSphere100} and \ref{fig:TwoSphere10}. We believe that this property is also a prerequisite for achieving a Schwarzschild-like geometry,  and that this property emerges from the specific charge distribution and positions of the black holes.
\item[•] Despite the ultra-compact nature of the bound states, the struts, which are necessary in four dimensions to prevent the gravitational collapse of Ernst solutions, do not become extremely tense as one might expect. Remarkably, the maximum strut tension remains of the order $-N$, where $N$ is the number of black holes in the bound state, and the total strut energy is of the order $-M/2$.
\end{itemize}

\subsection{Embedding in supergravity}

One major limitation of the solutions developed in this paper lies in the presence of struts. Struts are unphysical singularities that prevent us from regarding the bound states as genuine and regular configurations of black holes. However, previous studies in the literature have shown that struts, which are features of four-dimensional constructions, can be classically replaced when considering higher-dimensional theories of gravity \cite{Emparan:2001wk,Elvang:2002br}, particularly supergravity theories \cite{Bah:2021owp,Bah:2021rki,Heidmann:2021cms,Bah:2022yji,Bah:2022pdn,Heidmann:2022zyd,Bah:2023ows,Heidmann:2023thn,Heidmann:2023kry}.
 
First, it has been observed that the addition of extra compact dimensions does not alter the integrable structure of Einstein's equations when assuming staticity and axial symmetry \cite{Emparan:2001wk}. This observation has been extended to solutions with electromagnetic fields and the Ernst formalism \cite{Heidmann:2021cms}.

Second, deformation of extra compact dimensions can be used to replace a strut with a smooth Kaluza-Klein bubble, where an extra compact dimension degenerates. This approach imposes constraints on bound states of black holes, fixing the distances between the black holes in terms of the radius of the extra dimension \cite{Elvang:2002br,Heidmann:2023kry,Heidmann:2023thn}. While struts can have arbitrary tension to prevent collapse, a Kaluza-Klein bubble possesses a specific topological pressure for a given size, necessitating precise tuning to counterbalance gravitational attraction. Given that the characteristics of the struts in the bound states constructed in this paper are not extreme, we believe such a resolution is achievable without altering the main properties of the bound states in a similar manner as in \cite{Heidmann:2023kry,Heidmann:2023thn}.

Moreover, these gravitational constraints offer significant insights into the bound states. First, while families of solutions with struts are described by continuous parameters, the constraints lead to a discrete number of solutions \cite{Heidmann:2023thn}. Secondly, when embedded in suitable string theory frameworks where charges emerge as branes and anti-branes, these constraints establish a connection between microscopic degrees of freedom in brane/anti-brane systems derived at zero coupling \cite{Strominger:1996sh,Sen:1995in,Maldacena:1997de,Dijkgraaf:1996cv} and bound state entropy derived in supergravity,  as shown in \cite{Heidmann:2023kry}. \\

 A similar scenario is anticipated for our bound states. In future research, our aim is to embed the solutions in supergravity and regularize the struts by resolving them through Kaluza-Klein bubbles. This would lead to \emph{regular} bound states of (near)-extremal black holes, corresponding to localized branes and anti-branes in place of the Schwarzschild horizon. The entropy derived in gravity and the similarity to the Schwarzschild black hole up to the horizon scale should remain unaffected. The limit where the black holes are extremal is particularly interesting because we expect a similar scenario to \cite{Heidmann:2023kry}, where the bound state entropy can still be described by the microscopic entropies of the branes and anti-branes within the extremal black holes at weak coupling.

If successful, each bound state could provide an exact description of a distinct subset of states within the Schwarzschild black hole.  Each subset will contribute to a fraction of the total Schwarzschild entropy. Additionally, the solutions will offer a clear microscopic origin in terms of branes and anti-branes and a classical description as a geometry indistinguishable from Schwarzschild,  where  the Schwarzschild horizon resolves into a chain of charged black holes separated by smooth Kaluza-Klein bubbles. The resolution scale is expected to depend on the number of black holes $N$ as well as the infinitesimal Kaluza-Klein scales of the internal compact dimensions used to generate the topology.
 
Thus, our current results represent an initial step towards a fundamental understanding of the microstructure behind the Schwarzschild black hole and a brane/anti-brane description of the Schwarzschild black hole in string theory.  \\

Second, one can use the present four-dimensional solutions as building blocks to generate the first microstate geometries of the Schwarzschild black hole in supergravity. These will correspond to smooth, horizonless geometries that are indistinguishable from Schwarzschild up to an infinitesimal scale above the horizon but replace the latter using electromagnetic flux and topology. There are two potential routes for this project. The first involves leveraging the extensive families of microstate geometries of extremal black holes already constructed.  By (numerically) implementing them in the bound state of extremal black holes, one could resolve the extremal horizons into smooth horizonless structures. The second route involves using the generalization of the Ernst formalism in supergravity derived by one of the authors. This approach would use the same fields as those derived in four dimensions, but replace the horizons with KK bubbles, similar to how non-supersymmetric topological solitons and topological stars have been constructed \cite{Bah:2020ogh, Bah:2021owp, Heidmann:2021cms, Bah:2022yji, Bah:2023ows}.

\subsection{Thermodynamic description of the bound states}

In this paper, we have derived thermodynamic quantities of the bound states, such as temperature, entropy, and strut tensions, in terms of the ADM mass and some internal parameters. The most intriguing quantity was undoubtedly the temperature. While the entropy always scaled as a fraction of the Schwarzschild entropy, proportional to $M^2$, the temperature could be made arbitrarily close to zero by making the microscopic black holes extremal. This change did not affect the macroscopic properties of the bound states, allowing an effective temperature to emerge from the relation between mass and entropy \eqref{eq:EffectiveTemp}:
\begin{equation}
2M \,\=\,  T_\text{eff}\, S \,,\qquad T_\text{eff} \sim \left(1+a+\frac{3}{8} a^2+\frac{1}{2} a^4\right) \, T_\text{Schw}\,,
\end{equation}
where $a$ is the extremality parameter, ranging between $0$ (merging point) and $1$ (extremal).

This relation prompts many questions about the thermodynamic properties of the bound states. Firstly, it would be interesting to delve deeper into these properties by deriving the first law of thermodynamics for the class of solutions presented in this paper. This involves deriving the variation of the total mass in terms of a selected set of independent observables, typically containing entropy. This has been accomplished for various solutions consisting of collinear black holes by deriving the variation of the ADM mass in terms of internal parameters and relating them to variations in entropy, local charges, and strut tensions \cite{Gregory:2020mmi,Krtous:2019fpo}, or by deriving the Euclidean action \cite{Gibbons:1976ue} as in \cite{Garcia-Compean:2020gii,Ramirez-Valdez:2021zjs}.

Secondly, it would be interesting to investigate whether the effective temperature can be associated with radiation from the bound states akin to Hawking radiation \cite{Hawking:1975vcx}. Hawking radiation describes the decay of a black hole through the separation of virtual pairs that nucleate near the horizon from quantum fluctuations. The spectrum and characteristics of this effect follow a thermal profile with the same temperature as the value obtained from the surface gravity. While the black holes in our bound states should also radiate through the Hawking process near their horizon, another radiation channel could follow a thermal profile given by $T_\text{eff}$. This channel could involve the spontaneous creation of charged pairs not in the near-horizon region of the black holes but in the regions between the black holes. In these regions, the redshift is also very large, with intense electromagnetic fields. Thus, virtual pairs of charged particles could separate there, following the Schwinger mechanism \cite{PhysRev.82.914}, and one or both elements of the pairs could fall into nearby black holes, decreasing their charges and consequently reducing the energy of the entire bound states. Determining the emission rate for this decay channel could provide a physical interpretation of the effective temperature derived in this paper and a comprehensive thermodynamic description of the bound states.

\subsection{Gravitational signatures}

The bound states of Reissner-Nordström black holes introduce novel spacetime structures at the Schwarzschild horizon, which could manifest in observables. Therefore, investigating the gravitational signatures of these bound states is of great interest.

Concerning light scattering and imaging, the Schwarzschild black hole is primarily defined by its unstable photon ring at $r=3M$, where light rays crossing this boundary inevitably reach the horizon and are absorbed. Since the bound states are indistinguishable from Schwarzschild around $r=3M$, we anticipate a similar outer photon ring with equivalent scattering properties. However, incoming photons will intersect the region around $r\sim 2M$ and experience different trajectories compared to Schwarzschild, as they orbit around the microscopic black holes. Some photons may be absorbed by the black holes, while others might escape after an extended period due to the high redshift within the internal structure. Consequently, it would be interesting to apply techniques similar to those in \cite{Cunha:2018acu,Cunha:2015yba,Bacchini:2018zom,Hertog:2019hfb,Bacchini:2021fig,Heidmann:2022ehn,Staelens:2023jgr} to illustrate the effect of the internal structure at the horizon scale on photon scattering, directly comparing it with the Schwarzschild black hole. Additionally, further exploration could involve simulating the imaging of an accretion disk and the effect of internal structure using various techniques found in the literature (see \cite{Lupsasca:2024wkp,Guerrero:2021ues,Sui:2023yay} for a non-exhaustive list). Moreover, analyzing the scattering of charged particles that could be affected by the intense electromagnetic field entrapped around $r\sim 2M$ is another intriguing avenue.

Another research direction to explore is the quasi-normal modes and linear response of the bound states under perturbations. This involves deriving the  quasi-normal mode spectrum of scalar and, ultimately, gravitational perturbations. Initially, we might expect a spectrum very similar to the Schwarzschild spectrum, with additional echo modes from the internal structure as discussed in \cite{Bueno:2017hyj,Bena:2019azk,PhysRevD.106.024041,Maggio:2021ans}, or cavity effects in the damping time as explored in \cite{Heidmann:2023ojf}. These derivations can provide insights into the potential impact of new physics  emerging near the horizon of astrophysical black holes.

\section*{Acknowledgements}
We  would like to thank  Ibrahima Bah,  Iosif Bena,  Bogdan Ganchev, Marcel Hughes, Samir Mathur and Madhur Metha for useful discussions.  The work of PH is supported by the Department of Physics at The Ohio State University.

\appendix

\section{Collinear Reissner-Nordström black holes}
\label{app:ErnstSolNRN}

In this appendix, we review the solutions corresponding to $N$ collinear Reissner-Nordström black holes,  first derived in \cite{NoraBreton1998} as the static limit of collinear Kerr-Newman black holes \cite{Ruiz:1995uh}.  The solutions are uniquely defined by their axis data \eqref{eq:AxisData},   
\begin{equation}
e(z) = 1+ \sum_{i=1}^N \frac{e_i}{z-\beta_i},\quad f(z) = \sum_{i=1}^N \frac{f_i}{z-\beta_i},
\label{eq:AxisDataApp}
\end{equation}
which correspond to the field values above the last rod on the symmetry axis \eqref{eq:AxisDatavsFields}, but they also fix the solutions across the entire spacetime.  Thus,  the solutions are determined by $3N$ real parameters $(\beta_i,f_i,e_i)$. However,  as pointed out in \cite{NoraBreton1998},  it is more relevant to work with the $2N$ zeroes of the gravitational field,  denoted as $\alpha_{k}$,  which satisfy the algebraic equation:
\begin{equation}
e(z)+f(z)^2=0,
\end{equation}
and extract the values of $(f_i,e_i)$ from the $\alpha_k$ and $\beta_i$ as in \eqref{eq:DefFiEi}.  As such,  the $\alpha_k$ correspond to the rod endpoints and therefore to the positions of the black holes on the $z$-axis.

In Weyl-Papapetrou coordinates, the solutions can be expressed in terms of four matrix determinants dependent on the $3N$ parameters and the coordinates $(\rho,z)$, as shown in \cite{NoraBreton1998,Ruiz:1995uh}:
\begin{equation}
	E_{\pm} = \begin{vmatrix} 
		1 & 1 & \dots & 1 \\
		\pm 1 & \frac{r_1}{\alpha_1-\beta_1} &\dots & \frac{r_{2N}}{\alpha_{2N}-\beta_1}\\
		\vdots & \vdots &\ddots&\vdots\\
		\pm 1 & \frac{r_1}{\alpha_1-\beta_N} &\dots & \frac{r_{2N}}{\alpha_{2N}-\beta_N}\\
		0 & \frac{h_1(\alpha_1)}{\alpha_1-\beta_1} & \dots & \frac{h_1(\alpha_{2N})}{\alpha_{2N}-\beta_1}\\
		\vdots & \vdots & \ddots & \vdots \\
		0 & \frac{h_N(\alpha_1)}{\alpha_1-\beta_N} & \dots & \frac{h_N(\alpha_{2N})}{\alpha_{2N}-\beta_N}
	\end{vmatrix}
	,\qquad    F = \begin{vmatrix} 
		0 & f(\alpha_1) & \dots & f(\alpha_{2N}) \\
		-1 & \frac{r_1}{\alpha_1-\beta_1} &\dots & \frac{r_{2N}}{\alpha_{2N}-\beta_1}\\
		\vdots & \vdots &\ddots&\vdots\\
		-1 & \frac{r_1}{\alpha_1-\beta_N} &\dots & \frac{r_{2N}}{\alpha_{2N}-\beta_N}\\
		0 & \frac{h_1(\alpha_1)}{\alpha_1-\beta_1} & \dots & \frac{h_1(\alpha_{2N})}{\alpha_{2N}-\beta_1}\\
		\vdots & \vdots & \ddots & \vdots \\
		0 & \frac{h_N(\alpha_1)}{\alpha_1-\beta_N} & \dots & \frac{h_N(\alpha_{2N})}{\alpha_{2N}-\beta_N}
	\end{vmatrix}\,,
		\label{eq:MatrixDet1}
\end{equation}
and
\begin{equation}
	K = \begin{vmatrix} 
		
		\frac{1}{\alpha_1-\beta_1} &\dots & \frac{1}{\alpha_{2N}-\beta_1}\\
		\vdots &\ddots&\vdots\\
		\frac{1}{\alpha_1-\beta_N} &\dots & \frac{1}{\alpha_{2N}-\beta_N}\\
		\frac{h_1(\alpha_1)}{\alpha_1-\beta_1} & \dots & \frac{h_1(\alpha_{2N})}{\alpha_{2N}-\beta_1}\\
		\vdots & \ddots & \vdots \\
		\frac{h_N(\alpha_1)}{\alpha_1-\beta_N} & \dots & \frac{h_N(\alpha_{2N})}{\alpha_{2N}-\beta_N}
	\end{vmatrix}\,,\qquad 
	 I = \begin{vmatrix} 
		\sum_{k=1}^{N} f_k&0 & f(\alpha_1) & \dots & f(\alpha_{2N}) \\
		z & 1 & 1 & \dots & 1\\
		-\beta_1&-1 & \frac{r_1}{\alpha_1-\beta_1} &\dots & \frac{r_{2N}}{\alpha_{2N}-\beta_1}\\
		\vdots&\vdots & \vdots &\ddots&\vdots\\
		-\beta_N&-1 & \frac{r_1}{\alpha_1-\beta_N} &\dots & \frac{r_{2N}}{\alpha_{2N}-\beta_N}\\
		e_1&0 & \frac{h_1(\alpha_1)}{\alpha_1-\beta_1} & \dots & \frac{h_1(\alpha_{2N})}{\alpha_{2N}-\beta_1}\\
		\vdots&\vdots & \vdots & \ddots & \vdots \\
		e_N&0 & \frac{h_N(\alpha_1)}{\alpha_1-\beta_N} & \dots & \frac{h_N(\alpha_{2N})}{\alpha_{2N}-\beta_N}
	\end{vmatrix}\,.
	\label{eq:MatrixDet2}
\end{equation}
where we remind that $r_{k}=\sqrt{\rho^2+(z-\alpha_k)^2}$ are the distance to the rod endpoints,  and we  used:
\begin{equation}
	f(\alpha_k)=\sum_{n=1}^{N} \frac{f_n}{\alpha_k-\beta_n} ,\qquad h_k(\alpha_i)= e_k+2f_k f(\alpha_i)\,.
\end{equation}
The fields entering the metric and gauge field \eqref{eq:StaticErnstMetric4d} are given by:
\begin{equation}
	Z\=\frac{E_{-}}{\sqrt{E_{+}E_{-}+F^2}},\qquad e^{4\nu} \= \frac{E_{+}E_{-}+F^2}{K^2 \prod_{n=1}^{2N} r_n}\,,\qquad  A\=\frac{ F}{E_{-}}\,,\qquad	H=-\frac{ I}{E_{-}}\,.
	\label{eq:FieldsNRNApp}
\end{equation}

The solutions correspond to $N$ non-extremal Reissner-Nordström black holes on a line,  and we refer the reader to Section \ref{sec:SpacetimeStructure} for an analysis of the geometries.  As a side note, the limit in which one or several black holes become extremal (i.e., $\alpha_{2k-1} \rightarrow \alpha_{2k}$ if the $k^{th}$ black hole becomes extremal) is non-trivial because all the determinants become zero. Therefore, deriving such a solution requires reapplying the Sigbatulin method from scratch. The fields corresponding to $N$ extremal Reissner-Nordström black holes can be found in \cite{Heidmann:2023thn}.

\section{Black hole bound states and Schwarzschild limit}
\label{App:MicroBH}

In this appendix, we provide details on the derivations outlined in Section \ref{sec:AxisDataMicroBH} and Section \ref{sec:IndistinguishMicro2}. These sections demonstrate that the two classes of bound states constructed in the paper, each consisting of $N$ small Reissner-Nordström black holes, are indistinguishable from the Schwarzschild black hole. To achieve this, we used the indistinguishability argument introduced in Section \ref{sec:AxisData}, which allows us to focus on the axis data of the solutions.

In this section, we first demonstrate how to approximate the intricate values of the $f_i$ \eqref{eq:fparamNmicroRN}, which are part of the electric axis data. Then, we show that the axis data of the bound states constructed in this paper match those of Schwarzschild using a Riemann sum approximation. Finally, we provide an illustrative example showing that the analysis on the symmetry axis indeed extends to the entire spacetime.

\subsection{Approximation of the $f_i$}
\label{App:ApproxFi}

In Section \ref{sec:ClassMicroBH}, we introduced bound states of $N\gg 1$ small black holes, for which the positions and parameters $\beta_k$ are entirely fixed in infinitesimal patterns of length $\delta$ as follows:
\begin{equation}
	\alpha_{2k-1} \= (k-1) \,\delta,\qquad \alpha_{2k} \= (k-a) \,\delta\,,\qquad \beta_k \= \left(k-1-\frac{a}{2}\right)\, \delta\,,\quad k=1,\ldots,N\,.
	\label{eq:alphabetaApp}
\end{equation}
This fixes the parameters $f_i$ \eqref{eq:DefFiEi} to be:
\begin{equation}
	f_i^2  \= \frac{a(2-a)\, \delta^2}{4}\, \prod_{k \neq i} \left(1-\frac{a}{2\,(i-k)}\right)\left(1-\frac{2-a}{2\,(i-k)}\right)\,.
	\label{eq:Fi2def}
\end{equation}

To obtain a large-$N$ approximation of the $f_i$, we first consider the ratio:
\begin{equation}
	\frac{f_{i+1}^2-f_i^2}{f_i^2}= \frac{N}{(i-N)\,i}+\mathcal{O}\left(N^{-2}\right)\,.
	\label{eq:EOM fi}
\end{equation}
This can be interpreted as a discrete differential equation by introducing $y_i=i/N$:
\begin{equation}
	\frac{f_{i+1}^2-f_i^2}{f_i^2}= \frac{y_{i+1}-y_i}{y_i(1-y_i)},\quad	\leftrightsquigarrow \quad \frac{d(f^2)}{f^2} \= \frac{dy}{y(1-y)}\,.
\end{equation}
By integrating the relation, we find, at leading order in large $N$:
\begin{equation}
	\begin{split}
		f_i^2 \,\sim\,\cF\left(i+c_1\right)^2\,,\qquad  \cF(x) &\equi c_2 \,\sqrt{\frac{N-x}{x} }\,.
	\end{split}
\end{equation}
where $(c_1, c_2)$ are integration constants that we need to fix. By matching the large $N$ expression of $f_1^2$ and $f_{\frac{N}{2}}^2$ from \eqref{eq:Fi2def}, we obtain:
\begin{equation}
	c_1=-1+\frac{\sin(\frac{\pi}{2}\,a)}{\pi}\,,\qquad c_2= \frac{\delta\,\sin(\frac{\pi}{2}\,a)}{\pi}, 
	\label{eq:IntConstApp}
\end{equation}
which leads to the approximation given in \eqref{eq:ApproxFi}:
\begin{equation}
	\begin{split}
		f_i \,\sim\, \pm\,\cF\left(i-1+\frac{\sin(\frac{\pi}{2}\,a)}{\pi}\right)\,,\qquad  \cF(x) &\equi \frac{\delta \,\sin(\frac{\pi}{2}\,a)}{\pi}\,\sqrt{\frac{N-x}{x} }\,.
		\label{eq:ApproxFiApp}
	\end{split}
\end{equation}

\subsection{Approximation of the axis data}
\label{App:RiemannSum}

For both classes of solutions introduced in Sections \ref{sec:ClassMicroBH} and \ref{sec:BSRN2}, the parameters $(\alpha_k,\beta_k)$ can be interpreted as infinitesimal steps of order $\frac{k}{N}\ell$, where $\ell$ is the configuration size of order $2M$. Thus, the axis data \eqref{eq:AxisDataApp} can be approximated at large $N$ using the Riemann sum approximation, which states that for any function $g$ slowly varying between $0$ and $1$,
\begin{equation}
	\frac{1}{N}\sum_{k=1}^N g\left(\frac{k}{N}\right)\simeq\int_{0}^{1} g(x)\,dx\,.
	\label{eq:RiemannSumApp1}
\end{equation}
This can also be expressed in terms of a product:
\begin{equation}
\prod_{k=1}^N g\left(\frac{k}{N}\right) \simeq \exp \left[N \int_{0}^{1} \log(g(x))\,dx \right].
\label{eq:RiemannSumApp2}
\end{equation}
This approximation technique was first used in a similar context in \cite{Bah:2021rki}, and we refer interested readers to Appendix C of that paper for more details.

To approximate the axis data, we will not directly apply the Riemann sum approximation to $e(z)$ and $f(z)$, but rather to:
\begin{equation}
e(z)+f(z)^2 \= \frac{\prod_{k=1}^{2N}(z-\alpha_k)}{\prod_{k=1}^{N}(z-\beta_k)^2},\qquad  f(z) \=  \sum_{i=1}^N \frac{f_i}{z-\beta_i}.
\label{eq:AxisDataAppSol}
\end{equation}
These quantities correspond to the gravitational and electric fields on the symmetry axis above the last black hole \eqref{eq:AxisDatavsFields}.

\subsubsection{For the first class of solutions}

We analyze the first class of solutions defined by the parameters \eqref{eq:alphabetaApp}.

\begin{itemize}
\item[•] \underline{Gravitational axis data:}

First, we simplify the gravitational axis data\footnote{We used the relation \eqref{eq:RelationEllDelta} between $\delta$ and $\ell$ to make the $k/N$ dependence manifest.}
\begin{equation}
e(z)+f(z)^2 \= \prod_{k=1}^N \frac{\left( \frac{z}{\ell}-\frac{k-a}{N-a}\right)\left( \frac{z}{\ell}-\frac{k-1}{N-a}\right)}{\left( \frac{z}{\ell}-\frac{k-1-\tfrac{a}{2}}{N-a}\right)^2}.
\end{equation}
The argument in the product represents a slowly varying function of $x=k/N$ if and only if $z/\ell$ is not too close to a zero or a pole. By restricting to values of $z$ above the rod configuration, $z\geq \ell$, this requires $z\gtrsim \ell (1+\cO(\tfrac{1}{N}))$ where ``$\gtrsim$'' means an order larger than the right-hand side.\footnote{For instance, $z=\ell (1+N^{h-1})$ where $h>0$ is sufficient.}

Thus, in this range of the $z$ coordinate, we can apply the Riemann sum approximation \eqref{eq:RiemannSumApp2}, which leads to
\begin{equation}
e(z)+f(z)^2 \sim 1-\frac{\ell}{z}\,,
\end{equation}
at the leading order in the large $N$ limit. It has to be pointed out that this expression is not a large $z$ expansion. This is valid for any values of $z$ slightly above the rod configuration, $z\gtrsim \ell (1+\cO(\tfrac{1}{N}))$.

Using the relation \eqref{eq:AxisDatavsFields}, this shows that the gravitational redshift function $Z$ is identical to that of a Schwarzschild black hole on the symmetry axis right above the last black hole in the chain: $$Z\left(\rho=0,z\geq\ell\,(1+\cO(\tfrac{1}{N}))\right)^{-2} \sim  1-\frac{\ell}{z}.$$

\item[•] \underline{Electric axis data:}

The electric axis data \eqref{eq:AxisDataAppSol} involves the parameters $f_i$, approximated at large $N$ by \eqref{eq:ApproxFiApp}, for which the sign can be arbitrarily chosen.

By first assuming all $f_i$ to be positive, we have
\begin{equation}
f(z) \,\sim\,  \frac{\sin \left(\tfrac{\pi}{2}a \right)}{\pi (N-a)} \,\sum_{k=1}^N \,\frac{\sqrt{\frac{N-k-c_1}{k+c_1}}}{ \frac{z}{\ell}-\frac{k-1-\tfrac{a}{2}}{N-a}}\,,
\end{equation}
where $c_1$ is given in \eqref{eq:IntConstApp}. As before, the argument in the sum is a slowly-varying function of $x=k/N$  if we consider $z\gtrsim \ell (1+\cO(\tfrac{1}{N}))$. By applying the Riemann sum approximation in that range, we find
\begin{equation}
f(z)\, \sim\, 2 \sin\left(\tfrac{\pi}{2}\,a\right) \left(1-\sqrt{\frac{z-\ell}{z}}\right).
\end{equation}
Thus, the electric axis data is far from zero for $a\neq 0$, differing from the Schwarzschild axis data. Such a bound state of Reissner-Nordström black holes will therefore significantly deviate from a Schwarzschild black hole, despite having the same gravitational axis data. This could have been expected, as the $f_i$ are associated with the charges of the black holes, and having all the $f_i$ positive implies that the charges have the same sign, resulting in a nonzero net charge \eqref{eq:ConservedChargesMicroRN} given by $Q =\sum f_i = \sin\left(\tfrac{\pi}{2}\,a\right)\ell (1+\cO(N^{-1}))$, which aligns with the above expression.\\

To have a negligible electric axis data right above the last black hole, the charges need to change sign between two neighboring black holes, meaning the $f_i$ should alternate in sign. By doing so, we have
\begin{equation}
f(z) \,\sim\,  \frac{\sin \left(\tfrac{\pi}{2}a \right)}{\pi (N-a)} \,\sum_{k=1}^N \,\frac{(-1)^{k+1}\sqrt{\frac{N-k-c_1}{k+c_1}}}{ \frac{z}{\ell}-\frac{k-1-\tfrac{a}{2}}{N-a}}\,.
\end{equation}
To apply the Riemann sum approximation, we need to split the sum for even and odd values of $k$ and apply the approximation twice. We find that the leading order terms of both sums (of order $N^0$) cancel out, leaving a subleading $N^{-1/2}$ contribution:
\begin{equation}
f(z) \,\sim\, \frac{2\sin\left(\frac{a}{2}\right)\ell}{\sqrt{N}} \, \frac{1 }{z}\,, \qquad z\gtrsim \ell (1+\cO(\tfrac{1}{N}))\,.
\end{equation}
This corresponds to the axis data of an infinitesimal monopole charge in the large $N$ limit. However, it is important to note that this expression is not precise. It arises as the subleading contribution in the large $N$ expansion, which could be corrected by subleading contributions in the approximated values of $f_i$ \eqref{eq:ApproxFiApp}.

Nevertheless, this demonstrates that when the signs of the $f_i$ alternate, the electric axis data is given, for $z\gtrsim \ell (1+\cO(\tfrac{1}{N}))$, by
\begin{equation}
f(z) \,\sim\,  \frac{Q}{z}\,,\qquad Q= \cO\left( \frac{\ell}{\sqrt{N}}\right)= \cO\left( \frac{M}{\sqrt{N}}\right),
\label{eq:EMfieldAxisMicroApp}
\end{equation}
at the leading order in the large $N$ expansion. The electric axis data also corresponds to a monopole charge, but this time, the charge is negligible and parametrically smaller than the mass.

Thus, such a solution has the same gravitational and electric axis data as a Schwarzschild black hole at leading order in the large $N$ expansion and for $z\gtrsim 2M (1+\cO(\tfrac{1}{N}))$. This allows us to apply the indistinguishability property discussed in Section \ref{sec:AxisData} that shows that the solutions are indistinguishable in the whole spacetime for a similar range of radial distance $r\gtrsim 2M (1+\cO(\tfrac{1}{N}))$. We will analyze a specific example illustrating this extension to the whole space in Section \ref{App:OutOfAxis}.

\end{itemize}

\subsubsection{For the second class of solutions}

In Section \ref{sec:BSRN2}, we introduced a second class of solutions consisting of $N\gg 1$ small Reissner-Nordström black holes arranged along a line. This class serves as a refinement of the first family of solutions, ensuring they have no negligible net charge of order $M/\sqrt{N}$. To achieve this, we fix $3N$ parameters: the configuration length $\ell=2M$, the $f_k$ \eqref{eq:choiceFi}, the $\beta_k$, and $N-1$ rod edges, $\alpha_{2k-1}$ \eqref{eq:NewParam1}. However, because these parameters are close to those of the first class, the remaining $\alpha_1$ and $\alpha_{2k}$ are also similarly close for large $N$ \eqref{eq:NewParam2}.

The analysis of the axis data for these new solutions closely follows the previous one. For instance, the argument for the gravitational axis data remains identical, and we find similarly that $$Z\left(\rho=0,z\geq\ell\,(1+\cO(\tfrac{1}{N}))\right)^{-2} \sim  1-\frac{\ell}{z}.$$

However, the electric axis data changes significantly due to the alternation in sign of the $f_k$:
\begin{equation}
f(z) \= \frac{\sin \left(\tfrac{\pi}{2}a \right)}{N\,\pi } \,\sum_{k=1}^N \,\frac{(-1)^{k+1}\sqrt{\frac{N-2\lfloor \tfrac{k+1}{2} \rfloor +\tfrac{1}{2}-c_1}{2\lfloor \tfrac{k+1}{2} \rfloor -\tfrac{1}{2}+c_1}}}{ \frac{z}{\delta N}-\frac{k-1-\tfrac{a}{2}}{N}}\,.
\end{equation}
Similarly, one can apply the Riemann sum approximation by splitting the sum into even and odd values of $k$, and for $z$ not too close to the poles in the argument, $z\gtrsim \delta N (1+\cO(N^{-1})) = 2M(1+\cO(N^{-1}))$. We find:
\begin{equation}
f(z)\, \sim \,\frac{\sin (\frac{\pi a}{2})}{N} \,\frac{M^2}{z^2} \,\sqrt{\frac{z}{z-2M}}\,.
\end{equation}
This clearly indicates that the monopole term has been strictly canceled, as expected, and we are left with a function that is, at worst, of order $\cO(N^{-1})$ when $z\gtrsim 2M (1+\cO(N^{-1}))$. Moreover, the leading multipole moment is the dipole of order $M^2/N$, while all higher multipoles scale like $M^n/N$.

This demonstrates that the second class of solutions have axis data that match those of Schwarzschild even more closely than the first class of solutions for $z\gtrsim 2M (1+\cO(N^{-1}))$. Moreover, one does not necessarily need $N$ to be very large to achieve close values. The indistinguishability property discussed in Section \ref{sec:AxisData} shows that the solutions are indistinguishable in the whole spacetime for a similar range of radial distance $r\gtrsim 2M (1+\cO(\tfrac{1}{N}))$. This has been illustrated through a specific example in Fig.\ref{fig:OutOfAxis10BH}.

\subsection{Illustration out of the axis}
\label{App:OutOfAxis}

In the preceding section, we used the property of Ernst solutions to be entirely determined by their behavior on the axis of symmetry. Thus, by analyzing the field values on the symmetry axis just above the rod configurations, we demonstrated that configurations of $N$ aligned Reissner-Nordström black holes are indistinguishable from a Schwarzschild black hole from the asymptotic region up to an infinitesimal distance from the Schwarzschild horizon.

At first glance,  the extrapolation from the axis of symmetry to the entire spacetime might seem dubious. Clearly, this notion is fundamentally wrong for arbitrary solutions of Einstein's equations: one cannot obtain information about the entire spacetime solely by examining the field values on a single axial slice. However, the arguments outlined in Section \ref{sec:AxisData} establish that this holds true for any axially-symmetric and static solutions.

In this appendix, we support this argument by analyzing a specific solution and comparing it with a Schwarzschild geometry across the entire spacetime. We focus on a solution made of $N=100$ black holes, given by the parameters \eqref{eq:alphabetaApp} with $a=\frac{1}{2}$. The infinitesimal length $\delta$ is determined by the ADM mass and $N$ in \eqref{eq:ConservedChargesMicroRN}.

To compare the solution with a Schwarzschild black hole of the same mass, we introduce the  functions:
\begin{equation}
\Delta Z(r,\theta) \equi \left|\frac{Z^{-2}-\left(1-\frac{2M}{r}\right)}{1-\frac{2M}{r}} \right|\,,\qquad \Delta G(r,\theta) \equi \left| G-1\right|\,,
\end{equation}
which measure the deviations of the metric fields from their Schwarzschild counterparts. Regarding the electromagnetic gauge field, we focus solely on the electric potential $A$, as $A=0$ for Schwarzschild. The magnetic potential $H$ does not require specific attention as it is related to $A$ by electromagnetic duality \eqref{eq:MagDualErnst}.

\begin{figure}[t]
	\begin{center}
		\includegraphics[width=  \textwidth]{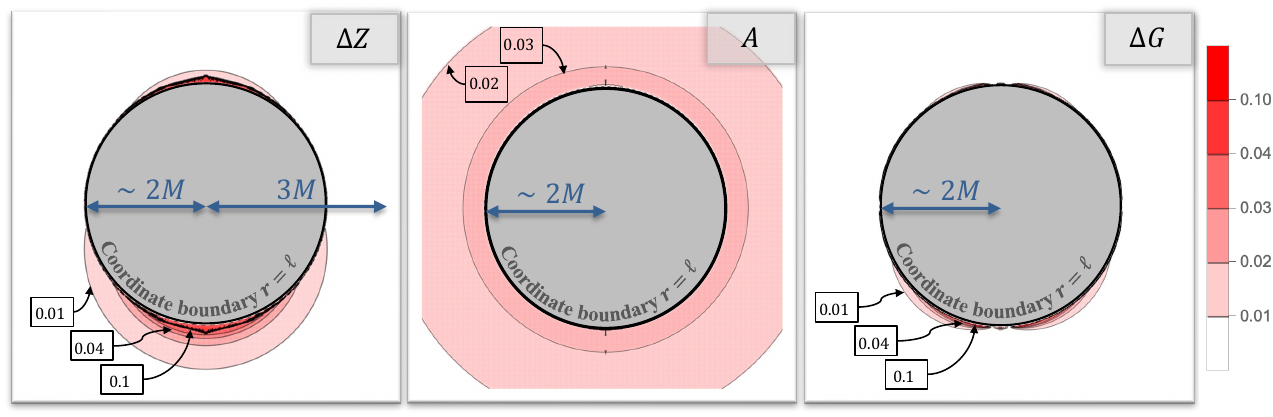}
		\caption{Comparison of the gravitational field,  $Z$,  electric potential,  $A$,  and base function, $G$ with the Schwarzschild values for a bound state made of $N=100$ black holes and $a=\frac{1}{2}$.  The contour plots are given in the polar $(r,\theta)$ plane from the coordinate boundary $r=\ell\sim 2M$ where the microscopic black holes are to $r=3M$.   From left to right: the deviation of $Z$,  $\Delta Z$,  $A$,  and the deviation of $G$,  $\Delta G= |G- 1|$.}
		\label{fig:OutOfAxis}
	\end{center}
\end{figure}

We present three distinct plots in Figure \ref{fig:OutOfAxis}. Each plot is depicted in the polar plane $(r,\theta)$, where the radial coordinate $r$ ranges from $r\geq \ell \sim 2M$,  the coordinate boundary,  up to $r=3M$. According to our analytical derivation, we anticipate the solution to deviate from Schwarzschild below $r\sim 2M(1+10^{-2})$, with corrections of order $N^{-1/2}=0.1$ at that radius.

The first plot, on the right-hand side, displays the difference with the Schwarzschild gravitational field $\Delta Z$. The discrepancy between our solution and Schwarzschild indeed becomes less than $N^{-1/2}$ at a proper distance from the horizon greater than $\frac{M}{N}$.

The second plot illustrates the electric potential. A residual monopole contribution is clearly manifest, as expected. However, in line with equation \eqref{eq:EMfieldAxisMicroApp}, the asymptotic charge is of order $\frac{M}{\sqrt{N}}$, resulting in the electric potential being, at worst, of order $N^{-1/2}$.

Finally, the last plot corresponds to $G$, which determines the three-dimensional base \eqref{eq:GenericMetricSpher}. The function aligns more closely with the Schwarzschild solution compared to the other plots. This can be attributed to the fact that a single Reissner-Nordström black hole has $G=1$, akin to a Schwarzschild black hole. Hence, our bound state closely resembles a single Reissner-Nordström black hole with an infinitesimal charge of order $\frac{M}{\sqrt{N}}$, with its primary deviation from a Schwarzschild black hole arising from this residual charge. This underscores the reason behind the second class of solutions, which are constructed to have a zero total charge by design, thereby better aligning with the Schwarzschild geometry, as illustrated in Fig.\ref{fig:OutOfAxis10BH} for $N=10$.

As a final observation, it's worth noting that during our investigation, we observed that as the bound states approach extremality ($a=1$), the deviation from Schwarzschild becomes more pronounced. However, it still remains a relatively accurate approximation overall. This implies that a physical quantity scaling as $\frac{1}{N}$ for a solution with small black holes far from extremality will continue to scale as $\frac{1}{N}$ in the near-extremal or extremal case. Nevertheless, the order-one constant is larger when we are closer to extremality.

\bibliographystyle{utphys}      

\bibliography{microstates}       


\end{document}